\documentclass[preprint, 5p, times, twocolumn]{elsarticle}

\pdfoutput=1

\usepackage{amssymb}

\usepackage{algorithm}
\usepackage{algpseudocode}

\usepackage[utf8]{inputenc} 
\usepackage[T1]{fontenc}    
\usepackage{lmodern}        
  
\usepackage[utf8]{inputenc} 
\usepackage{amsmath}
\usepackage{graphicx}
\usepackage{subfigure}
\usepackage{amsmath, amsthm, amssymb, url, color, enumitem}

\usepackage{listings}
\usepackage[colorlinks=true,linkcolor=blue,anchorcolor=blue,citecolor=blue,urlcolor=blue]{hyperref}   

\newcommand{\acronym} {ExaGRyPE}
\newcommand{\peano}   {Peano}
\newcommand{\Peano}   {Peano}
\newcommand{\exahype} {ExaHyPE}

\def\vq{\vec{Q}}
\def\Lquad{\quad \quad \quad \quad \quad}
\def\half{\frac{1}{2}}
\def\tiga{\tilde{\gamma}}
\def\tiA{\tilde{A}}
\allowdisplaybreaks

\hypersetup{
           breaklinks=true,   
           colorlinks=true,   
           pdfusetitle=true,  
        }

\newcommand{\revision}[1]{\textcolor{black}{#1}}
\newcommand{\revisionTwo}[1]{\textcolor{black}{#1}}

\usepackage{xcolor}
\usepackage[draft]{changes}
\definechangesauthor[name=Suggestion, color=olive]{TW}

\journal{Computer Physics Communications}

\begin{document}

\begin{frontmatter}


\title{\acronym: Numerical General Relativity Solvers Based upon the
Hyperbolic PDEs Solver Engine \exahype}

\author[inst1,inst2]{Han Zhang}
\author[inst2]{Baojiu Li}
\author[inst1]{Tobias Weinzierl}
\author[inst3]{Cristian Barrera-Hinojosa}

\affiliation[inst1]{
  organization={Department of Computer Science},
  addressline={Upper Mountjoy, Stockton Rd}, 
  city={Durham},
  postcode={DH13LE}, 
  country={Great Britain}}

\affiliation[inst2]{
  organization={Institute for Computational Cosmology, Department of Physics},
  addressline={Lower Mountjoy, South Rd}, 
  city={Durham},
  postcode={DH13LE}, 
  country={Great Britain}}

\affiliation[inst3]{
  organization={Instituto de F\'{i}sica y Astronom\'ia, Universidad de Valpara\'iso},
  addressline={Gran Breta\~na 1111}, 
  city={Valpara\'iso},
  country={Chile}}

\begin{abstract}
  \acronym\ describes a suite of solvers and solver ingredients for numerical relativity that are based upon \exahype\ 2, the second generation of our Exascale Hyperbolic PDE Engine. Numerical relativity simulations are crucial in resolving astrophysical phenomena in strong gravitational fields and are fundamental in analyzing and understanding gravitational wave emissions. The presented generation of \acronym\ solves the Einstein field equations in the standard CCZ4 formulation under a 3+1 foliation and focuses on black hole space-times. It employs a block-structured Cartesian grid carrying a higher-order order Finite Difference scheme with full support of adaptive mesh refinement (AMR), it facilitates massive parallelism combining message passing, domain decomposition and task parallelism, and it supports the injection of particles into the grid as static data probes or as moving tracers.  We introduce the \acronym-specific building blocks within \exahype\ 2, and discuss its software architecture and compute-n-feel.

For this, we formalize the creation of any specific astrophysical simulation with \acronym\ as a sequence of lowering operations, where a few abstract logical tasks are successively broken down into finer and finer tasks until we obtain an abstraction level which can directly be mapped onto a C++ executable. The overall program logic is fully specified via a domain-specific Python interface, we automatically map this logic onto a more detailed set of numerical tasks, subsequently lower this representation onto technical tasks that the underlying \exahype\ engine uses to parallelize the application, before eventually the technical tasks in turn are mapped onto small task graphs including the actual astrophysical PDE term evaluations, initial conditions, boundary conditions, and so forth. These can be injected manually by the user, or users might instruct the solver on the most abstract user interface level to use out-of-the-box \acronym\ implementations. We end up with a rigorous separation of concerns which shields \acronym\ users from technical details and hence simplifies the development of novel physical models. We present the simulations and data for the gauge wave, static single black holes and rotating binary black hole systems, demonstrating that the code base is mature and usable. However, we also uncover domain-specific numerical challenges that need further study by the community in future work.

\end{abstract}


\end{frontmatter}


\section{Introduction}
\label{sec:intro}

%
%
More than one hundred years ago, Einstein wrote his famous equations of General Relativity \cite{einstein:1916:GR}.
Revealing the intrinsic coupling of space and time, those equations today lay
the foundation of our understanding of the space-time geometry of the Universe.
The Einstein equations have a rather compact mathematical formulation
manifesting in ten highly non-linear tensor equations. Due to their
complexity, analytical solutions are notoriously
difficult to find once the \revision{initial condition of the system} lacks strong symmetries,
i.e.~once we tackle setups of practical relevance.
We thus need numerical techniques.

%
%
The field of numerical relativity has been under active development for many
decades, and numerical simulations are now omnipresent when studying
astrophysical phenomena involving strong gravitational fields.
By recasting the Einstein field equations, numerical relativity reduces
the underlying physics to a Cauchy initial value problem, which
we can solve via standard numerical schemes:
The space-time is cut into a sequence of three-dimensional hypersurfaces, labeled by a global temporal parameter. 
Every hypersurface represents one ``time
step'', which is a snapshot or slice through the actual
space-time~\cite{Rezzolla_Zanotti:2013:Relativitstic_Hydrodynamics}.
Despite being a well-established approach, the number of
mature and actively developed open-source codes available to the community is
limited \revision{(e.g., {\sc BAM}~\cite{BAM}, {\sc gh3d2m}~\cite{gh3d2m}, {\sc Einsteintoolkit}~\cite{einsteintoolkit}, {\sc GRChombo}~\cite{GRChombo}, {\sc SpEC}~\cite{SPEC}, {\sc Dendro-GR}~\cite{Dendro-GR}, {\sc GR-Athena++}~\cite{GR-Athena}, {\sc SENR/NRPy+}~\cite{SENR_NRPY} and {\sc SACRA}~\cite{SACRA}. A more comprehensive list can be found in Table 1 of \cite{nr_code_overview}).}

\begin{figure*}[htb]
 \begin{center}
  \includegraphics[width=0.8\textwidth]{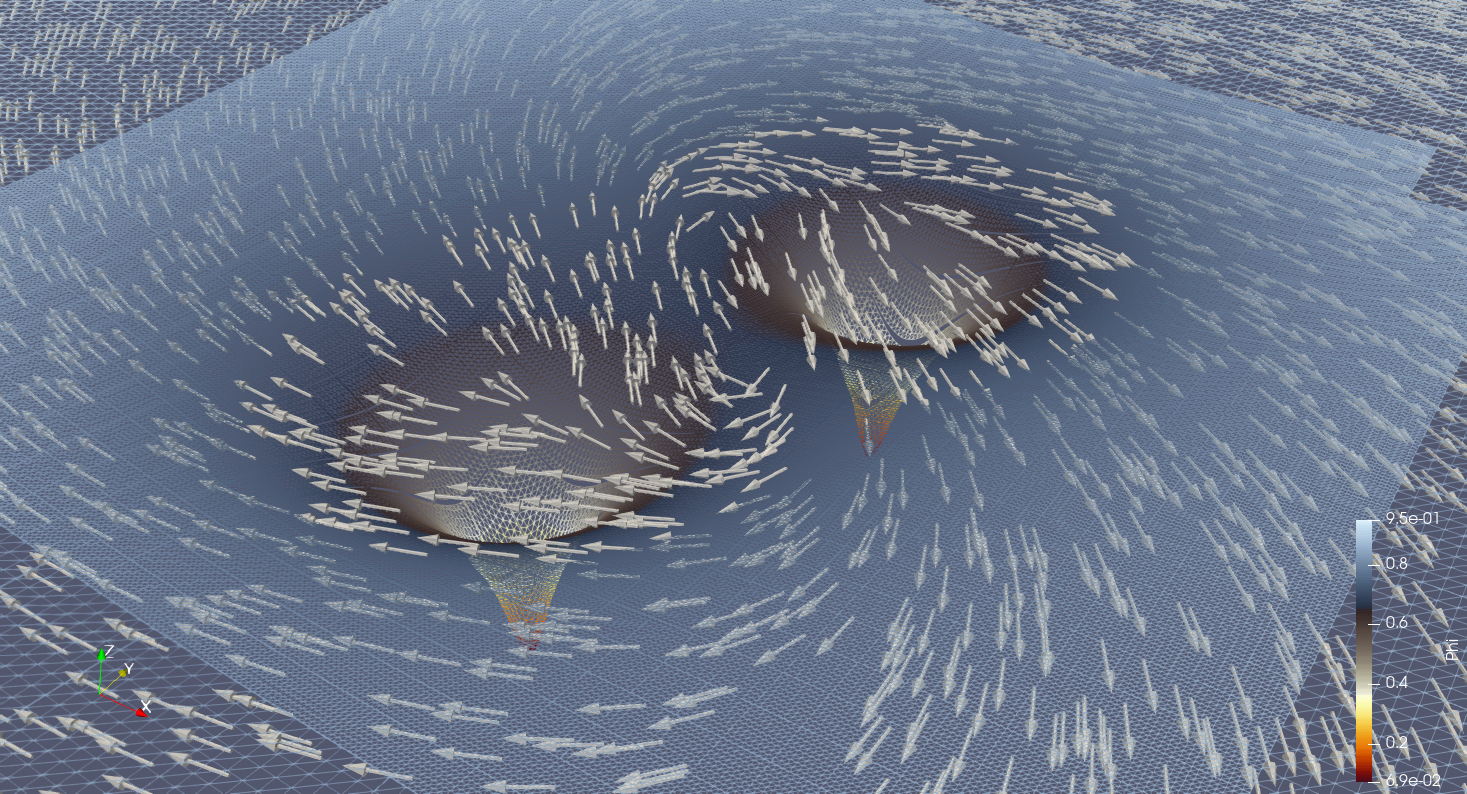}
 \end{center}
 \caption{
  Snapshot of two black holes orbiting around each other. \revision{The colored frame represents the conformal factor field $\phi$, while the arrows show the distribution of the shift vector field $\beta^i$.}
  \label{figure:introduction:two-black-holes}
 }
\end{figure*}
%
%
The software packages share some similarities and have to tackle similar
technological and numerical challenges.
First, numerical relativity codes have to be able to simulate a vast span of
spatial scales (spatial is to be read in the sense of numerical discretization of a
space-time snapshot).
They have to simulate over a huge computational domain, while some
features---around black holes---require a very detailed spatial resolution
(Figure~\ref{figure:introduction:two-black-holes}).
At the same time, explicit time-stepping remains the state-of-the-art in the
field,
even though 
many astrophysical phenomena of interest such as the merger of
black holes require us to simulate them over a long time span while explicit
time-stepping yields tiny time-step sizes.
Second, \revision{numerical relativity codes usually cast Einstein's equations into hyperbolic formulations under proper gauge choices,} and its combination with many nonlinear terms implies that we have to pick our numerical techniques carefully to avoid numerical instabilities and
numerical inaccuracies introducing long-term pollution of the results.
Higher-order methods thus have established themselves as state-of-the-art,
while several further ingredients, such as appropriate boundary conditions, have
to be chosen carefully to fit the experimental challenge.
Finally, the complex equations lead to complicated compute kernels
involving potentially thousands of lines of code.
The translation of PDE terms into code is not simple.
As a result of these three characteristics, competitive numerical relativity
codes have to combine sophisticated software technology, e.g., good (strong) scalability plus efficiency, with advanced numeric and a clear
strategy how to master the arising code complexity.

%
%
In this paper, we present \acronym\ (The General Relativity Solver on \exahype).
It currently contains a suite of numerical relativity solvers and related code blocks focusing on black hole spacetimes.
\acronym\ is based upon a from-scratch rewrite, i.e.~a second generation of
\exahype\ (An Exascale Hyperbolic PDE Engine)~\cite{ExaHyPE1}, which in turn uses a complete rewrite of its
underlying meshing framework Peano~\cite{peano}.
Besides the core solvers, which are really just instances of \exahype\ solvers,
\acronym\ adds many ingredients to \exahype,
and it compasses domain-specific utilities.
Astrophysical simulations have been conducted before with the first-generation
\exahype~\cite{Dumbser18}.
However, \acronym\ is a complete rewrite, brings all ingredients together to
simulate systems of black holes, and it comprises previously unavailable
numerical ingredients.
It also implements a new, rigorous separation of concerns.

The evolving system adopted in our code is a recast version of
CCZ4 (Conformal and covariant formulation of the Z4)~\cite{Alic12}, which transforms the equations into a pure first-order
formulation via a set of auxiliary variables. 
This idea is first proposed in \cite{Dumbser18}, serving as the foundation for
our code. Besides this original formalism where auxiliary
variables are introduced into the evolving system alongside the primary
variables, \acronym\ comes along with a second flavor: 
here, auxiliary variables are derived from the primary
variables in a post-processing stage after every time step.
This feature enables users to implement a second-order formulation similar \revision{to many other numerical relativity codes}.
With lower-order and higher-order implementations of both formulations at hand,
\revision{in \acronym\, we implement both lower-order and higher-order schemes (solvers) for both formulations, they provide the infrastructure to investigate the numerical details and modifications of the underlying physics.}
We even provide an ecosystem to couple various solvers, \revision{allowing us to utilize different schemes according to the qualitative features of the solution.}

%
%
\acronym's solvers combine several state-of-the-art numerical building blocks
realized over patch-based AMR~\cite{Dubey:16:SAMR}, but it
rigorously strips users from the opportunity to ``program'' their workflow. 
Instead, we introduce a high-level, Python-based API (Application Programming Interface) in which users specify
``what'' they want to compute.
Our code then maps this specification onto a C++ code base, i.e.~determines
``how'' this is realized.
This strict separation-of-concern follows the philosophy introduced with Peano~\cite{peano} and adopted by the first-generation ExaHyPE code, too
\cite{ExaHyPE1}.
Eventually, the abstract specification yields one relatively simple C++ base
class, into which users can inject their actual PDE terms.
\acronym\ provides pre-defined realizations for them.
We refer to this mapping of a high-level specification onto a running code as
\emph{lowering}~\cite{lattner:2020:MLIR}.
As a whole, solvers built with \acronym\ differ from other mainstream codes in
the field:
at no point can the user control the simulation flow or orchestrate
calculations; at no point, do users have the opportunity to loop over resolution levels or
mesh entities, and at no point can they  
decide how to distribute tasks between ranks or
accelerators, and so forth.
These decisions are made by the actual engine under the hood.
Our present contribution formalizes
this transition from the specification into a ready-to-run simulation as a
sequence of lowering steps yielding a hierarchy of abstraction levels.
Each level serves a particular purpose or represents a particular
knowledge domain---numerical schemes or parallelization for example---and each
level can in itself be represented as a task graph.

%
%
Our driving vision behind this software design is that physicists can puzzle
their application together on a very abstract level, i.e.~pick solvers,
domain size, simulation timespan, minimal resolution, and so forth. 
We favor Python on this level.
Scientific computing experts enlarge the set of available components where
appropriate and necessary.
Our own \acronym\ development work for example contributes higher-order Finite
Difference (FD) schemes, novel GPU-focused task
parallelism~\cite{loi:2023:sycl,wille:2023:gpuoffloading} or bespoke boundary conditions.
Performance experts finally tune and optimize
various code building blocks on the appropriate abstraction
level (cmp.~\cite{Li:22:ISC}, e.g., for an example in the context of \acronym).
We consider this strict separation of responsibilities or
roles~\cite{Gallard:2020:RoleOrientedCodeGeneration} and the hiding of the program
complexity a blueprint for future simulation software stacks.
%
%

The remainder of this paper is organized as follows: 
Section~\ref{sec:theory} introduces the physics scenario we aim to simulate, served as an overview of the domain challenge driving
the development of \acronym.
In Section~\ref{sec:code}, we describe the technical ingredients that are newly
developed for \acronym\ and are essential for achieving stable and accurate
simulations in numerical relativity.
We continue with an in-depth discussion of the lowering of abstraction levels 
and highlight how they enable us to
deliver fast and efficient code, before we summarize the overall code usage,
i.e.~the user's look-n-feel (Section~\ref{sec:tasking-decomposition}).
Numerical results (Section \ref{sec:result}) demonstrate the validity of the
approach and allow us to close the paper with a summary of insights and
identified future work and challenges (Section \ref{sec:conclusion}).
An extensive appendix provides reproducibility information, details regarding
the underlying physics and implementation remarks.

\section{Problem Statement}
\label{sec:theory}

The space-time evolution equations used in \acronym\ stem from the
Z4 formulation of the Einstein field equations, where an additional dynamic field $Z_a$ and a corresponding damping term are
added to the system to enhance its stability (see \ref{app:spacetimeFoliation} for the convention)~\cite{bona:03:Z4original,
Gundlach:2005:Z4damping}:
\begin{flalign}
    R_{ab}-&\frac{1}{2} g_{ab}R+\nabla_{a} Z_{b}+\nabla_{b} Z_{a}-g_{ab}\nabla^c
    Z_c \\
    &-\kappa_{1}[n_{a} Z_{b}+n_{b} Z_{a}+\kappa_{2} g_{ab} n_{c}
    Z^{c}]  =  8\pi T_{ab}. \nonumber
\end{flalign}

\noindent
To solve these equations, our code uses the standard ADM (Arnowitt–Deser–Misner) 3+1 space-time
foliation \cite{arnowitt:2008:ADM}. It cuts the space-time into a sequence of three-dimensional space-like
hypersurfaces labeled by a global scalar field $t(x^a)=\text{const}$. The
value of this scalar field is later treated as the time parameter $t$ in the
3+1 simulation. The induced spatial metric on the hypersurfaces is denoted by $\gamma_{ab}$,
and it is linked to the space-time line element via
\begin{equation}
    d s^{2}=-\alpha^{2} d t^{2}+\gamma_{i j}\left(d x^{i}+\beta^{i} d t\right)\left(d x^{j}+\beta^{j} d t\right),
\end{equation}

\begin{figure}[htb]
 \begin{center}
  \includegraphics[width=0.35\textwidth]{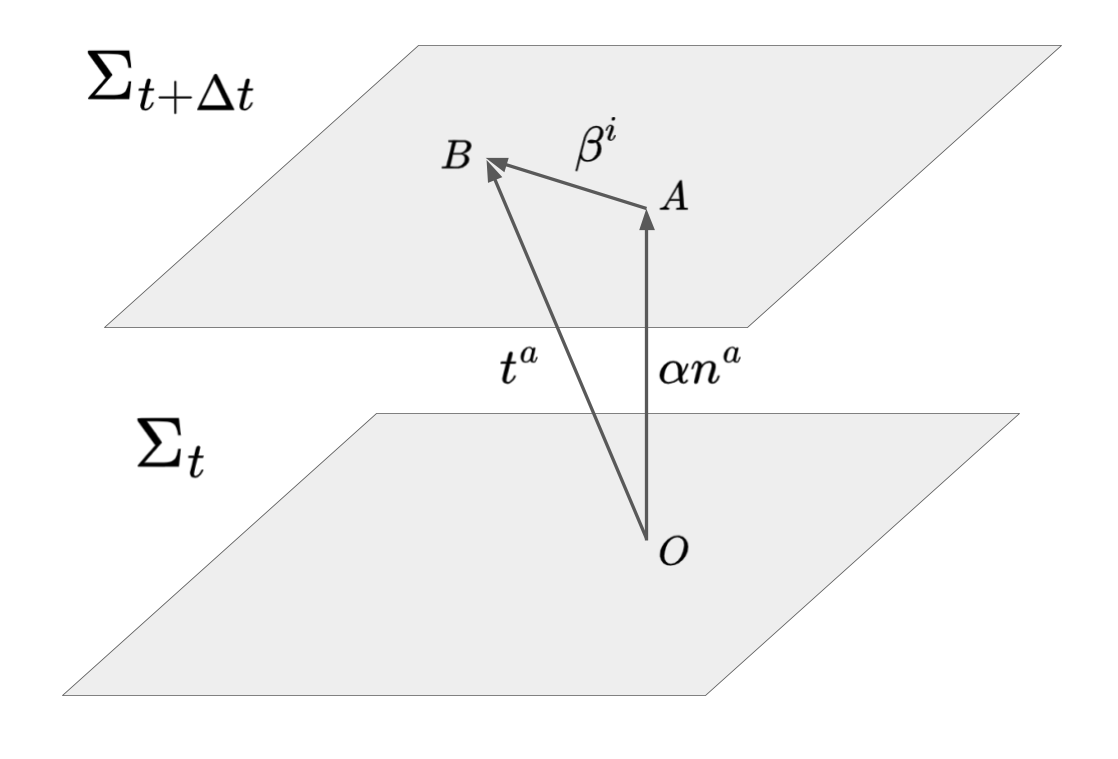}
 \end{center}
  \vspace{-0.8cm}
 \caption{
  Illustration of the lapse function and shift vector. Given two
  neighbouring hypersurfaces labeled as $\Sigma_t$ and $\Sigma_t+\Delta t$, the
  four-dimensional time vector $OB$ can be decomposed into $OA$ and $AB$, which
  yield the proper time $\alpha$ passed for a normal observer at O in the
  coordinate time $\Delta t$, and the coordinate shift between the two hypersurfaces. 
  $B$ and $O$ have the same spatial coordinates while the spatial coordinates of
  $A$ is different by $-\beta^i$.
  \label{figure:theory:lapse_shift}
 }
\end{figure}

\noindent
where the $\gamma_{ij}$ are the spatial parts of $\gamma_{ab}$.
$\alpha$ and
$\beta^i$ are respectively the lapse function and shift vector, representing the four
degrees of freedom of coordinates during the evolution
(Figure~\ref{figure:theory:lapse_shift}).

One can construct simulations with a formulation only including $\gamma_{ij}$, the spatial component $K_{ij}$ of the extrinsic curvature
$K_{ab}:=-\frac{1}{2}\mathcal{L}_{\mathbf{n}} \gamma_{ab}$, $\alpha$,
$\beta^{i}$ and the $Z^a$ field, as they already form a complete evolving system.
\revision{We further follow \cite{Alic12} and introduce a conformal factor $\phi$ and a conformal metric
$\tilde{\gamma}_{ij}$, such that}

\begin{equation}
    \label{eq_NR_conformal_decomposition_gamma}
    \tilde{\gamma}_{ij}:=\phi^2 \gamma_{ij},~\phi=[\det(\gamma_{ij})]^{-1/6},
\end{equation}

\noindent
and further define $\tiA_{ij}$, $K$, $\Theta$, $\hat{\Gamma}^i$ following a similar approach in the original formulation.
Further details can be found in~\ref{app:spacetimeFoliation}. \revision{We note that this formulation is just one of several candidates for a strongly hyperbolic evolution system. Other options include the generalized-harmonic formalism \cite{general_harmonic_gauge1, general_harmonic_gauge2}, the BSSN formalism \cite{BSSN1, BSSN2}, and other variants within the ``Z4 family'', such as the original Z4 \cite{hyperbolic_z4} and Z4c \cite{z4c}. For \acronym, we have chosen to utilize CCZ4 to achieve better control over singularities and gauge conditions in black hole spacetimes.}

\subsection{Gauge Conditions}
\label{sec:theory:gauge}

Our four degrees of freedom are determined by the evolution
equations of the lapse field $\alpha$ and shift vector $\beta^i$, which are
called gauge conditions. \acronym\ allows users to define their bespoke gauge
conditions, but common choices for the evolving system
above are available off-the-shelf.

The default evolution equation of $\alpha$, which is also known as the
\emph{time slicing} condition \cite{bona:1995:gauge}, is
\begin{equation}
    (\partial_t-\beta^i \partial_i )\alpha=-\alpha^2 g(\alpha) K,
\end{equation}

\noindent
where $g(\alpha)$ is a positive scalar function depending only on the lapse. In
simulations of black hole spacetimes, a popular choice of this function is
$g(\alpha)=2/\alpha$, as it has a good property of avoiding singularities. The resulting
gauge equation is the \emph{1+log} condition given as
\begin{equation}
    \partial_t\alpha = \beta^k A_k-\alpha^2 g(\alpha)\left(K-K_0-2c\Theta\right)
\end{equation}

\noindent
in our notations. The source term gets modified accordingly as we implement the CCZ4 formulation~\cite{Alic12}.

The default gauge equation we adopt for the shift vector $\beta^i$ in \acronym\ is the \emph{gamma-driver} condition~\cite{alcubierre:2003:gauge}:
\begin{eqnarray}
    \partial_{t} \beta^{i} &=& \beta^{k} B_{k}^{i}+ f b^{i},
    \\
    \partial_{t} b^{i}- \beta^{k} \partial_{k} b^{i} &=& \partial_{t} \hat{\Gamma}^{i}-\beta^{k} \partial_{k} \hat{\Gamma}^{i}-\eta b^{i}.
\end{eqnarray}

\noindent
Here, $b^i$ is a helper vector field that enhances the hyperbolicity of the system and also appears in the evolving system, cf.~Eqs.~\eqref{equation:theory:second-order-variables} and \eqref{equation:theory:first-order-variables}. 
This gauge is also called \emph{shifting shift}.
One can
remove all advection terms above to switch to the \emph{no-shifting shift}
variant, which is also available in \acronym.

\subsection{First- and Second-order Formulations}

Collecting all derived quantities and the helper variables $b^i$ from the gauge
condition yields an evolving system over 24 variables
\begin{equation}
  \vq(t) = 
  \left( \tilde{\gamma}_{ij}, \alpha, \beta^i, \phi, \tilde{A}_{ij},K, \Theta,
  \hat{\Gamma}^i, b^i \right)(t),
  \label{equation:theory:second-order-variables}
\end{equation}

\noindent
with a first-order time-derivative and second-order spatial derivatives defined in the 
hypersurface. Let
\begin{equation}
   A_{i}:=\partial_{i} \alpha, \ B_{k}^{i}:=\partial_{k} \beta^{i}, \ D_{k i
   j}:=\frac{1}{2} \partial_{k} \tilde{\gamma}_{i j}, \
   P_{i}:=\partial_{i} \phi
   \label{equation:theory:auxiliary-variables}
\end{equation}

\noindent
be auxiliary variables. Adding them to the system and treating them as further
evolving variables yields a system of 58 independent variables with 
\begin{equation}
  \vq(t) = 
  \left(
   \tilde{\gamma}_{ij}, \alpha, \beta^i, \phi, \tilde{A}_{ij},K, \Theta, \hat{\Gamma}^i, b^i, A_k, B^i_k,
    D_{kij}, P_k \right).
  \label{equation:theory:first-order-variables}
\end{equation}

\noindent
This system is first-order in both space and time. 
Both formulations can be phrased over 58 quantities.
The first-order formulation (FO) from Eq.~\eqref{equation:theory:first-order-variables}
evolves all quantities.
The second-order formulation (SO) Eq.~\eqref{equation:theory:second-order-variables}
evolves only 24 out of 58 entries, then computes (or ``reconstructs'')
Eq.~\eqref{equation:theory:auxiliary-variables} from the new solution in a
post-processing step, and continues with the next time step which is fed by both
the primary and the reconstructed auxiliary quantities.

As a result, both the second-order and first-order formulations can be schematically written 
in a first-order hyperbolic formulation of partial differential equations (PDEs):
\begin{eqnarray}
  \partial_t \vq +\nabla_i F_i(\vq) +B_i (\vq) \nabla_i \vq
  =S(\vq),
\label{equation:theory:pde}
\end{eqnarray}

\noindent
with $\vq: \mathbb{R}^{3+1} \mapsto  \mathbb{R}^{n}$, $n=24$ in SO or $58$ in FO.
We can use the same numerical building blocks subject to an additional
reconstruction step for Eq.~\eqref{equation:theory:auxiliary-variables} for both
formulations.
In Eq.~\eqref{equation:theory:pde}, $\vq$ represents the array of 58 evolving
variables.
$F_i(\vq)$, $B_i(\vq)$ and $S(\vq)$ are the flux, non-conservative product (NCP) and source term respectively. 
In the implementation, we let the NCP absorb the flux term and hence have
$\nabla_i F_i(\vq)=0$ (cmp.~\ref{app:spacetimeFoliation} for details on the
numerics and physics).

\subsection{Initial Conditions}
\label{sec:theory:scenario}

For simple test scenarios where an analytical solution exists, \acronym\ can be
initialized by directly setting every evolving variable to a hard-coded initial
value.
We use this approach for the gauge wave setup, for example.

For systems involving black holes, we adopt the \emph{Bowen-York solution} \cite{bowenYork:1980:BHIC} to initialize the scenario. 
It assumes an initial space-time hypersurface with maximal slicing $K=0$ and
conformal flatness $\bar{\gamma}_{ij} = \eta_{ij}$. The curvature on this
hypersurface can be calculated analytically by the momentum $P^i$ and spins
$S_i$ of the black holes. The initial metric of the black hole system is then
derived from the \emph{moving puncture approach} \cite{brugmann:2008:movingP},
where a second-order elliptic equation is solved to obtain the conformal factor
numerically. The current version of \acronym\ links against 
the \texttt{TwoPunctures} module from the numerical library {\sc
Einsteintoolkit} \cite{einsteintoolkit}, using the numerical
approach introduced in \cite{ansorg:2004:solvingPuncture}.

\subsection{Gravitational Wave Extraction}
\label{sec:theory:GW}

\acronym\ extracts the gravitational wave signal from the evolving system using
the \emph{Newman-Penrose approach}~\cite{Newman:2004:Psi4}:
we calculate the complex scalar field $\psi_4$ as
\begin{equation}
    \label{eq_NR_psi4_from_weyl}\psi_4= ^{(4)}C_{abcd} k^a \bar{m}^b k^c \bar{m}^d.
\end{equation}

\noindent
In this expression, $^{(4)}C_{abcd}$ is the four-dimensional Weyl tensor, and
$k^a$ and $\bar{m}^a$ are two members of a null tetrad (\ref{app:psi4}).
$\psi_4$ can be further decomposed into a superposition of modes with the base $s=-2$ spin-weighted spherical harmonics ${}_{-2}Y_{lm}$:
\begin{equation}
    \psi_4(t,r,\theta,\phi)=\sum^\infty_{l=2} \sum^l_{m=-l}\psi^{lm}_4(t,r){}_{-2}Y_{lm}(\theta,\phi),    
\end{equation}

\noindent
where the coefficients (the mode strength) are found by the inner product via spherical integrals
\begin{equation}
    \psi^{lm}_4=\int^{\pi}_{0}d\theta \int^{2\pi}_{0} { }_{-2} Y^*_{lm} \psi_4 \sin \theta d\phi.
\end{equation}

\noindent
In \acronym, the spherical integration is done by the spherical t-design
scheme~\cite{brauchart:2015:tdesign}, \revision{where we interpolate the values of the integrated function from the closest grid points onto a set of $N$ sample points on the sphere of interest}. It is shown that the average of those sample values is also the average of the integrated function itself, i.e.
\begin{equation}
    \int^{\pi}_{0}d\theta\int^{2\pi}_{0} { }_{-2} Y^*_{lm} \psi_4 \sin \theta d\phi=
    \frac{4\pi}{N}\sum^{N-1}_0 { }_{-2} Y^*_{lm} \psi_4.
    \label{eq:tdesign_spherical_integral}
\end{equation}

\noindent
The method is most effective when the function of interest has a degree of
freedom that is close to or lower than $t$ (i.e., it can be roughly described by a polynomial up to the degree of $t$). $N$ needs to be increased if we want higher accuracy by raising $t$. \revision{In the current version of \acronym, the default set of sample points is $t=43$ with $N=948$ from \cite{womersley:2015:tdesigndata} for a balance of accuracy and performance. In practice, sets of different orders can be selected in the code, based on the accuracy requirements of individual simulations.}

\section{Simulation Building Blocks}
\label{sec:code}

\begin{figure*}[ht]
  \begin{center}
    \includegraphics[width=0.8\textwidth]{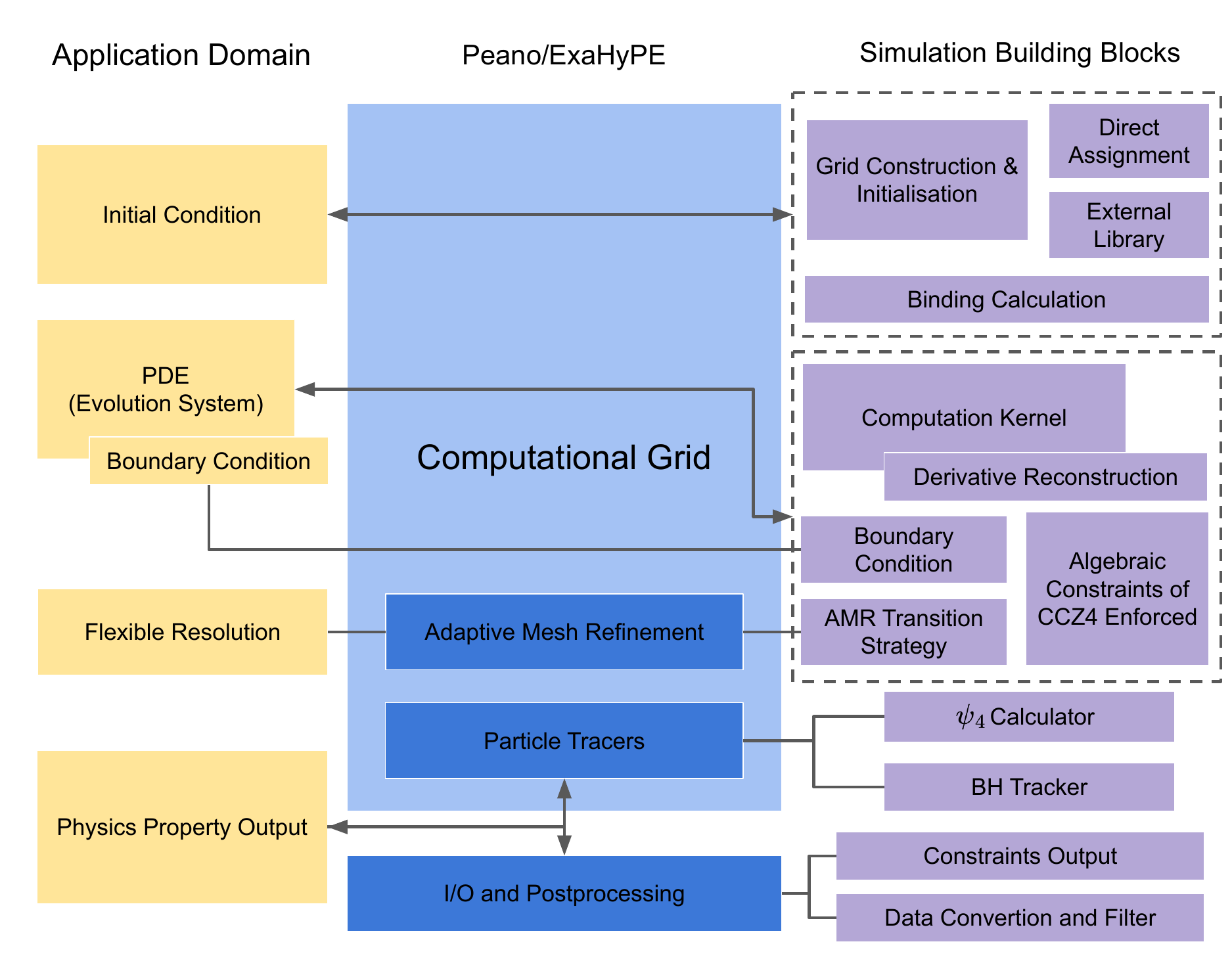}
  \end{center}
  \caption{
    The overview of the code ingredients of \acronym, showing how the physics
    problem (application domain) is translated into the simulation code (building blocks), bridged by the code base \Peano\ and \exahype: the targeted physics system can be seen as a Cauchy initial value problem with a set of PDEs, an initial
    condition and a boundary condition. They are deployed on a computational
    grid following certain spatial discretization schemes and are handled by
    different code modules. Our code supports adaptive mesh refinement to achieve flexible resolution. It also provides output of standard snapshot I/O with filtering and particle tracer features, the latter is utilized as data probes.
    \label{figure:building-blocks:exagrype_structure}
  }
\end{figure*}

\acronym\ is based upon a rewrite of \exahype. 
To the user, an \acronym\ solver is a Python script which creates an \exahype\
solver and uses bespoke ingredients from \exahype.
All of these building blocks
designed for \acronym\ have been integrated into the \exahype\
software base.
They are hence available to other solvers from other application domains, too.
However, they uniquely map the requirements arising from astrophysical challenges
onto code building blocks
(Figure~\ref{figure:building-blocks:exagrype_structure}).

\subsection{Spatial Discretisation}

As \exahype\ solvers, \acronym\ relies on a block-structured Cartesian mesh \cite{Dubey:16:SAMR}
constructed through a spacetree formalism \cite{Weinzierl:2011:Peano}:
the whole (three-dimensional) domain is embedded into a single cube and split
into three equal parts along each coordinate axis. 
This yields $3^3=27$ smaller cubes. 
We continue recursively, i.e., decide for each cube whether to refine it into 27
subcubes again.
Our code base provides a plug-in point for users to guide the refinement.
The process yields an adaptively refined Cartesian grid of cubes.

\begin{figure}[htb]
  \begin{center}
    \includegraphics[width=0.4\textwidth]{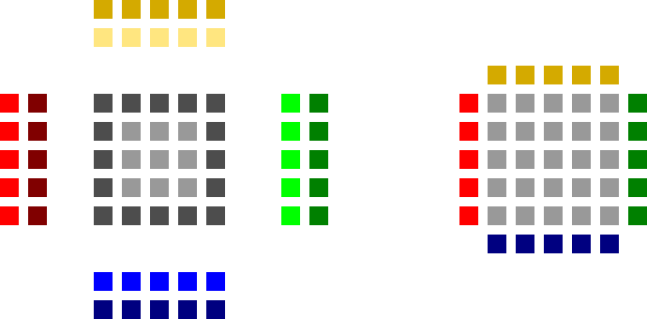}
  \end{center}
  \caption{
    Two-dimensional, schematic illustration of patch and face data arrangements
    in \exahype\ 2:
    A two-dimensional patch with $5\times 5$ mesh elements (volumes) and a halo
    of width $K=1$ are used. 
    Left: The faces host the halo data of both adjacent patches.
    Right: Prior to the kernel invocation, we merge the cell patch data and the
    data from the $2d$ adjacent faces into one $(5+2) \times (5+2) $ patch
    augmented with its halo.
    \label{figure:building-blocks:patch-face-reconstruction}
  }
\end{figure}

Every cube in this grid hosts a small Cartesian mesh (\emph{patch}).
Every patch consists of $p \times p \times p$ mesh elements (again small cubes),
and each element holds either a piecewise constant solution of the evolving
system, i.e. defines one Finite Volume, or we read its center as a sample
point of the solution in a Finite Difference sense.
$K$ additional layers of elements are logically attached to the faces of the
patches (Figure \ref{figure:building-blocks:patch-face-reconstruction}).
Those extra layers are called \emph{halos}. 
They provide data from neighboring patches which are needed to update the solution in the current patch, thus allowing the latter to advance one step in time without any further data input.
The patch carrying the actual evolving solution is hence the minimal atomic unit
of computation and represents the finest discretization level.

Technically, faces hold copies of the closest volumes from both
adjacent patches.
They hold data from $ 2K \times p \times p $ mesh elements.
Prior to the update of a cell, we glue the respective half of the face data and
patch data together to obtain our logic patch-plus-halo input data.
After the update, we write the new relevant data back into the $2d$ ($d$ is the number of space dimensions)
adjacent faces of a patch.

Each solver picks its $K$ such that it can update its cell elements
independently of all neighbor patches.
For the traditional 3-point stencil, $K=1$ is sufficient.

\subsection{Numerical Solvers}
\label{sec:code:solver}

The update of a patch is performed by a \emph{compute kernel}.
\exahype's baseline offers a generic Finite Volume (FV) kernel employing a
Rusanov Riemann solver with explicit time stepping, implementations of Runge-Kutta
Discontinuous Galerkin (RKDG) solvers, and ADER-DG (arbitrary-high-order-method-using-derivatives DG).
For \acronym, we add a fourth-order Finite Differences
(FD4) scheme that can be combined with Runge-Kutta (RK) time stepping to fourth-order, too.
It mirrors state-of-the-art solvers from the field (cmp.~{\sc GRChombo}).
We use exclusively FD4 in \acronym, while FV is also used in some benchmark and test implementations, as well as other applications based on \exahype\ (e.g. \cite{zhang:2022:ssinfall}).

\paragraph{Finite Volumes (FV)}

Our FV solver is a straightforward implementation of a generic Rusanov scheme in an explicit Euler time integrator:
each volume is updated according to the flux through its six adjacent volume interfaces.
Let each volume host a constant function of $Q \in \mathbb{R}^{n}$.
The time stepping scheme of FV from Eq.~(\ref{equation:theory:pde}) can be written as

\begin{equation}
    \vq(t+\delta_t)=\vq(t)+\delta t S(\vq) + \sum_{i \in d}\text{Flux}^{\pm}_{i}(\vq)\Big| _{\partial v},
\end{equation}

\noindent
with the flux from the corresponding two faces in $i$ dimension:

\begin{eqnarray}
  \text{Flux}_i^-  &= & \frac{\delta t}{\delta x} \Bigg[
    -\half \left(\vq-\vq^- \right)B_i \left(\half (\vq^+ + \vq^-) \right) 
    \\
    & - &\eta \ \text{max}
    \left(\lambda_\text{max}(\vq),\lambda_\text{max}(\vq^-) \right) \left(\vq-\vq^- \right)\Bigg],\nonumber
      \\
  \text{Flux}_i^+ & = & \frac{\delta t}{\delta x} \Bigg[
    -\half \left(\vq^+-\vq^- \right)B_i \left(\half (\vq^+ + \vq^- ) \right) \\
    & +  &
    \eta\ \text{max} \left(\lambda_\text{max}(\vq^+),\lambda_\text{max}(\vq)
    \right) \left(\vq^+ -\vq^- \right)\Bigg], \nonumber
\end{eqnarray}

\noindent
where $B_i$ is the NCP term along the $i$th dimension,$\vq^+$ and $\vq^-$ are the volume solutions left and right of the considered volume, $\delta x$ is the volume size, and $\eta$ is a problem-specific constant with a default value of $0.5$. We have used $\partial_t \vq = \left(\vq(t+\delta t)-\vq(t)\right)/\delta t$ as the Euler time integration scheme. This yields a 7-point stencil coupling each volume to its face-connected neighbors: $K=1$ is sufficient to realize this scheme, and we need the \acronym\ specification to provide a function that evaluates the
non-conservative product $B_i$ for a given input, as well as one that returns the
maximum eigenvalue $\lambda _{\text{max}}(\vq)$ along a coordinate axis for the flux calculations. 
For the current CCZ4 equations, \acronym\ follows the conservative upper estimate for this eigenvalue from the work \cite{Dumbser18}.

\paragraph{Finite Differences (FD4)}

To implement the fourth-order finite difference solver in \exahype, \acronym\ samples the solution in the center of the mesh elements. Therefore, the FD4 scheme can use the same storage scheme as the FV scheme besides a larger ($K=2$) halo.

On this grid over sample points, we use central finite differences with a ``tensor product'' approach. For $K=1$, we could construct a seven-point stencil by cutting the Taylor expansion's second-order term. In \acronym, we adopt the fourth-order accuracy scheme which requires $K \geq 2$ (we use $K=3$ \revision{in the practice of the code} for KO dissipation, see below). Its time-stepping scheme is 
\begin{align}\label{eq_exa_fd4_ts}
    \vq(t+\delta t)&=\vq(t)+\delta t  S(\vq) 
    - \\-
    &\sum_{i \in d} B_i \frac{\delta t}{12 \delta x_i}\left[8\left(\vq^{i+}-\vq^{i-}\right)-\left(\vq^{i++}-\vq^{i--}\right)\right], \nonumber 
\end{align}
where the $i+$ and $i-$ upper indices of quantities indicate the relative position of the indexed quantities with the current grid point along $i$ direction, and the number of signs specifies the distance. Again we have used $\partial_t \vq = \left(\vq(t+\delta t)-\vq(t)\right)/\delta t$. As all the approximation happens within the truncation of the Taylor expansion, it is sufficient for the user code to provide an implementation of $B_i$ to the compute kernel. As our default kernel of \acronym, readers are referred to \ref{app:solver} for more details on the FD4 solver. 

The fourth-order differences can be combined with an arbitrary-order time step
integrator.
\acronym\ indeed provides Runge-Kutta schemes up to order four, where we would
naturally hit the Butcher barrier.
However, by default, we use a first-order Runge-Kutta scheme, i.e.~an
explicit Euler as higher-order time integration yield to be non-economical given its extra computational cost, see Section \ref{sec:result:gw}.

\paragraph{Time step size calculation}

The admissible time step size of all explicit schemes is subject to the
Courant–Friedrichs–Lewy (CFL) condition/constraint with
\begin{equation}
 \label{eq_exa_cfl_condition}
 \Delta t < C \frac{|\delta x|}{(2R-1)\lambda_{\rm max}},
\end{equation}

\noindent
where $C<1$ is a user-specified safety parameter that may vary in different simulations and $R$ is the order of the Runge-Kutta scheme ($R=1$ for Euler). 
Our scheme employs a global time-stepping scheme and thus uses the smallest
global volume length $|\delta x|$ to guide Eq.~(\ref{eq_exa_cfl_condition}). For all present experiments, it remains invariant over time as we fix the finest resolution (static refinement) in our simulations which naturally clusters
around the black hole.

The time step size calculation again requires the solver to evaluate $\lambda_{\rm
max}$ over each patch after it has been updated and we continue to use the same evaluation as we introduced in the Finite Volume solver above. This quantity encodes the fastest propagation speed of the quantities
in our system.
As it changes over time we have to restrict it globally and adopt the 
time step size accordingly.

\paragraph{Kreiss-Oliger damping}

Astrophysical solutions with explicit time stepping are notoriously vulnerable
to numerical instabilities.
It therefore is helpful to involve a penalty term in the evolution equation, as it smooths out solutions affected by numerical inaccuracies in certain regions of the domain, such as those near physical or AMR boundaries.

This smoothing is often done using the so-called Kreiss-Oliger (KO) dissipation \cite{kreiss:1973:KOterms}, which traditionally
evaluates a higher-order term (a ``fourth-order curvature'' for example) for second-order PDEs. Its order needs to exceed that of the kernel updating scheme to avoid introducing larger numerical errors in smoothing. As we work with the fourth-order accuracy in FD4, \acronym\ employs the KO term with $N=3$ (i.e., the term use value of cells up to three points away) in the code and makes this feed into our evaluation equation \eqref{eq_exa_fd4_ts} with a user-defined parameter to suppress
oscillations (\ref{app:solver}).
The KO term can be calibrated or switched off upon demand.

\paragraph{Explicit gradient reconstruction}

In the first-order formulation, all 58 variables of Eq.~\eqref{equation:theory:first-order-variables} are
updated by the numerical scheme on a patch (Algorithm \ref{alg:fd4-fo}).
In the second-order formulation, however, only the primary variables Eq.~\eqref{equation:theory:second-order-variables} are evolved. After that, the remaining auxiliary variables
Eq.~\eqref{equation:theory:auxiliary-variables} are calculated from those primary variables.
We update the solution, compute the auxiliary variables, i.e., their gradients,
in a postprocessing step, and feed them into the subsequent time step again. With this scheme, we can realize a second-order formulation even though all PDEs are written down as first-order
(Algorithm \ref{alg:fd4-so}).
For the auxiliary variables, we use a central fourth-order finite difference scheme again \eqref{eq_exa_fd4_central_difference}.

\subsection{Domain Decomposition}

\exahype\ splits the computational domain spanned by the spacetree along the
\Peano\ space-filling curve (SFC) \cite{peano}. 
The embedded patches are atomic, i.e.~never split.
Segments along the SFC then can be deployed among ranks and threads, i.e., we
facilitate domain decomposition between ranks and on ranks.
It is a non-overlapping domain decomposition.

Non-overlapping here is to be read logically:
Each volume or degree of freedom within any patch has a unique owner, i.e., a unique thread on a unique rank that is responsible for updating it. The patch data replicated within the faces establish an effective overlap of $2K$.

The copy from patch data into faces happens after every time step. We use this copied data in the neighboring patch in the subsequent mesh sweep. Therefore, no direct, synchronous data exchange of face data is required. Instead, we can send out data in one mesh traversal and receive it prior to the subsequent one. With this non-blocking data transfer, our code allows for some overlap of communication and computations. 

\begin{algorithm*}[htb]
  \caption{
    Time stepping scheme (explicit Euler) for the fourth-order finite difference
    solver of \acronym\ in the first-order formulation.
    \label{alg:fd4-fo}
  }
  \begin{algorithmic}
    \For {each patch in Domain}
    \For {each volume in patch}
    \State compute source term $S(\vq)$
    \State $RHS_{\vq} \leftarrow \Delta t \cdot S(\vq)$
    \For {$i=x,y,z$}
        \Comment Run over all elements in the patch
    \State compute fourth-order Finite Difference $(\Delta \vq)_i$
        \Comment Eq. \eqref{eq_exa_fd4_central_difference}
    \State compute non-conservative product $B_i(\vq)(\Delta \vq)_i$
    \State compute KO term $(KO_{\vq})_i$ 
        \Comment Eq. \eqref{eq_exa_fd4_ko}
    \EndFor
    \State $RHS_{\vq} \leftarrow RHS_{\vq}-\sum_i \Delta t \cdot B_i(\vq)(\Delta \vq)_i$
    \State $RHS_{\vq} \leftarrow RHS_{\vq}+\sum_i \Delta t \cdot (KO_{\vq})_i$
    \State $\vq \leftarrow \vq+RHS_{\vq}$
    \State $t \leftarrow t+\Delta t$
    \State compute $\lambda_{max}$ to inform next time step
    \EndFor
    \EndFor
  \end{algorithmic}
\end{algorithm*}

\begin{algorithm*}[htb]
  \caption{
    Time-stepping scheme with an explicit Euler for the fourth-order finite
    difference solver of \acronym\ in the second-order formulation.
    \label{alg:fd4-so}
  }
  \begin{algorithmic}
    \For {each patch in Domain}
    \For {each volume in patch}
    \State compute Source $S(\vq_{pri})$
    \State $RHS_{\vq_{pri}} \leftarrow \Delta t \cdot S(\vq_{pri})$
    \For {$i=x,y,z$}
    \State compute fourth-order FDs $(\Delta \vq_{pri})_i$
        \Comment Eq. \eqref{eq_exa_fd4_central_difference}
    \State compute NCP $B_i(\vq_{pri})(\Delta \vq_{pri})_i$
    \State compute KO term $(KO_{\vq_{pri}})_i$
        \Comment Eq. \eqref{eq_exa_fd4_ko}
    \EndFor
    \State $RHS_{\vq_{pri}} \leftarrow RHS_{\vq_{pri}}-\sum_i \Delta t \cdot B_i(\vq_{pri})(\Delta \vq_{pri})_i$
    \State $RHS_{\vq_{pri}} \leftarrow RHS_{\vq_{pri}}+\sum_i \Delta t \cdot (KO_{\vq_{pri}})_i$
    \State $\vq_{pri} \leftarrow \vq_{pri}+RHS_{\vq_{pri}}$
    \For {$i=x,y,z$}
    \State compute fourth-order FDs of the primary $(\Delta \vq_{pri})_i$
    \State Assign the auxiliary variables $\vq_{aux}\leftarrow (\Delta \vq_{pri})_i$
    \EndFor
    \State $t \leftarrow t+\Delta t$
    \State compute $\lambda_{max}$ to inform next time step
    \EndFor
    \EndFor
  \end{algorithmic}
\end{algorithm*}

\subsection{Task Parallelism and GPU Offloading}

Patch updates in \exahype\ are atomic units.
Patches are not subdivided further.
Therefore, \exahype\ models them as tasks, and \peano\ provides one abstract interface which
allows us to spawn them into either Intel's TBB, OpenMP, or plain C++ tasks.
We also offer an experimental SYCL back-end.

The task creation patterns realize the paradigm of enclave tasking
\cite{Charrier:2020:EnclaveTasking}: 
tasks that do not feed into MPI or are adjacent to adaptive mesh resolution transitions are spawned into our tasking backend of choice. \exahype\ refers to them as enclave tasks.
Other tasks are executed directly with high priority.
Once they terminate, we trigger all MPI data exchange and AMR mesh inter-grid transfer operators, while the tasking back-end now handles the enclave tasks.

This principle can be used to offload batches of enclave tasks in one rush to an accelerator:
we collect the enclave tasks without any side effects, i.e., those that do not
alter the global solver state, in a separate queue.
Once the number of buffered tasks in this queue exceeds a prescribed threshold,
we bundle them into one meta task and ship them off to the GPU
\cite{wille:2023:gpuoffloading,Li:22:ISC}.
The remaining enclave tasks are handed over to our task back-end of choice.
For the GPU offloading, we support OpenMP, SYCL and C++ offloading.

\subsection{Boundary Treatment}
\label{sec:code:bc}

Boundary conditions in \exahype\ 2 are implemented through the halo layers of faces that are adjacent to the global domain boundary.
Such faces hold elements $\vq_\text{out}$ and $\vq_\text{in}$.
After each mesh sweep or the initialization step, the array $\vq_\text{in}$ holds a
valid copy of the data within the adjacent patch.
To facilitate an update of this patch, we have to manually set the
$\vq_\text{out}$, as there is no adjacent patch here, and we can use this
``reconstruction'' step to impose appropriate boundary conditions.
 
\begin{figure}[htb]
  \begin{center}
    \includegraphics[width=0.49\textwidth]{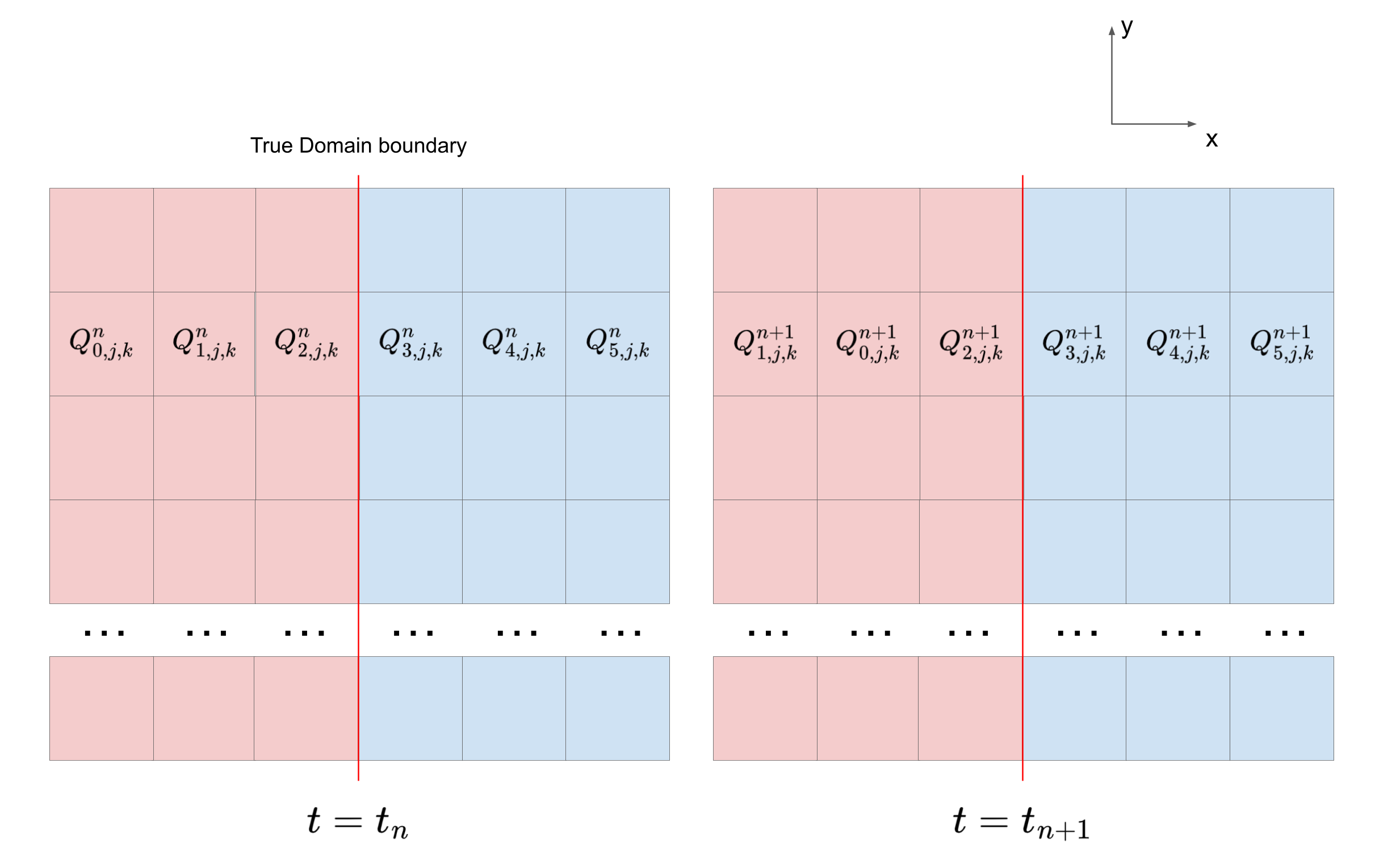}
  \end{center}
  \vspace{-0.8cm}
  \caption{
    Schematic illustration of boundary initialization for $K=3$:
    Half of the face's solution data is outside the domain (right, light blue).
    We need to initialize the outside layers of the boundary ($Q^{n+1}_{3,j,k}$,
    $Q^{n+1}_{4,j,k}$, $Q^{n+1}_{5,j,k}$) from values inside the domain to
    impose proper boundary conditions and allow the adjacent patch to
    continue its time stepping with valid halo data.
    We first assign the values $Q_{3,j,k}$ use $Q_{2,j,k}$ by treating them as
    $\vq_\text{out}$ and $\vq_\text{in}$ in the boundary condition function, then
    assign the values $Q_{4,j,k}$ using $Q_{3,j,k}$, and finally assign the
    values $Q_{5,j,k}$ use $Q_{4,j,k}$ to fill all three layers.
    We work our way outwards.
    More sophisticated schemes setting all values in one go are feasible within
    \exahype\ yet not used within \acronym.
    \label{fig_exa_fd4_bc}
  }
\end{figure}

To implement a homogeneous \revision{Neumann} boundary condition, we simply set $ \vq_\text{out} = \vq_\text{in}$ for example. 
For solvers that host multiple halo layers for
their time-stepping scheme, i.e.~$K \geq 2$, we apply the initialization scheme to specify the values in the ghost volumes layer by layer:
we move from the element layer closest to the domain boundary outwards (Figure~\ref{fig_exa_fd4_bc}).
 
While \exahype\ 2 ships a variety of pre-defined boundary treatments, one
particular type of boundary condition required by \acronym\ is 
the Sommerfeld (radiative) boundary condition \cite{alcubierre:2003:gauge} to
mitigate the effect of open-domain boundaries.
It eliminates spurious modes arising from 
homogeneous Newmann boundary conditions, by effectively suppressing wave-like
reflection and enforces that the wave at the boundary only travels out of the domain. The mathematical formulation of the Sommerfeld boundary condition reads as
\begin{equation}
    \partial_t Q+c \frac{r}{x_i}\partial_i Q=-\frac{c}{r}(Q-Q_\text{ini}) \bigg|_{\partial\Omega},\quad  i \in \{1,2,3\} \label{eq:SommerfeldBC_v1}
\end{equation}

\noindent
where $r=\sqrt{x^2+y^2+z^2}$ is the coordinate distance from the system's origin to the considered boundary position and $x_i \in \{x,y,z\}$ is the coordinate perpendicular to the considered boundary. $Q_\text{ini}$ is the value of evolving quantities at the boundary in the initial condition, as it serves as a ``background'' and thus does not enter the wave-like equation. $c$ is the wave speed, i.e., the eigenvalue of the characteristic matrix of the evolution system.

In the code, the derivatives are replaced by corresponding finite differences. 
To initialize layer $3$ of a FD4 setup with $K=3$ along a vertical boundary face (cmp.~Figure~\ref{fig_exa_fd4_bc}) we compute
\begin{equation}
  \label{eq_exa_sommerfeld_inter}
    \frac{1}{\delta t}\left(Q^{n+1}_3-Q^{n}_3\right)+c \frac{r}{x_i \delta x_i}\left(Q^n_3-Q^n_2\right)=
    -\frac{c}{r}\left(Q^n_3-Q_\text{ini}\right),
\end{equation}

\noindent
with an element size of $\delta x_i$. The indices for the
$y$ and $z$ directions are omitted as they are all the same in this case. 
The upper indices $n$ and $n+1$ represent the previous and current time step. 
As we work our way outside of the domain to fill all layers of the boundary data iteratively, $r$, $c$, and $x_i$ in Eq.~\eqref{eq_exa_sommerfeld_inter} need to
be adjusted accordingly.

\subsection{Adaptive Mesh Refinement}
\label{sec:code:amr}

Adjacent patches are coupled with each other through their shared faces.
On a regular grid, keeping face data consistent with its adjacent cells requires trivial one-to-one copies.
Special care is needed along mesh refinement transitions, i.e., along the boundary where the resolution changes between patches. 
However, the principle remains the same: 
adjacent patches communicate through their shared faces.

Logically, we can read the adaptive meshing over a spacetree structure as a mechanism to span ragged, non-overlapping Cartesian meshes: the patches of one resolution form one regular mesh which has holes where the code decides to use a finer or coarser resolution,
respectively.
This means we encounter ``boundary'' faces on this level, where the 
$ \vq_\text{out} $ are not set by an adjacent patch.

The $ \vq_\text{out} $ data on such hanging \revision{faces} are initialized from the coarser
volumes prior to the mesh traversal.
We \emph{interpolate}.
The counterpart is the \emph{restriction}.
Both schemes are implemented exclusively over face data.
Alternative implementations \cite{berger:1989:patchamr} realize the interpolation over
the patches themselves, i.e.~use a volumetric restriction.
While \exahype\ facilitates such volumetric coupling in principle
(cmp.~discussion below), we do not use expensive volumetric coupling
in \acronym's use cases presently.
Furthermore, we write the interpolation rules down with a tensor-product approach which
ignores diagonal coupling, i.e., interpolation of elements that are not aligned
along a coordinate axis, and we use exclusively a trilinear scheme.

Denoting the quantities of the volumes in the coarse and fine faces are $Q_{i,j,k}^c$ and
$Q_{i,j,k}^f$, the interpolation map can be written down as
\begin{equation}
    Q^f_{i,j,k}=P^{x}_{il}P^y_{jm}P^z_{kn} Q^c_{l,m,n},
\end{equation}

\noindent
where $P^x$, $P^y$, and $P^z$ are the matrices responsible for the corresponding one-dimensional map. 
Analogously, we obtain the matrix $R^i$ 
\begin{equation}
    Q^c_{i,j,k}=R^{x}_{il}R^y_{jm}R^z_{kn} Q^f_{l,m,n}
\end{equation}

\noindent
for the restriction (see \ref{app:amr} for the exact formulation of those matrices). \acronym\ currently defaults to averaging for the restriction.

\exahype's face-centric scheme is convenient to implement, as it allows users to solely focus on face-to-face interactions. 
We also can use it to construct interpolations and restrictions over multiple levels of refinement automatically: we recursively apply the intergrid transfer operator.
\acronym\ does not enforce any 2:1 balancing~\cite{sundar:2008:21balance}. One might want to call 3:1 balancing for \peano\ as we use tri-partitioning, which enforces that the cells adjacent to any face differ in size at most by one level of refinement, i.e., a factor of three in \Peano.
No such constraint exists, although users might want to impose it manually.

The logic where to refine or coarsen can be realized by the user. \acronym\ ships with examples of how to use feature-based refinement for example, where the maximal difference between any two adjacent elements within a patch triggers either a refinement or coarsening in the subsequent time step. This allows for solution-driven, dynamic adaptive mesh refinement.
However, the present experiments with \acronym\ all use a static refinement: we know where the black holes reside or the orbits on which they rotate initially, and prescribe concentric spheres of refinement levels around these areas.
Dynamical adaptive mesh refinement is not free and induces load-balancing challenges.
The static refinement circumnavigates these.

There are three open research questions to study:
First, codes without 2:1 balancing are prone to strong reflections along the resolution transitions, as high-frequency waves cannot escape regions of high resolution \cite{baker:2005:reflectiononAMRBoundary}.
Second, higher-order interpolation might be required to match the higher-order spatial discretion of the solvers, and we have already observed improvement in solution by using the second-order scheme in certain benchmarks below.
Finally, our resolution transition scheme does not guarantee the conservation of the interpolated and restricted physical quantities, which may unexpectedly affect the long-term evolution of the spacetime.

\subsection{Volumetric Coupling of Multiple Solvers}
\label{sec:code:solver_coupling}
In \exahype, the mesh can carry multiple solvers at the same time. 
In such a case, they run parallel while the code traverses the mesh.
\acronym's solvers can be composed, i.e., they can be told to run in-sync.
This way, solvers with different time step sizes (e.g., different $\lambda _{\text{max}}$), solvers with different time integrators, or solvers with different mesh
resolutions are kept synchronous: we can, for example, disable them for some mesh traversals or manually prescribe their time step size.

This is a synchronization or coupling of the global solvers' states.
\acronym\ can also couple different solvers per patch: if a spacetree cube holds two patches---of an FD4 and an FV solver, e.g.---we can
hook into a patch update postprocessing step and manually overwrite either of them after each time step.
This way, it is possible for \acronym, for example, to run an FV solver around a black hole and to use FD4 further away, we conduct a simulation test based on this and report in Section \ref{sec:result:sbh}.
The feature is typically combined with a localization of the solvers, where each solver holds data only for some cubes of the underlying spacetree.

\subsection{Particle Tracers as Data Probes}
\label{sec:code:tracer}

\exahype\ features particles which are embedded into the computational mesh.
This feature originally has been introduced to support Particle-in-Cell schemes~\cite{weinzierl:2016:ExaParticle}.
In \acronym, we use it to realize tracers. Particles are inserted into the mesh for data probing and object tracking. They are not evolved independently, but instead either follow evolving variable fields in PDEs---the quantities in $\vq$ which represent a velocity field for example---or are spatially invariant and simply record the values in $\vq$ at this point.

Particles allow users to track the solution's evolution at certain points in the domain without high I/O overhead, as they sample the solution in one point and do not require us to dump large output files. If they move, they yield trajectory information of important objects in the domain.  Adding tracers to an \acronym\ simulation involves three stages:

\begin{itemize}
    \item \emph{Initialization}. The users specify the number and initial coordinates of the tracers. \exahype\ offers two approaches for initializing the tracer coordinates: one can set the coordinates
    explicitly by writing them down in the \acronym\ specification script from where they are passed into \exahype, or we can provide a coordinate file. Besides the coordinates, users also need to specify what evolving variables the tracers need to plot. 
    \item \emph{Data Projection}. Users then decide which and how the solver's values on the patches are projected onto tracer attributes. \acronym\ currently uses trilinear interpolation, i.e., it uses the values within the closest $2^d$ elements (Finite Volumes or Finite Difference sample points) to determine
    the value at the tracer particles. Along with the projection instructions, users can declare three quantities which serve as velocity field $v_i$ for the tracers. If specified, the tracers' positions are updated using an explicit Euler time integrator
    \begin{equation}
        x^{t+\delta t}_i=x^{t}_i+\delta t v^t_i.
    \end{equation}

    \noindent
    where $x^{t}_i$ is the position of the considered tracer at time $t$, and $v^t_i$ is the velocity reconstructed from the solver solution. It allows tracers to follow the flows within the simulation. $\delta t$ can be set to the solver's time step size.

    \item \emph{Data dumping}. Tracers record ``their'' values in each and every time step. \acronym\ can instruct them to dump this chronological data into output files in CSV format. For large-scale simulations, \exahype\ 2 offers various thresholds to control the dumping: users can ask tracers to dump in fixed code time intervals, or to dump when recorded variables or tracking positions have changed substantially. To reduce the memory footprint, the
    recording can be lossy, i.e., only record new values whenever their relative difference to previously recorded values exceeds a certain threshold.
\end{itemize}

\noindent
The tracer module is used in \acronym\ extensively as the ``extractor'' of gravitational wave signals and for black hole puncture trackers.
\section{Functional Decomposition and Software Architecture}
\label{sec:tasking-decomposition}

An \emph{\exahype\ project} forms the starting point of any \acronym\ endeavor.
It is eventually translated into a set of C++ classes defining the actual executable.
It is lowered in multiple steps into the actual execution code.
In \acronym, we rely on Python bindings for an \exahype\ project, while the actual code is represented by a \peano\ project that can be translated one-to-one into C++ code.
The starting point is hence an instance of \texttt{exahype2.Project} in Python.

Once an \exahype\ project, aka instance of the Python class, is set up, we lower it into a sequence of four tasks: initial mesh generation, solution initialization, time stepping and I/O steps. 
Every \acronym\ project consists of these fundamental steps, where plotting and time stepping are iteratively executed.
The lowering is a mere task sequence instantiation. From hereon, we focus on the time step.

An \exahype\ project hosts an arbitrary number of solvers.
The solvers' time-stepping scheme determines how many mesh traversals we need to implement in one time step, as well as the semantics per traversal.
Let a solver be a Runge-Kutta scheme of order $k \leq 4$.
\exahype\ lowers a time step into four mesh traversals.
For $k>4$, we hit the Butcher barrier \cite{butcher:2016:ODEbook} and hence require additional
mesh sweeps.
This lowering follows the solver's time stepping
specification.
Besides the translation into a sequence of mesh traversals, a solver carries meta information such as maximum mesh size, patch dimensions $p$, the halo size $K$ or the number of unknowns.
This yields a specification of the required data
structures: an \acronym\ FD4 for example makes the lowering fit a $p \times p \times p$ patch into each cell, equip the faces with an overlap of three, i.e.~a $2\cdot 3 \times p \times p$ patch, and store helper data for the selected
Runge-Kutta scheme such as temporary storage for the right-hand sides.
The solver also informs the lowering process what global parameters the solver has to host---such as $\lambda _{\text{max}}$---and how they are exchanged and kept consistent.
For multiple solvers, our translation into a lower abstraction level ensures automatically that solvers are kept in sync.

\begin{figure}
  \begin{center}  
    \includegraphics[width=0.45\textwidth]{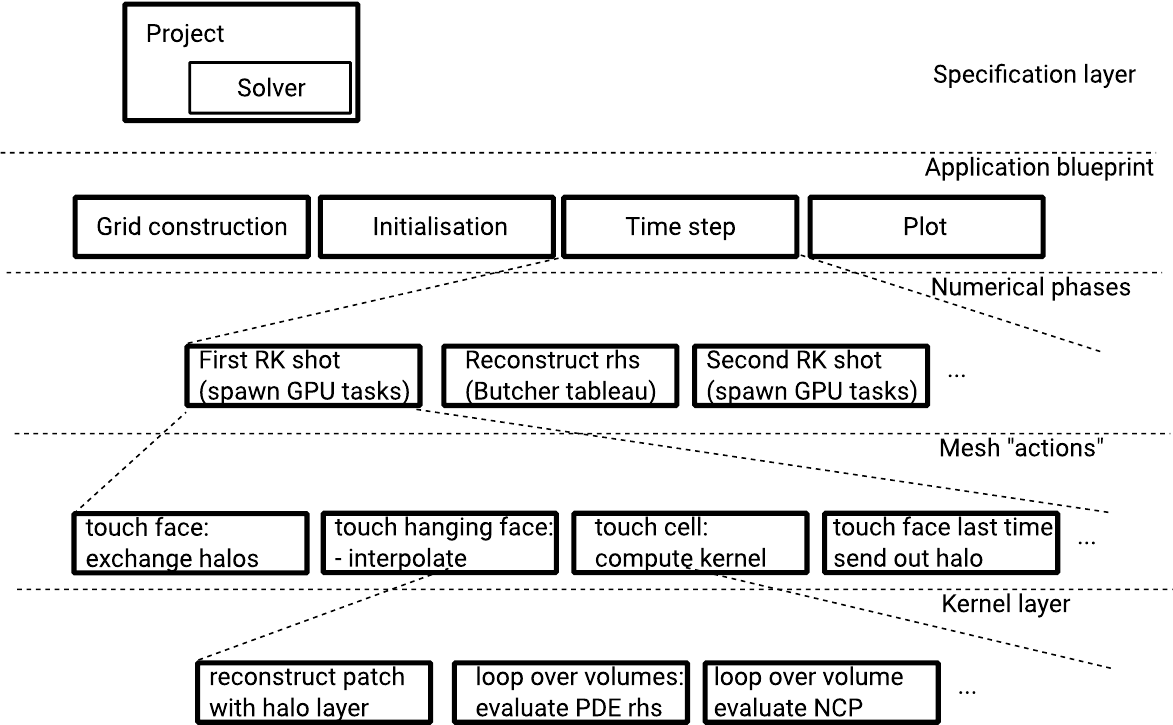}
  \end{center}
  \caption{
    Starting from an abstract specification (which solvers to use, initial
    mesh resolution, total runtime, time in-between plots, \ldots), we generate
    a tailored series of tasks in multiple steps. We lower the representation
    from a high-level description into something that reflects the C++
    implementation one-to-one.
    \label{figure:tasking-decomposition:lowering}
  }
\end{figure}

Once the nature of a mesh sweep is known, we can lower it once more into fundamental tasks over mesh entities:
Per cell, we know how to equip the current time step or Runge-Kutta guess data with halo data information, i.e., how to reconstruct the $(p+2K) \times (p+2K) \times (p+2K)$ patch, which compute kernel (update of a patch) to invoke,
which halo data to exchange, which AMR criteria to evaluate, which patch post-processing to call, and so forth.
We take the numerical recipe and lower it onto a task graph over the actual mesh in use.

The compute kernels themselves can be lowered one final time:
a time step in a Finite Volume scheme for example maps onto a sequence of $d$-dimensional loops over all volumes within a patch \cite{loi:2023:sycl}.
Per volume, we evaluate the actual PDE terms.
This is the lowest level within \acronym's cascade of abstraction levels: within this final lowering step, we create a C++ class per solver.
It allows the user to hook into mandatory steps of the simulation such as the refinement control or the patch initialization, but it notably also provides a
method stub per PDE term in use.
This empty stub has to be filled by the user.
Alternatively, users can define in the top-level specification which functions have to be called for the PDE terms, the initialization, the refinement criterion, and so forth.
If specified, the final lowering step inserts these function calls into the stub.
\acronym\ provides pre-defined functions for all features used in the present manuscript.

Our description of \acronym's abstraction levels relies on multiple levels of abstraction layers.
Users provide an abstract description in Python.
From this \emph{specification layer}, we can break down the workflow into finer and finer logical representations.
This lowering is completely hidden from the user---it is realized in the \exahype\ engine layer---and it relies
on a task language (Figure~\ref{figure:tasking-decomposition:lowering}).
On the higher abstraction levels, the tasks describe sequences of activities that are performed to produce the simulation output (\emph{application blueprint} or workflow).
Once we break down these activities into activities per
mesh entity (mesh \emph{actions})~\cite{peano}, we obtain a real task graph over the mesh: what has to be done for a cell exclusively relies on the fact that all adjacent
hanging faces have been interpolated completely and the halo layers of the adjacent faces of the $2d$ faces of this cell have been made consistent.
Data along a face that is to be exchanged via MPI to keep the halos for the subsequent time step consistent can be sent out immediately after the corresponding adjacent cell has finished its kernel.
This list is not comprehensive.

All lowering down onto mesh traversals and activities per mesh entity is automated.
We implement it within the \exahype\ engine once our \acronym\ description has built up the program specification.
It is only the very last step that links back to actual user code and might require manual intervention unless the used implementations are inserted into the specification script as well.

\subsection{Static Software Architecture}

\begin{figure}[htb]
  \begin{center}
    \includegraphics[width=0.48\textwidth]{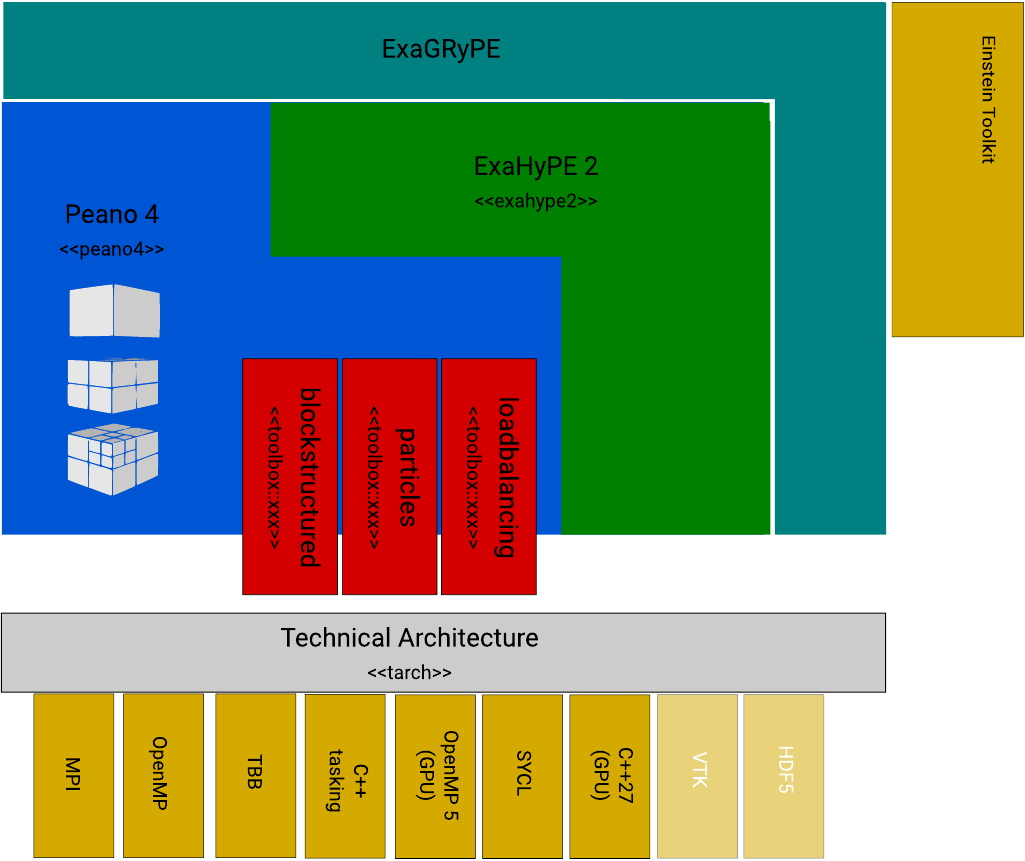}
  \end{center}
  \caption{
    The schematic architecture of the C++ libraries used within \acronym. 
    The Python layer on top mirrors the core architecture, but can omit
    representations of the technical architecture and third-party tools in
    Python.
    \label{figure:tasking-decomposition:static-architecture}
  }
\end{figure}

As \acronym\ is built on top of \exahype, it inherits its multi-layer static
architecture (Figure~\ref{figure:tasking-decomposition:static-architecture}).
A technical architecture wraps away various external libraries such as OpenMP,
TBB and MPI, but also I/O facilities such as HDF5.
In \acronym, we can hence swap different multithreading backends in and out.
It is written in plain C++ although some subcomponents use extensions such as
SYCL or OpenMP pragmas.

The technical architecture is a utility collection used by all of our code.
\peano\ is the actual AMR and mesh traversal component.
It is written in C++.
Its design principles follow the ``\textit{don't call us, we call you}'' paradigm introduced with its third generation~\cite{peano}, i.e.~this component owns the
mesh and its traversal.
While it runs through the mesh, it issues events such as ``\textit{I read this vertex for the first time}''~\cite{Weinzierl:2011:Peano}. Codes can hook into
these events to realize the actual solver semantics.

\exahype\ is such a code. Per solver and solver flavor (different optimization variants of each solver exist for example), it provides classes and functions that plug into these events.
It is all written in C++ and also uses routines from \Peano\ as well as the technical architecture.
As many features within \exahype\ are used by multiple solvers or even other applications built on top of \peano, they are modeled as separate toolboxes: inside the C++ core of \peano\ plus \exahype\ we functionally decompose the code again.

\acronym\ adds additional domain-specific features on top of \exahype, but also makes contributions to \exahype's core functionality.
The latter includes a solver for fourth-order Finite Differences that utilizes the block-structured routines or bespoke boundary conditions for example.
Other features such as the implementation of the CCZ4 PDE terms or the KO damping are domain-specific and kept separate from the \exahype\ core.
Here, \acronym\ is a strict extension of \exahype.
These extensions comprise also routines to initialize the simulation setup.
While \peano\ and \exahype\ are independent of third-party libraries or, through the technical architecture, shielded from them, the initialization routines within \acronym\ link to the {\sc Einsteintoolkit} \cite{einsteintoolkit}.

All of \acronym, \exahype\ and \peano\ are written in plain C++ with a technical architecture which separates them from extensions such as SYCL.
While we do use metaprogramming in many places, a lot of these C++ ingredients are rather independent and not tied together.
Instead, \peano\ offers a Python wrapper that can represent a whole \peano\ application including its \texttt{Makefile}, global variables/settings and so forth.
This Python wrapper comprises all the glue code required to link certain components to each other.
For example, it exactly defines which memory is to be allocated per spacetree cube, which functions are to be called whenever we run through a face throughout the mesh traversal, and so forth.
The Python component represents the lowest level of our automatic code lowering and can be translated one to one onto C++ glue code, i.e., a working executable linking to all static \peano, \exahype\ and \acronym\ libraries comprising all building blocks.

The C++ architecture is mirrored by the Python API.
We do not need a representation for the technical architecture or any third-party library underneath.
The most basic representation is the Python \peano\ layer which represents a \peano\ C++ application exactly plus all required build information
(\texttt{makefile}).
All toolboxes have their Python equivalent such that we can add them to a \peano\ project and configure them on the Python level throughout the lowering process. 
\exahype\ adds an additional Python abstraction on top, where individual objects represent solvers, data structure definitions, and so forth.
On this level, we also explicitly host the main routine.

On top of the \exahype\ Python layer, we install the \acronym\ layer, which reflects the program specification and offers many domain-specific helper routines.
In this code design, all lowering can be done exclusively in Python.
Objects within the lower Python layers however often correspond one-to-one to C++ objects.

\subsection{Programming Workflow}


\begin{algorithm*}[htb]
  \caption{
    Baseline structure of any \acronym\ solver within \exahype.
    \label{code:project}
  }
  \begin{algorithmic}[1]
    \State project = exahype2.Project(namespace = ["benchmarks",
    "exahype2", "ccz4"], name = "CCZ4", executable="test ")
    \State project.set\_global\_simulation\_parameters(dimensions = 3,
    offset = [-0.5, -0.5, -0.5], domain\_size = [1, 1, 1],
    periodic\_boundary\_conditions = [True, True, True], end\_time = 1.0,\ldots
    )
    \State project.set\_Peano4\_installation(directory=\ldots,
    build\_mode=peano4.output.CompileMode.Release)
    \State
      \Comment{Parse Peano's configuration}
    \State \ldots
      \Comment{Actual solver construction}
    
    \State peano4\_project = project.generate\_Peano4\_project()
    \State
      \Comment{Lowering}
    \State peano4\_project.generate()
    \State
      \Comment{Spill out C++ code plus Makefile}
    \end{algorithmic}
\end{algorithm*}

We run through the structure and programming of a \acronym\ solver by means of a simple gauge wave benchmark.
The example highlights how the individual simulation steps manifest in Python instructions and illustrates how they inform the lowering.
Every \exahype\ project follows the same structure
(Algorithm~\ref{code:project}): we create a project and hence implicitly define the application blueprint
(Figure~\ref{figure:tasking-decomposition:lowering}) with its four distinct phases.

The project is manually lowered down into a \peano\ project, which is the one-to-one equivalent of C++ source code. 
Consequently, we dump this project then onto disk.
We assume that all \acronym, \peano\ and \exahype\ libraries have been built before using either the \texttt{autotools} or \texttt{CMake}.
The Python script automatically picks up the used configuration arguments and creates an appropriate \texttt{Makefile} alongside the generated C++ code.
Load balancing, additional third-party libraries, or global I/O settings all can be added to the \peano\ project before we write to disk. \revision{While \exahype\ allows users to define their own load-balancing strategy, the current version of \acronym\ utilizes the default built-in option. We plan to conduct a full investigation of load balancing strategies in future work along with the performance analysis.}

\subsubsection{Solver Configuration}

\begin{algorithm*}[htb]
  \caption{
    Creating an actual \acronym\ solver is a two-liner.
    \label{code:solver}
  }
  \begin{algorithmic}[1]
    \State my\_solver = 
    \State exagrype.api.CCZ4Solver\_FD\_GlobalAdaptiveTimeStep(
        name="CCZ4FD",
        patch\_size=9,
        min\_volume\_h=\ldots
    )
    \State project.add\_solver(my\_solver)
    \end{algorithmic}
\end{algorithm*}

\algdef{SE}
[CLASS]
{Class}
{EndClass}
[1]
{\textbf{class} \textsc{#1} \{}
{\textbf{\};}}

\algdef{SE}
[METHOD]
{Method}
{EndMethod}
[1]
{\textbf{virtual} \textsc{#1} (}
{\textbf{) override;}}

\begin{algorithm}[htb]
  \caption{
    By default, \exahype\ creates one (empty) class per solver. Users can fill
    this one manually. The class signature resembles the example given as
    a pseudo-code below.
    The (generated) superclass holds generic information passed into the Python
    representation such as the used patch size $p$.
    \label{code:cppstub}
  }
  \begin{algorithmic}[1]
    \Class{...::exahype2::ccz4::CCZ4FD: public
    ...::exahype2::ccz4::AbstractCCZ4FD}
    \Method{RefinementCommand refinementCriterion}
      \State double Q[58],
      \State const Vector$<$Dimensions,double$>$\& x,
      \State \ldots
    \EndMethod
    \Method{void initialCondition}
      \State double Q[58],  // out
      \State const Vector$<$Dimensions,double$>$\& x,
      \State \ldots
    \EndMethod
    \Method{void NCP}
      \State double Q[58],
      \State const Vector$<$Dimensions,double$>$\& x,
      \State \ldots
      \State int                                          normal,
      \State double BgradQ[58] // out
    \EndMethod
    \State \ldots
    \EndClass
    \end{algorithmic}
\end{algorithm}

The actual \acronym\ solver has to be added to the \exahype\ project before triggering the lowering. 
While the creation is close to trivial (Algorithm~\ref{code:solver}), different variants are on the table about how to connect the vanilla PDE solver to the actual physics (Figure~\ref{figure:building-blocks:exagrype_structure}).
Without any further information, the lowering yields an empty C++ class that users have to fill (Algorithm~\ref{code:cppstub}).
After that, the lowering process will create almost all
simulation code without any further user interference. Only on the very lowest level, when we actually evaluate the PDE terms or the refinement criterion or require initial values, it will call routines from the PDE solver.

As an alternative to the manual filling of the routines in C++, all solvers in \exahype\ offer a \texttt{set\_implementation} routine.
With this one, users can inject C++ code directly into the generated code from the Python API.
\acronym\ provides wrappers around the generic \exahype\ solvers.
There is a set of bespoke Finite Volume solvers, bespoke FD4 solvers, and so forth.
They already set some default implementations, configure halo layer sizes, or set the correct number of unknowns.

\paragraph{Grid construction}

As the first step of the simulation, \peano\ constructs the computational grid and builds the corresponding spacetree. 
In this stage, it also already splits up the computational domain following the advice of the load balancing metrics.
\peano\ offers plug-in points to guide the mesh refinement, and \exahype\ automatically runs through all the \acronym\ solvers searching for the minimum of their maximum mesh sizes. It ensures that the initialization starts from a sufficiently fine grid.

Once we have a reasonably fine initial mesh, \exahype\ starts to query the solver objects if a certain area should be refined.
In this stage, the solvers do not yet have a solution at hand.
Either we hard-code the initial refinement pattern, or we make the algorithm run through a decision logic which analyses ``\textit{if we had a solution here, would we
refine}''.
This implements a feature-based initial refinement.
The present case studies all use hard-coded patterns, adopting static grids where the finest resolution covering the whole interested region. The refinement criterion can be either written into the C++ template or injected the implementation of the C++ code directly from
Python via \texttt{my\_solver.set\_implementation(refinement\_criterion\\ = "my
fancy C++ code snippet")}. \revision{The dynamic grid refinement during evolution is also available in \acronym\ and can tag volumes for refinement either based on coordinates, similar to the “Tracker-based AMR” in~\cite{AMR_in_NR_2}, or based on variables, similar to the “tagging on gradient” in~\cite{GRChombo2022}. This dynamic feature is not used for the current benchmarks, and a full investigation is left for future work.}

\paragraph{Grid initialization}

As soon as the initial grid is initialized, i.e., as soon as \exahype's main logic realizes that the grid does not change anymore, and as soon as the initial load balancing has terminated, the code switches into the initialization phase.
\exahype\ now embeds the actual compute data structure, i.e., the patches, into the spacetree cubes and then runs through each element (Finite Volume or FD sample point) and sets the initial value. This happens by calling the
solvers' \texttt{initialCondition} routine.

\begin{algorithm}[htb]
  \caption{
    The initial conditions for the Gauge wave setup can be set directly through
    the Python interface. It injects C++ code snippets into the generated
    solver that call a \acronym\ C++ factory function setting the Gauge wave
    data.
    \label{code:initial-condition}
  }
  \begin{algorithmic}[1]
    \State my\_solver.set\_implementation(
        initial\_conditions="""
    \State \phantom{xx} for (
    \State \phantom{xxxx} int i=0;
    \State \phantom{xxxx} i$<$NumberOfUnknowns+NumberOfAuxiliaryVariables; 
    \State \phantom{xxxx} i++
    \State \phantom{xx} ) Q[i] = 0.0; 
    \State ::applications::exahype2::ccz4::gaugeWave(Q, x, 0); 
    \State """)
  \end{algorithmic}
\end{algorithm}

Our present \acronym\ solvers realize two different approaches for this: for simple setups where we know the analytical solution, the function body
of \texttt{initialCondition} is hard-coded and assigns the initial values of evolving variables directly.
This happens for the gauge wave benchmark, e.g., for which \acronym\ provides a helper function (Algorithm~\ref{code:initial-condition}).

For more complex setups such as the black hole system where a numerical solution is required that obeys the Hamiltonian constraints, we link against the
\textsc{TwoPuncture} module from {\sc Einsteintoolkit}.
It calculates the spacetime quantities $\gamma_{ij}$, $K_{ij}$ and $\alpha$ from the specified physical parameters like masses and momenta of the black holes.
\acronym\ provides a wrapper which converts the basic spacetime quantities ($\gamma_{ij}$, $K_{ij}$) from the external library into the ones that \acronym\ utilizes
($\tiga_{ij}$, $\tiA_{ij}$, $\phi$, $K$).

If combined with FD4, \acronym\ offers a fourth-order finite difference helper function to compute the first derivative feeding into the auxiliary variables \eqref{equation:theory:auxiliary-variables}.
This is a post-processing step which can be combined with any initial condition.

\paragraph{Evolution equations (Timestepping)}

The actual PDE evolution is pre-configured in all \acronym\ solvers, i.e.~they call \texttt{set\_implementation} for the non-conservative product and source code throughout the construction, and they also provide an implementation of the $\lambda _{\text{max}}$ calculation.
Users can redefine and alter it, but our present experiments all rely on the default settings.
The three core routines are re-implementations in C++ of the Fortran implementations of \cite{Dumbser18}, although we modified some Fortran code entries to align them with the CCZ4 formulation as outlined in the Appendix.

An explicit time stepping over the CCZ4 formulation yields solutions which violate the CCZ4 constraints \cite{Dumbser:2020:curlcleaning}. 
The default \acronym\ solvers therefore add a post-processing of $\vq$ to each time step's computational kernel.
It enforces the two \emph{algebraic constraints} 
$\det(\tiga_{ij})=1$ and ${\rm tr}(\tiA_{ij})=0$ point-wisely: the corresponding evolving variables in $\vq$ are over-written to satisfy the constraints, i.e.~the traceless property is re-achieved by removing the trace
from every component of the tensor, and the unity of determinant is re-achieved by rescaling every component.

If an FD4 solver with a second-order formulation is instantiated, \acronym's solvers add an additional mesh sweep to each solver which serves as an epilogue to each time step.
It reconstructs the derivatives (Section \ref{sec:code:solver}).
The additional mesh sweep is necessary, as the reconstruction requires some inter-patch/MPI data exchange to get all halo data consistent.

\paragraph{Further features}

Boundary conditions can be injected via \texttt{set\_implementation} but are set to Sommerfeld by default.
The routine to control dynamic adaptive mesh refinement defaults to \exahype's \textsc{Keep} command, i.e., the mesh is neither refined nor coarsened.
More advanced setups might want to alter this behavior.

\subsubsection{I/O and Post-Processing}

\exahype\ dumps data into a bespoke patch format, which can dump patches with cell- or vertex-associated degrees of freedom, higher-order polynomials or mesh metadata with low overhead and without any constraints on dynamic adaptivity.
The mesh can change from each and every data dump.
The dump is realized through a writer in the technical architecture.
\peano\ ships a Python tool to merge and convert them into the standard \texttt{VTU} files, which is compatible with various visualization software.
The realization relies on the VTK/Paraview software.

Given the scale of the simulations, dumping all 58 equations over the whole computational domain results in large output files.
Therefore, we incorporated a \emph{data filter} into our code. 
This filter enables users to selectively extract slices and clips from the domain.
One can also specify what variables are to be included in the output files.

This in-situ filtering reduces the I/O pressure significantly.
Auxiliary variables used as helper quantities in the evolution for example are often omitted in the snapshots.
If phenomena are known to show intrinsic symmetry or invariance along certain coordinate axes---cmp.~the static single black hole scenario which has rotational symmetry or the gauge wave setup which effectively yields a one-dimensional evolution of interest---we can boil down the output file sizes significantly.

Besides mere filtering, we can also compute accuracy metrics in situ: if enabled, \emph{Hamiltonian and Momentum constraints}
\eqref{eq_hamilton_constraints}--\eqref{eq_momentum_constraints} are calculated after each time step and attached to the evolving quantities $\vq$ as ``additional variables''. 
As the calculation of the extra quantities is separated from the computational kernel, \acronym\ offers a big freedom to pick the computational rule or cardinality of the output. We can calculate intermediate quantities, e.g., the Ricci tensor $R_{ij}$ or the conformal connection function $\tilde{\Gamma}^i$, for testing purposes.

\subsubsection{Tracers}

\begin{algorithm}[htb]
  \caption{
    A one-liner adds tracers to an \acronym\ solver.
    \label{code:tracers}
  }
  \begin{algorithmic}[1]
    \State  my\_solver.add\_tracer(
    \State \phantom{xx} name="Tracer",
    \State \phantom{xx} coordinates=[[-0.4251, 0, 0], [0, 0, 0], [0.4251, 0, 0]], 
    \State \phantom{xx} project=project,
    \State \phantom{xx} number\_of\_entries\_between\_two\_db\_flushes = 100,
    \State \phantom{xx} data\_delta\_between\_two\_snapsots = 1e-8,
    \State \phantom{xx} \ldots
    \State )
  \end{algorithmic}
\end{algorithm}

Each \acronym\ solver is pre-prepared to host tracers
(Algorithm~\ref{code:tracers}).
If they are added, the solver automatically picks an appropriate interpolation scheme and informs \peano\ that appropriate particle handling is to be added. Multiple tracer types can be added to each solver.

Tracers write their outcome into a plain CSV file for all unknowns from $\vq$ specified plus the tracers' position.
\acronym\ also automatically assigns each tracer a unique ID, such that post-processing scripts can track their evolution over many time spans.
The I/O capabilities of a tracer can be configured such that updates to the CSV file are written if and only if enough particles have altered their state, i.e., we can constrain the output frequency.
We can furthermore configure after which relative change of a particle quantity we consider the particle worth writing a trajectory update into the database.

Tracers unfold significant potential once we couple it with the in-situ post-processing: for the \emph{$\psi_4$ calculation} for example, we calculate the $\psi_4$
quantities following Section~\ref{sec:theory:GW} and attach them to the solution vector $\vq$.
Rather than plotting, we interpolate it onto the tracers and let them record the $\psi_4$ evolution over time. 
Once the tracers are arranged over a sphere, they enter the integral \eqref{eq:tdesign_spherical_integral} to extract the gravitational waves.

Puncture trackers are different tracers.
They move. 
We drop the virtual particles into the initial locations of the black holes and specify the opposite shift vector $-\beta^i$ to serve as their velocity. 
They feed into the explicit Euler time integrator (Section~\ref{sec:code:tracer}). 
The trajectories of those particles represent the movement of the black holes.

\subsubsection{Coupling}
\label{sec:workflow:coupling}

Coupling of two \acronym\ solvers starts from the creation of two solvers within the Python script which are both added to the \exahype\ project.
Without further steps, these solvers run completely independent of each other.
Different to all steps previously, coupling requires users to run through a series of steps:

First, one solver has to determine the other solver's time step sizes. Each solver's generated C++ class allows users to overwrite the routine \texttt{startTimeStep} which is the canonical place to let one discard
its own time step size determined by the $\lambda _{\text{max}}$, and instead use the other solver's step size.
From here on, the two solvers run in-sync.

Second, it might be reasonable to restrict the computational domain of one solver, unless we want both solvers to cover the whole domain and to couple
volumetrically everywhere.
For this, \exahype\ solvers can redefine an internal function \texttt{\_store\_cell\_data\_default\_guard} and similar routines for faces such that the solver's area, i.e.~where it exists, is narrowed down.
The routines return a C++ statement which has to be evaluated into a boolean expression.

Finally, the solvers have to be coupled. 
For doing this, all solvers host an attribute
\texttt{\_action\_set\_preprocess \_solution} which is by default \texttt{None}.
We reset it to an instance of
\texttt{exahype2\-.solvers\-.rkfd\-.actionsets .PreprocessReconstructedSolutionWithHalo}, e.g., and inject the interpolation/restriction from one solver to the other here.
Again, a simple if statement can localize the operations.

\section{Numerical Results}
\label{sec:result}


In this section, we present \acronym\ simulation results for various physical scenarios. 
We begin with a standard gauge wave benchmark and then examine a black hole space-time setup, discussing both the single Schwarzschild black hole and the binary black hole merger. 
The tests are used to validate the correctness of the code implementation. While they affirm that the code base is working properly and reasonably robust for numerical astrophysics applications, they also reveal required optimizations and challenges that will be addressed in the future.

\subsection{Gauge Wave}
\label{sec:result:gw}


%
%
The gauge wave scenario is one of the standard test cases for numerical
relativity codes \citep{alcubierre:2003:NRtestbed}. 
A flat Minkowski spacetime is considered and no actual ``physical'' phenomenon
occurs in the system, i.e. solution characteristics should stay invariant
under a consistent and robust numerical scheme. 
For this, we slice the static space-time by performing a time-dependent
coordinate transformation:
\begin{eqnarray}
    \hat{t}&=&t-\frac{A}{2k\pi}\cos [k\pi(x-t)],\\
    \hat{x}&=&x+\frac{A}{2k\pi}\cos [k\pi(x-t)],
    \qquad \text{and} 
    \\
    \hat{y}&=&y,\ \hat{z}=z,
\end{eqnarray}

\noindent
which transcribes the original Minkowski metric
$ds^2=-d\hat{t}^2+d\hat{x}^2+d\hat{y}^2+d\hat{z}^2$ into 
\begin{eqnarray}
    ds^2 & =& -H(x,t)dt^2+H(x,t)dx^2+dy^2+dz^2, 
    \\
    H(x,t) & =& 1-A\sin [k\pi (x-t)].
\end{eqnarray}

\noindent
It yields a gauge wave propagating along the x-axis (positive-wards) with an amplitude of $A$. As
we know the complete four-dimensional metric here, and the related quantities can be read straightforwardly:
\begin{equation}
    \alpha=\sqrt{H},~\beta^i=0,~\phi=H^{-1/6},
    \quad \text{and}
\end{equation}
\begin{eqnarray}
  K_{xx} & =& -\frac{\partial_t H}{2\alpha}\nonumber
    \\
    & = &
    -\frac{k\pi A}{2}\frac{\cos [k\pi(x-t)]}{\{ 1-A\sin[k\pi(x-t)]\}^{1/2}},
    \\
  K_{ij,{\rm others}} & =& 0,
    \\
  K=\gamma^{ij}K_{ij} & =& \frac{K_{xx}}{H} \nonumber
    \\
    & = &
    -\frac{k\pi A}{2}\frac{\cos [k\pi(x-t)]}{ \{ 1-A\sin[k\pi(x-t)]\}^{3/2}}.
\end{eqnarray}

\noindent
The static zero shift $\beta^i=\partial_t \beta^i=0$ and the harmonic slicing
$f(\alpha)=1$ are adopted as the gauge conditions for this scenario. Despite its
simplicity, this test poses a highly nontrivial challenge to the formulation and implementations of numerical relativity codes, as the periodicity, wave shape and amplitude have to be preserved over a long time span \cite{babiuc:2008:NRtestbed}.

The first simulation test we report in this subsection utilizes a
computational domain $\Omega=[-0.5,0.5]^3$ on a regular grid subject to periodic boundary conditions. Periodic boundary conditions are also a newly added feature released with \exahype\ 2 as opposed to the first-generation code. We divide the domain into 162 volumes per dimension, thus leading to a volume size of $0.006173$. AMR is not enabled for this test.

We pick $A=0.1$, $k=2$ for the physical parameters. For the running parameters, we follow previous literature, setting $\kappa_1=1.0,\kappa_2=0,\kappa_3=0$ and $e=c=\tau=1.0$.  $\mu=0.2$ keeps the impact of the constraints small (cmp.~reports in \cite{Dumbser18}). As for the solver-specific parameters, the coefficient of the KO dissipation $\epsilon$ is set to be $8.0$ and a CFL ratio of $C=0.1$ is adopted. The parameters $f$ and $\eta$ are irrelevant as the shift vector does not evolve in the gauge wave scenario.

\begin{figure}[htb]
   \centering
   \includegraphics[width=0.5\textwidth]{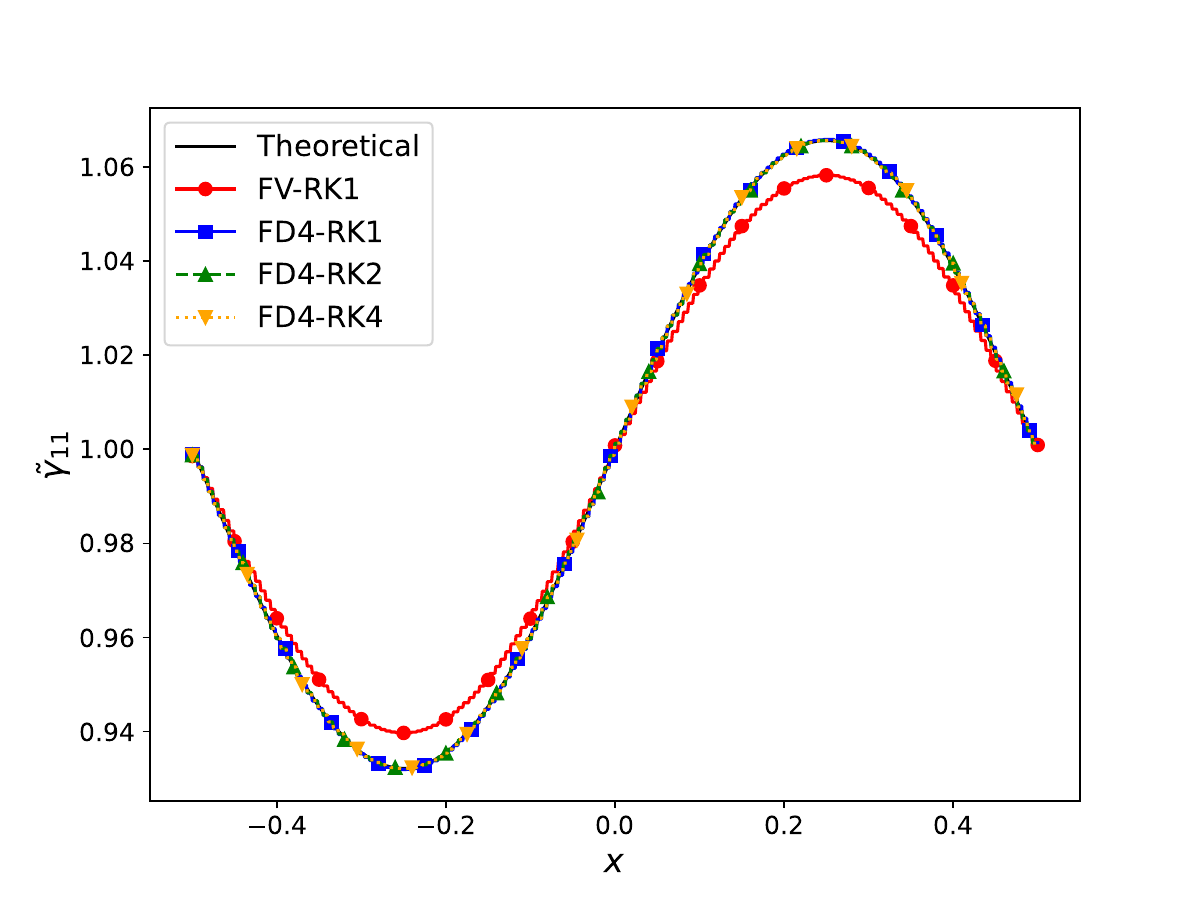}
   \caption{The profile of $\tiga_{11}$ along $x$-axis at code time $0.5$ for different solvers: FV-RK1 (red solid), FD4-RK1 (blue solid), FD4-RK2 (green dashed) and FD4-RK4 (orange dotted).
     \label{fig_gw_wavepattern}
   }
 \end{figure}

We illustrate the simulation result at code time $T=0.5$ in
Figure~\ref{fig_gw_wavepattern}.
It shows the profiles of $\tiga_{11}$ along the $x$-axis for
different kernels (FV, FD4-RK1, FD4-RK2 and FD4-RK4), employed with the first-order formulation. All the solvers yield the correct traveling speed and phase of the wave. However, the Finite Volume solver suffers from severe amplitude damping, and a similar dissipation is also observed in the black hole benchmarks. Consequently, we have chosen to use the FD4 scheme as our default solver.

 \begin{figure}[htb]
   \centering
   \includegraphics[width=0.5\textwidth]{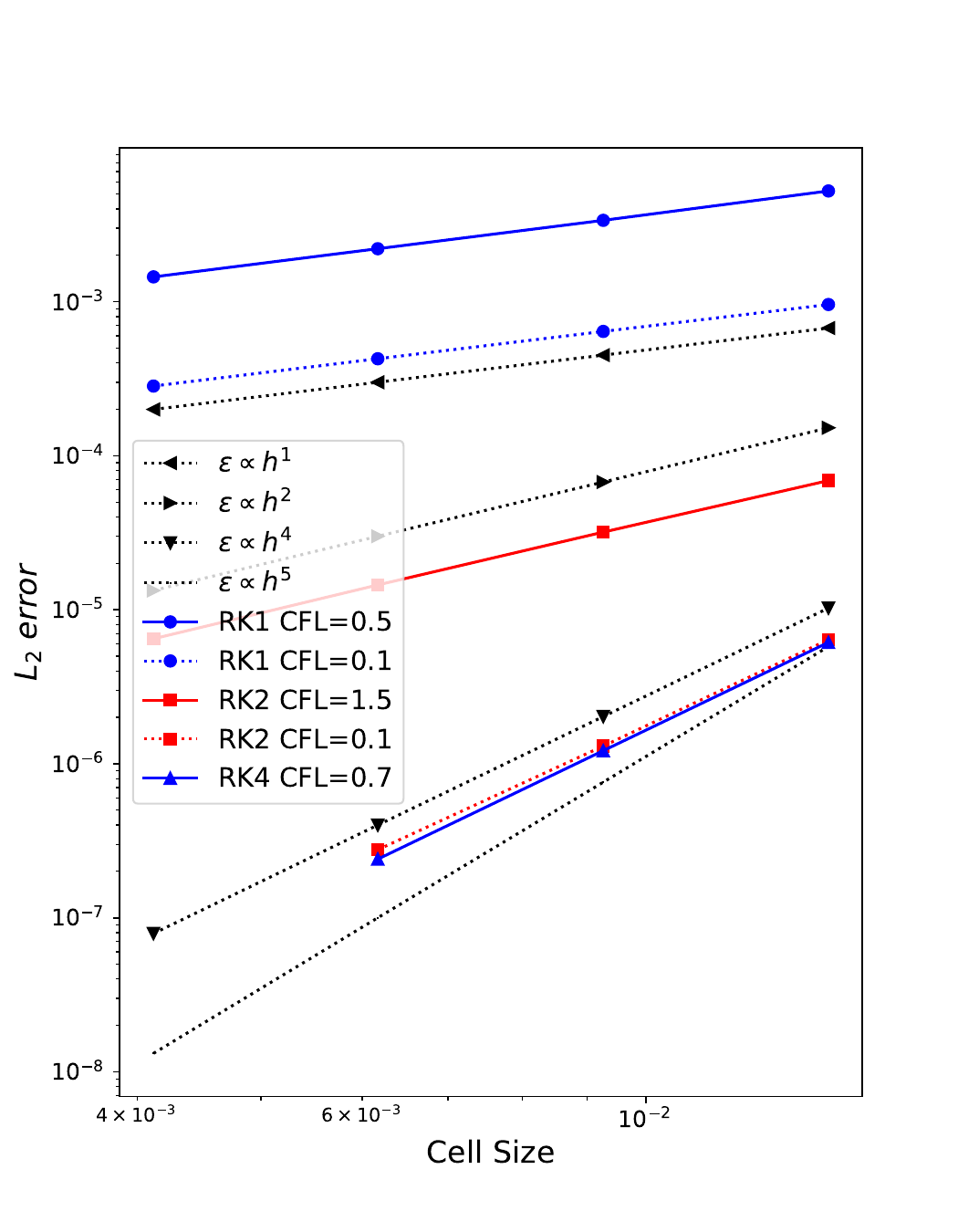}
   \vspace{-1.2cm}
   \caption{ 
       Convergence behavior of FD4 with three different time integrators for
       different CFL constants. The convergence of the final error depends on the relative magnitude of the contributing terms, i.e., spatial and temporal discretization errors. See the text for detailed explanations. 
     \label{fig_gw_convergence}
   }
 \end{figure}

\revision{
On the other hand, different orders of the RK schemes give rather consistent results against the analytical solution. To further investigate different effects from the temporal discretization scheme, we calculate the $L_2$ error of the Hamilton constraint as
\begin{equation}
    L_{2, H}:=\sqrt{\frac{\sum_{i}(H^i_{\rm simulated}-H_{\rm theoretical})^2}{n}},
\end{equation}
where $n$ is the total number of grid points, and we always have $H_{\rm theoretical}=0$. We conduct a convergence study for different orders of the RK scheme based on this accuracy metric at code time $t=0.4$ and report the result in Figure~\ref{fig_gw_convergence}. A similar study can be conducted for any other reasonably small time stamp, and long-term effects will be the subject of study later. All physical and running parameters of tests here are identical to the first benchmark above except for the KO coefficient, which is set to 0.1. }

\revision{
In our explicit time-stepping scheme, three terms contribute to the $L_2$ error:
\begin{enumerate}
    \item The spatial discretization error of the physical quantities, which is the
 same for all setups and scales with the mesh size in a convergence speed $\sim h^4$.
    \item The time discretization error due to the Runge-Kutta scheme of choice. Through Eq.~(\ref{eq_exa_cfl_condition}), the chosen time step size correlates
 linearly to the mesh size, thus following the convergence line $\sim h^k$ where $k$ is the order of the RK scheme.
    \item The influence of the KO term, which is an artificial construct and hence
 constitutes an error in itself. It has a convergence speed $\sim h^5$ (see \ref{app:solver}).
\end{enumerate}
How the $L_2$ error in our tests converges depends on which term takes the lead in magnitude. As shown in Figure~\ref{fig_gw_convergence}, for the RK1 scheme, the temporal discretization error is quite large and results in an expected convergence speed $\sim h^1$ in the Hamiltonian constraint violation (blue lines with circle markers). RK2 and RK4 have relatively small temporal errors and if we have a too-large KO coefficient, then term 3 would overtake the leading term in error and yield an incorrect fifth-order convergence speed (not shown). When a proper (small) KO coefficient is set, the RK4 prediction also shows an expected fourth-order convergence (blue line with triangle markers). However, if the orders of spatial and temporal schemes do not match, the convergence of the final error will depend on the comparison of the magnitudes of each term. To demonstrate this, we plot two lines of the RK2 scheme with different ${\rm CFL}$ ratios, i.e., different sizes of the time steps. When the timestep is small (${\rm CFL}=0.1$, red dotted line with square markers), the RK2 scheme shows nearly identical error magnitude and convergence speed with the RK4 line, indicating we are in a regime dominated by spatial discretization errors. On the other hand, in the big timestep case (${\rm CFL}=1.5$, red solid line with square markers), the temporal discretization error is big enough and takes the lead, thus the line shows $\sim h^2$ convergence speed as we expected. We also identify more complex convergence behavior for the RK2 scheme when an intermediate time step size is used ($0.1<{\rm CFL}<1.5$). We hope to investigate the convergence properties of \acronym's numerical schemes in more detail in future work.}

 \begin{figure}[htb]
   \centering
   \includegraphics[width=0.5\textwidth]{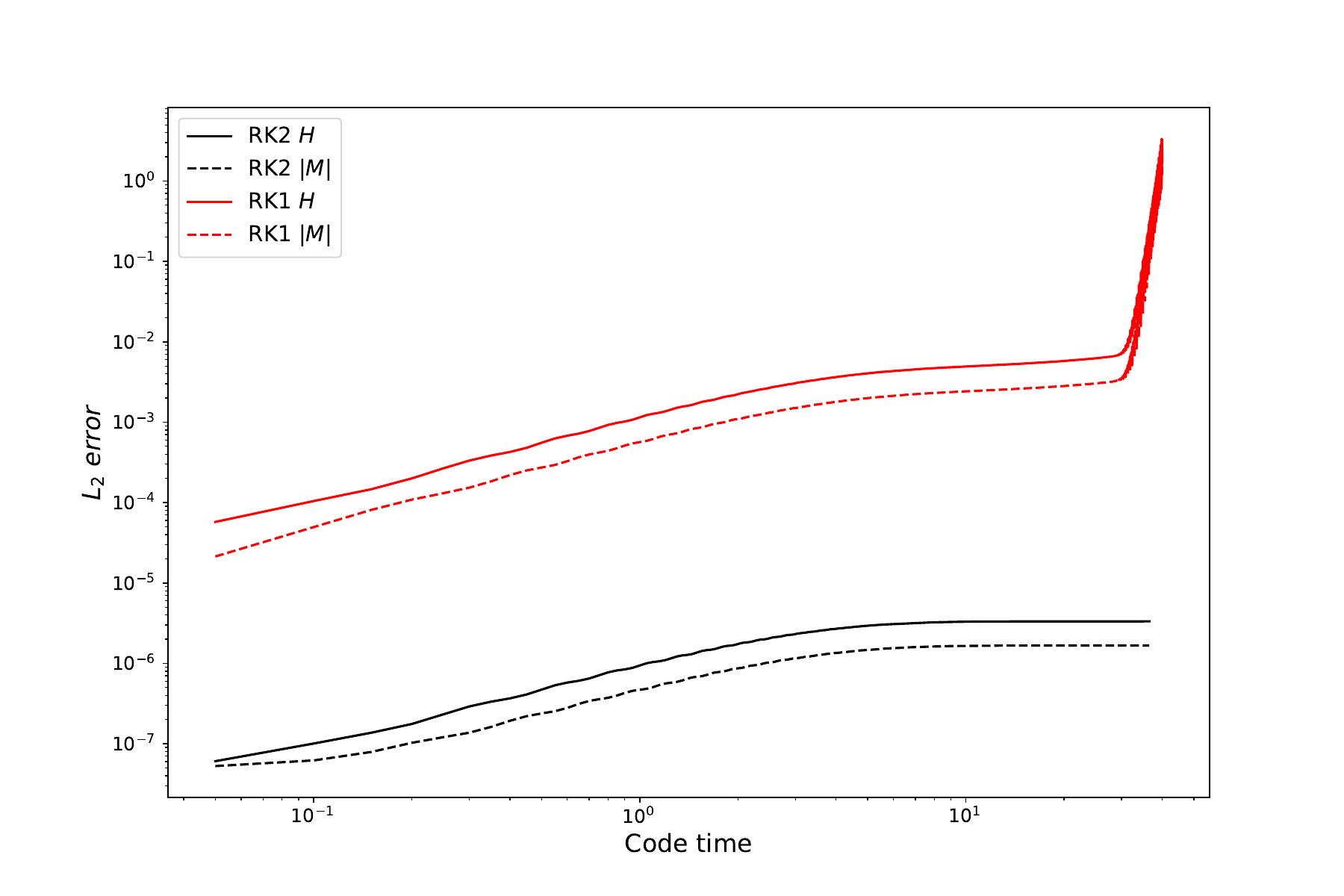}
   \caption{ 
       \revisionTwo{Long-term evolution of the $L_2$ error with first-order (RK1) and second-order (RK2) time-integrator for the same gauge wave setup. The CFL constraint and KO term weight are chosen to dampen the constraint violation effectively. The low-order temporal scheme exhibits significant error accumulation, eventually leading to a crash, while the second-order scheme maintains a more stable error evolution throughout the simulation.}
     \label{fig_gw_long_term_evolution}
   }
 \end{figure}

\revision{
 For a code with known convergence behavior and a reasonably chosen damping
 through the KO term and the CFL factor, we expect to be able to run long-term
 simulations.
 Simulations without a damping term are in turn expected to become unstable
 after a while \cite{Alic12}.
 This hypothesis is subject to our final gauge wave experiment.
 Unfortunately, our data suggest that a mere application of the KO correction
 plus a reasonably small CFL factor is not sufficient to construct a long-term
 stable \revisionTwo{solution} (Figure~\ref{fig_gw_long_term_evolution}):
 Even though the error seems to plateau after a while, it ``suddenly'' explodes within the plateauing region.
 Plots of the wave profile just before the error explosion reveal that the wave
 the pattern becomes skewed (not shown):
 The original \revisionTwo{sinusoidal} pattern approaches a smoothed-out sawtooth profile,
 i.e.~we observe some stiffening of the numerical solution.
 This is still a smooth effect, i.e.~difficult for an artificial KO term to
 flag.}
 
\revision{
 Convergence studies at a certain sample point consequently are 
 insufficient to make statements on the long-term applicability of a code base.
 They might suggest that low-order time integrators are capable of long-term evolution of the system as long as we are willing to accept the resulting
 temporal discretization error.
 However, in practice, errors in the time evolution accumulate and eventually render the
 simulation unstable, i.e.~make the scheme consistent but not converging.
 \revisionTwo{We need to employ a high-order temporal discretization scheme (RK2 or RK4) that matches our high spatial discretization order to construct stable schemes for the gauge wave. Our results demonstrate that a second-order temporal scheme has significantly reduced the $L_2$ error and stabilizes the simulation, as shown by the black lines in Figure~\ref{fig_gw_long_term_evolution}.}}
 

\revision{
 The exact interplay of the KO term, CFL condition, Hamiltonian constraint violation, and
 long-term stability has to be the subject of future studies.
 We notably cannot make statements yet if this behavior results from an error
 accumulation that only arises for particular stationary or quasi-stationary
 solutions. Empirical data suggest that this is the case and that notably outflow
 conditions allow the error to escape from the domain (see the single black hole benchmark below). From hereon we pick reasonably small CFL factors and assume that
 the solution remains stable and consistent, while the arising error is
 dominated by the temporal discretization order.}

\subsection{Single Schwarzschild Black Hole}
\label{sec:result:sbh}

%
%
We next examine a single Schwarzschild black hole. This is yet another stationary test, where all numerical ingredients have to be carefully balanced against each other such that the initial setup effectively yields a long-term stable, time-invariant solution. The black hole is placed at the origin of the three-dimensional domain, with an ADM mass $M=1$ and zero spin $S=0$. All initial conditions for this test stem from the \texttt{TwoPuncture} module of the {\sc Einsteintoolkit} library and \revision{the initial lapse is set to}
\begin{equation}
    \alpha=\frac{1}{2}\left(\frac{1-\frac{1}{2}\left(M / r\right)}{1+\frac{1}{2}\left(M / r\right)}+1\right).
\end{equation}

\noindent
\revision{Other initial profiles and lapses (e.g., a power of conformal factor) are also used in other numerical codes. However, convergence from different setups has been observed: different initial lapses evolve into nearly the same static solution.} The initial shift vector is set to zero. Also, the extrinsic curvature $K_{ij}$ vanishes in the initial condition: its trace is zero as we assume the maximal slicing $K=0$, and its residual is zero as no linear or angular momentum is presented in this system.  Finally, we also assume conformal flatness in the initial condition, i.e.~$\tiga_{ij}=1$.

Besides the lapse field $\alpha$, the only non-trivial quantity in the initial condition of this scenario is the conformal factor $\phi$, which can be solved analytically as its source term now vanishes. The solution of its equation is
\begin{equation}
    \phi \equiv \psi^{-2}=\left(1+\frac{M}{2r} \right)^{-2}.
\end{equation}

%
%

The domain in this test has a size of $[-9M,9M]^3$. We use three levels of \revision{SMR (Static Mesh Refinement)} based on the radial distance from the origin, which refines the domain at radius $r=7M$ and once again at $r=3M$. As we adopt a three-partition \cite{peano}, the volume size of each level are $[0.333M,0.111M,0.037M]$. Sommerfeld conditions (Section~\ref{sec:code:bc}) suppress reflections along the domain boundary. No volume center coincides with the origin to avoid resolving numerical infinity at the puncture location.

For setup and running parameters, we adopt $\kappa_1=0.1, \kappa_2=0, \kappa_3=0.5, e=c=\tau=1.0$ and $\mu=0.2$ for a similar evolving system as the original CCZ4 system\cite{Alic12}. The fully functional gamma driver condition is employed for the evolution of the shift vector $\beta^i$ in the black hole system, and we set the parameters as $f=0.75$ and $\eta=1$ in Eqs.~\eqref{eq_NR_foccz4_shift_gauge} and \eqref{eq_NR_foccz4_b^i}. The KO coefficient and the CFL ratio remain at $8.0$ and $0.1$, respectively, and the simulation runs until a code time of $T=240$ \revisionTwo{with the RK1 temporal scheme}. All figures below utilize $x:=r/M$ as the unit of distance for simplicity.

\begin{figure}
  \centering
  \includegraphics[width=0.49\textwidth]{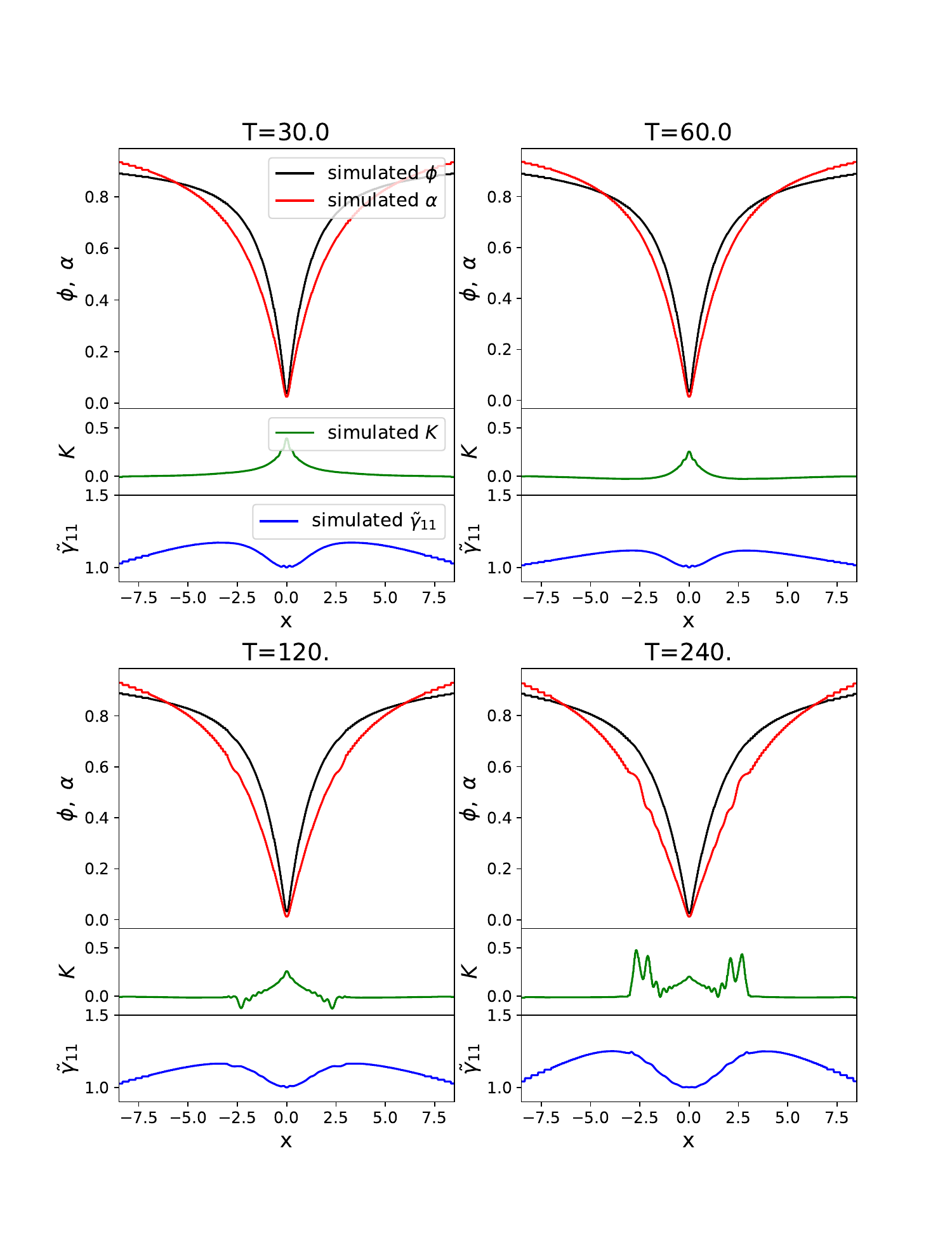}
  \vspace{-1.6cm}
  \caption{
    Profiles of $\phi$, $\alpha$, $K$ and $\tiga_{11}$ along the $x$-axis at four-time stamps $T \in \{30,~60,~120,~240\}$ for a single Schwarzschild black
    hole subject to a second-order formalism.
   \label{fig_sbh_s02}
  }
\end{figure}

\begin{figure}
  \centering
  \includegraphics[width=0.5\textwidth]{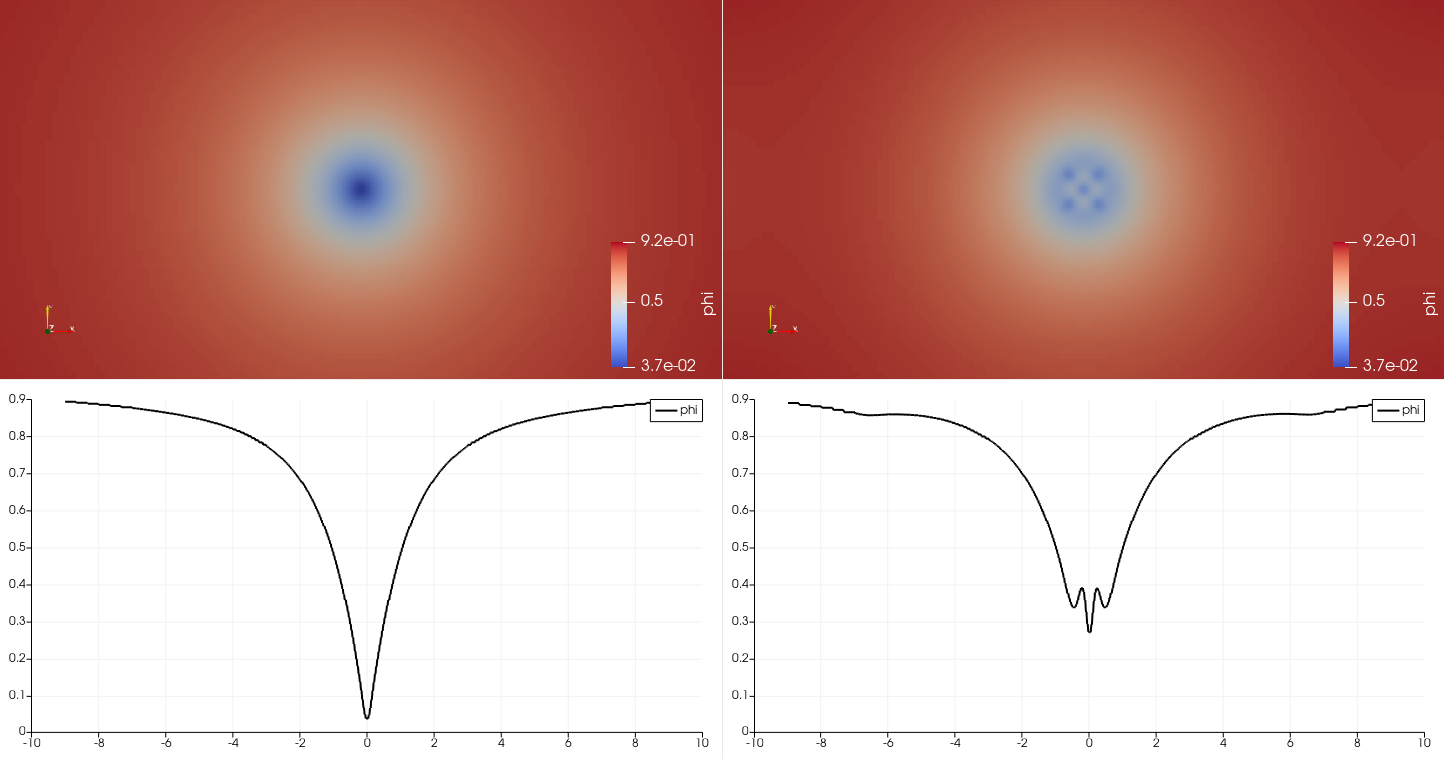}
  \caption{
    Comparison of the $\phi$ field of the single Schwarzschild black hole
    at timestamp $T=36$ in the second-order (left) and first-order
    formulation (right). 
    Color maps illustrate a 2d cutting through the black hole center (upper
    panels), while the profiles track $\phi$ along the $x$-axis (lower panels).
    \label{fig_sbh_foso}
  }
\end{figure}

%
%
\paragraph{Semi-singularity around black hole}
Over an initial ramp-up phase, our code yields a stable evolution of the space-time for \acronym's second-order formulation (Figure~\ref{fig_sbh_s02}). The solution approaches the static solution as expected \cite{hannam:2007:puncture}. However, a first-order formulation for the same setup is not stable, the simulation crashes due to an unphysical solution (Figure~\ref{fig_sbh_foso}). While the second-order formulation yields a rather smooth and long-term stable solution, the
first-order formulation develops instabilities around the center, which eventually lead to the crash. The same experiment with a higher-order time integrator brings the crash forward in time for the first-order formulation.

One possible explanation for the instability of the first-order formulation is that the derivatives of the auxiliary variables become large compared to those of the primary variables. Their errors are hence relatively large, too. For example, a primary quantity with a magnitude of $0.0$5 (a typical value for the conformal factor around puncture) would have a derivative around unity given a volume size of $0.04$. Even a $1\%$ numerical error within the derivative evolution therefore has a high impact. 

The second-order formulation avoids this issue as the first-order derivatives are re-calculated every time step. They do not decouple from the primary variables (i.e., evolving upon their own evolution equations) and amplify themselves. It is not clear if a recalibration of additional damping terms could tackle the first-order instabilities which are inherent due to the semi-singularity of the solution close to the black hole. The derivatives are physically large. Adaptive mesh refinement can possibly mitigate this effect, but it may only serve to postpone the numerical difficulties in time.

\paragraph{Challenges arising around refinement boundaries}

\begin{figure*}
  \centering
    \includegraphics[width=0.8\textwidth]{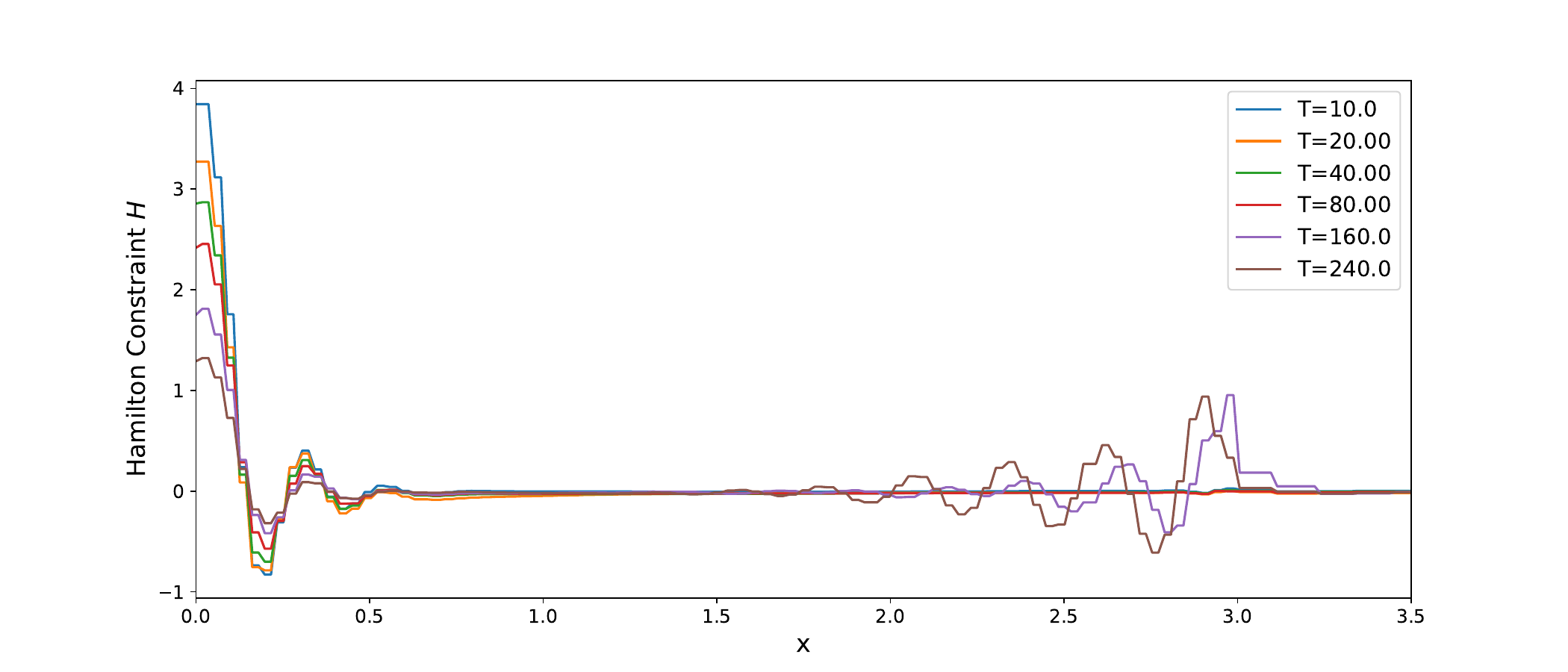}
    \vspace{-0.4cm}
  \caption{
    Violation of the Hamiltonian constraints $H$ for the Schwarzschild black
    hole along the positive $x$-axis at $T \in
    \{10,~20,~40,~80,~160,~240\}$.
    \label{fig_sbh_ham}
  }
\end{figure*}

\begin{figure*}
  \centering
    \includegraphics[width=0.8\textwidth]{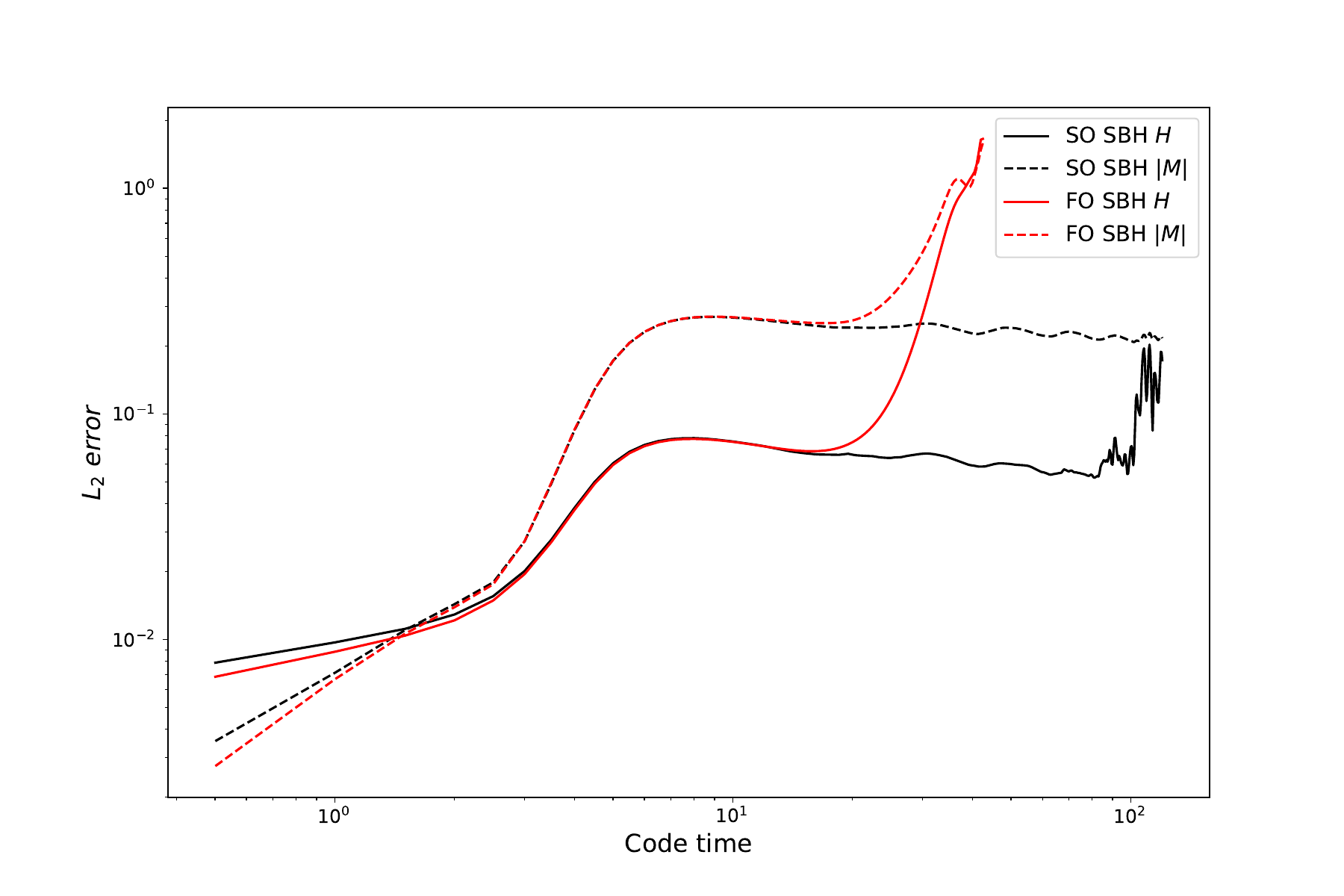}
  \caption{
  Evolution of the $L_2$ error of the Hamiltonian constraint $H$ and the magnitude of Momentum constraint $|M|$ for the single black hole scenario in two formulations. First-order and second-order formulations give nearly identical evolution initially, but eventually the former becomes unstable and crashes. The second-order formulation can run stably until the errors raised on the refinement boundary become non-negligible.
    \label{fig_sbh_L2_evo}
  }
\end{figure*}

%
%
Aggressive SMR introduces its specific challenges: with SMR, the evolution of a second-order formulation with KO damping remains stable until a distortion appears around the resolution boundary ($r=3M$), where the fine grid transitions into a coarser resolution. There, $K$ and $\alpha$ show a discontinuity of the gradient which gets amplified over time. $\tiga_{11}$ also exhibits fluctuations around the resolution boundary. The pattern is more clearly shown in Figure \ref{fig_sbh_ham}, where the profiles for violation of the Hamiltonian constraint $H$ are plotted along the positive x-axis, at six timestamps. Any violation around the black hole located at the center of the domain gets damped. However, the fluctuations along the resolution boundary ($r=3M$) continue to grow and start to propagate inwards. The simulation is likely to become unstable when the fluctuations reach the puncture. \revision{This is confirmed by Figure \ref{fig_sbh_L2_evo}, where we plot the evolution of the $L_2$ error of the Hamiltonian constraint up to code time $120$. The $L_2$ error in the first-order formulation increases quickly, departs from the stable second-order counterpart around 20 code time, and leads to a crash later. In contrast, the second-order formulation can provide a stable solution for the puncture all the time and shows continuous damping of the constraint violation until the errors at the refinement boundary start to take the lead and amplify the violations.}

%
%
Any adaptive mesh resolution transition on non-conformal meshes is known to trigger reflections of outgoing waves:
We have hanging vertices and hanging faces, i.e.~fine grid vertices with less than \revision{$ 2^d $} adjacent cells on the finest mesh level and faces where the left and the right adjacent cell are of different size.
Let a wave travel from a fine mesh patch through a face into an adjacent coarser one. 
High-frequency modes stemming from the fine-resolution domain cannot be
represented on an adjacent coarser grid and hence are reflected back into the the direction they are coming from. For dynamic simulations as encountered for moving black holes, these spurious reflections typically do not cause major stability issues. For the stationary black hole, however, we observe that the reflection waves are caught within the fine resolution domain and eventually amplify each other over time.
They accumulate.
A stationary black hole instead can create standing error waves that grow larger over time.

%
%
\acronym's out-of-the-box linear interpolation and restriction schemes are not powerful enough to handle our highly non-linear PDE system over a long
simulation span if we encounter a stationary black hole resolved with mesh refinement. However, as long as black holes move, the KO dissipation is sufficient to compensate for any spurious reflections. \revision{A more general solution is to implement higher-order interpolation and restriction schemes, which can provide higher accuracy around the refinement boundary. We have already observed a significant improvement in reflection suppression when we apply a second-order scheme in our recent tests of single black hole spacetimes (not shown). The refinement transition strategies play an important role in achieving high-quality numerical relativity simulations (e.g., see discussion in \cite{AMR_in_NR_2} and \cite{GRChombo2022}), and we plan to report more comprehensively on how different strategies affect the simulation behavior of \acronym\ in future work.}


\paragraph{Coupled numerical schemes}

\begin{figure}
  \centering
  \includegraphics[width=0.35\textwidth]{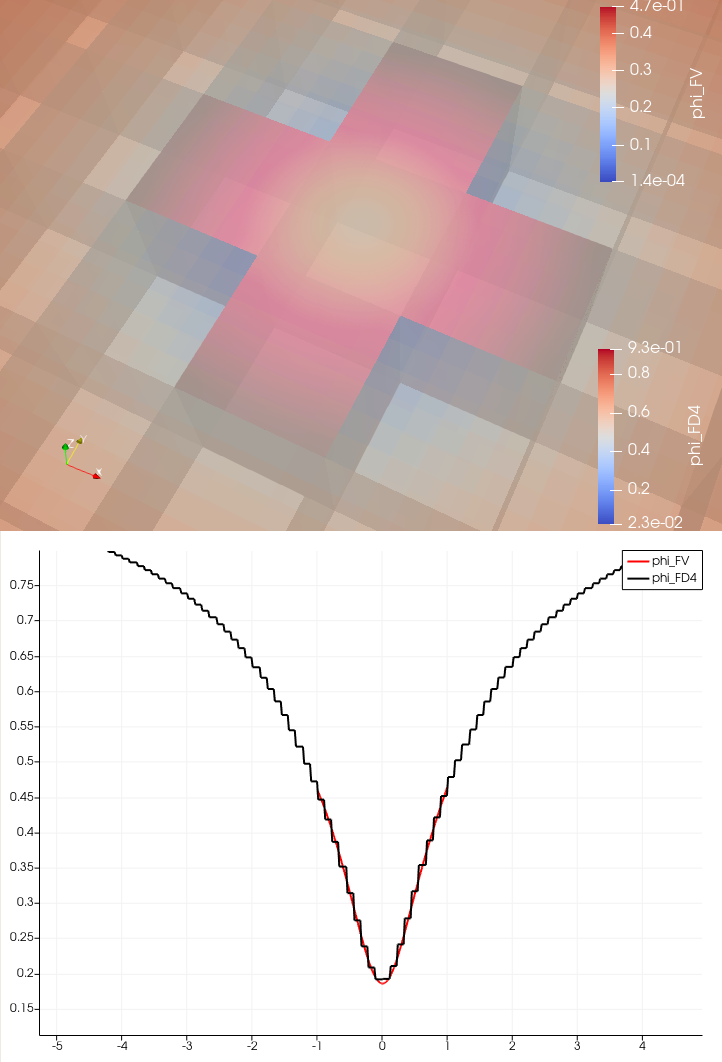}
  \caption{
    An FV \acronym\ solver is embedded within an FD4 scheme such that it covers
    the stationary black hold (top). 
    The FV solution on a significantly finer mesh yields a smooth representation
    of the solution around the black hole and overwrites the behavior of FD4
    there. $T=3$ (bottom).
    \label{fig_sbh_coupling}
  }
\end{figure}

%
%
Finite Volumes rule itself out as a global solver due to their high dissipation and high computing cost and memory footprint vs.~accuracy. We need a very fine mesh to compete with a higher-order scheme. However, we recognize that the KO term used to stabilize FD4 injects numerical diffusion, i.e., mimics what we suffer from with FV, is in itself a ``magic'' stabilization term lacking a physical motivation.

It is hence a natural idea to combine different schemes. We use FD4 throughout the domain. However, we also use the FV scheme around the black hole (Figure~\ref{fig_sbh_coupling}, upper panel). It is a localized solver typically covering seven patches around the puncture. Its Dirichlet boundary values are projected from the FD4 solution. Within the overlap region, the FV scheme is considered to be valid and overwrites the corresponding FD4 solution after each time step, served as a limiter. 
We assume the CFL condition is dominated by

\begin{equation}
  {\delta t} \leq C\frac{h}{p^2},
\end{equation}

\noindent
where we have a polynomial degree of $p=4$ for the FD4 solver and $p=1$ for FV.
To match the time step sizes, the FV patch size is $4^2$
times larger (per dimension) than its FD4 cousin, \revision{i.e., contains $16^3$ times more volumes per patch}.

%
%
The profile of the $\phi$ field of the two solvers shows a stronger smoothness of the FV scheme around the black hole (Figure \ref{fig_sbh_coupling}, lower panel). As the FV solution overwrites FD4 in the center, while the FD4 solution determines the halo data of the FV solution, the solutions appear to overlap. \revision{However, only the FV scheme evolves without interference.}

%
%
The solver coupling (Sections~\ref{sec:code:solver_coupling} and
\ref{sec:workflow:coupling}) complements spatial adaptivity with numerical
adaptivity, where the numerical scheme changes throughout the domain.
It is a realization of the classic a posteriori limiter concept~\cite{Dumbser18}.
The Finite Volume patches serve as a tool to resolve the
semi-singularities of the solution as well as a sponge sucking up any high-frequency error modes.
Its overly dissipative character is now used as a feature rather than a challenge.

While the coupling is an appealing concept to create smooth, stable solutions without the need to use an artificial KO damping, i.e.~solely due to first-order numerical principles, several challenges remain: tests with \acronym\ suggest that a direct overwrite of the FD4 solution can sometimes yield spurious waves similar to the reflections from the refinement boundary. \acronym\ hence offers averaging and lets the ``validity'' of the FV scheme fade in:
Directly around the black hole, the FV scheme dominates.
Further away, FD4 yields a valid solution to and the FV solution is not even computed.
In-between, we have a buffer zone.
Both solvers run and the eventual result is a weighted average between the two solutions.
The averaging shifts from ``all FD4'' over ``predominantly FD4'' over to ``predominantly FV''.
While this seems to yield higher-quality data, the precise effects have yet to be studied.
It is also an open question of how big the FV region around the black hole has to be chosen overall. If it is too small, the damping effect of the FV region does not cover the unstable domain and we suffer from instabilities again. If it is too large, the numerical dissipation of FV pollutes the result in the region we are interested in. The biggest challenge arising from the coupled approach, however, is the likely difficult load balancing, as almost all computing effort is spent on a few FV patches of a very high resolution. We aim to investigate and explore more the potential of this coupling scheme in future work.
\subsection{Rotating Binary Black Hole Merger}
\label{sec:result:bbh}

%
%
We finally switch to an equal-mass rotating black hole merger simulation. To construct the correct initial condition that allows for quasi-circular orbits, we utilize one set of the computed parameters from \cite{tichy:2004:bbhIC} through \acronym's ported TwoPuncture component. The binary black holes both have a bare (puncture) mass of $m_+=m_-=0.46477$ and zero spin, $S_+=S_-=0$. 
Their initial distance is $d=4M$. The black holes are located at the coordinates $[2M,0,0]$ and $[-2M,0,0]$ respectively, and rotate on the $x$-$y$ plane, with the initial linear momentum $P^i_\pm=[0,\pm 0.19243M,0]$. This setup corresponds to an ADM mass of $M=0.5$ for each black hole and a total ADM mass of the system $M_\text{tot}=0.98074$.

The average initial lapse 
\begin{equation}
    \alpha=\frac{1}{2}\left(\frac{1-\frac{1}{2}\left(m_{-} / r_{-}\right)-\frac{1}{2}\left(m_{+} / r_{+}\right)}{1+\frac{1}{2}\left(m_{-} / r_{-}\right)+\frac{1}{2}\left(m_{+} / r_{+}\right)}+1\right),
\end{equation}

\noindent
with vanished initial shift is used. Otherwise, we preserve the black hole settings, i.e.~$\kappa_1=0.1,\kappa_2=0,\kappa_3=0.5, e=c=\tau=1.0$ and $\mu=0.2$. The gamma driver gauge condition uses the parameters $f=0.75$ and $\eta=1$. Compared to the FD4 setup, no modifications are made to the KO coefficient ($8$) and the CFL ratio ($0.1$). \revisionTwo{We also use the RK1 temporal scheme for this scenario, and convergence tests have confirmed that we achieve the expected order ($\sim h^1$) of convergence across different resolutions with it (test not shown).}

We employ a domain of $[-12M,12M]^3$ with three levels of \revision{SMR}. The domain gets refined at radius $r=9M$ and again at $r=5M$ around the center of the coordinate system, where we assume the black holes to merge. We obtain FD spacings at each level of $[0.333M,0.111M,0.037M]$. Sommerfeld boundary conditions are used.
We run both the first-order and second-order formulation for this test, taking
into account previously described stability challenges for the former once the
black holes have merged.

\begin{figure*}
  \centering
    \includegraphics[width=0.9\textwidth]{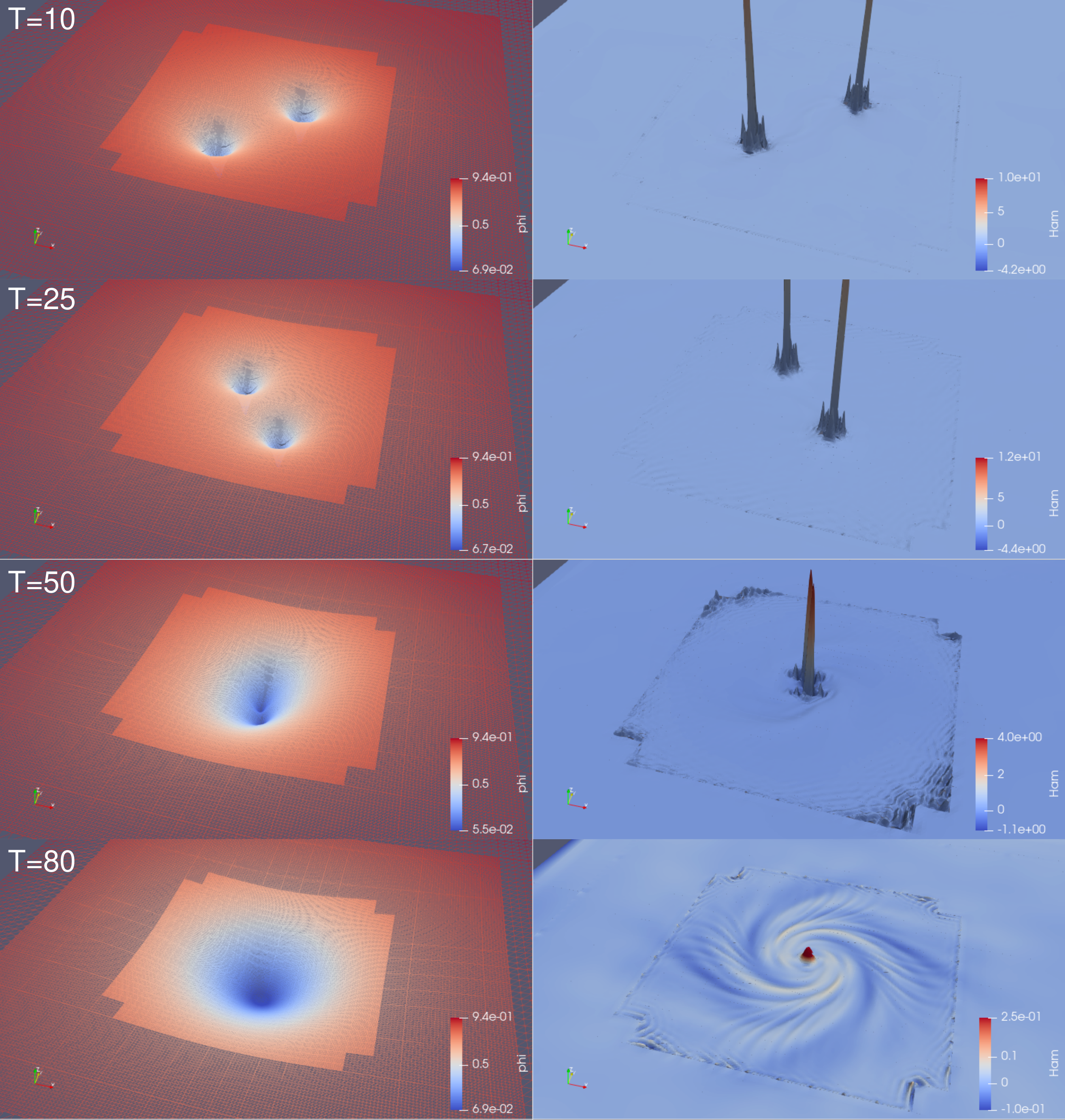}
  \caption{
     The snapshots of the conformal factor $\phi$ and the violation of
     the Hamiltonian constraints $H$ at time stamps $T \in \{10,~25,~50,~80\}$
     over three-dimensional warped wireframes and surfaces, respectively. 
    \label{fig_bbh_snapshot}
  }
\end{figure*}

\begin{figure*}
  \centering
  \includegraphics[width=0.70\textwidth]{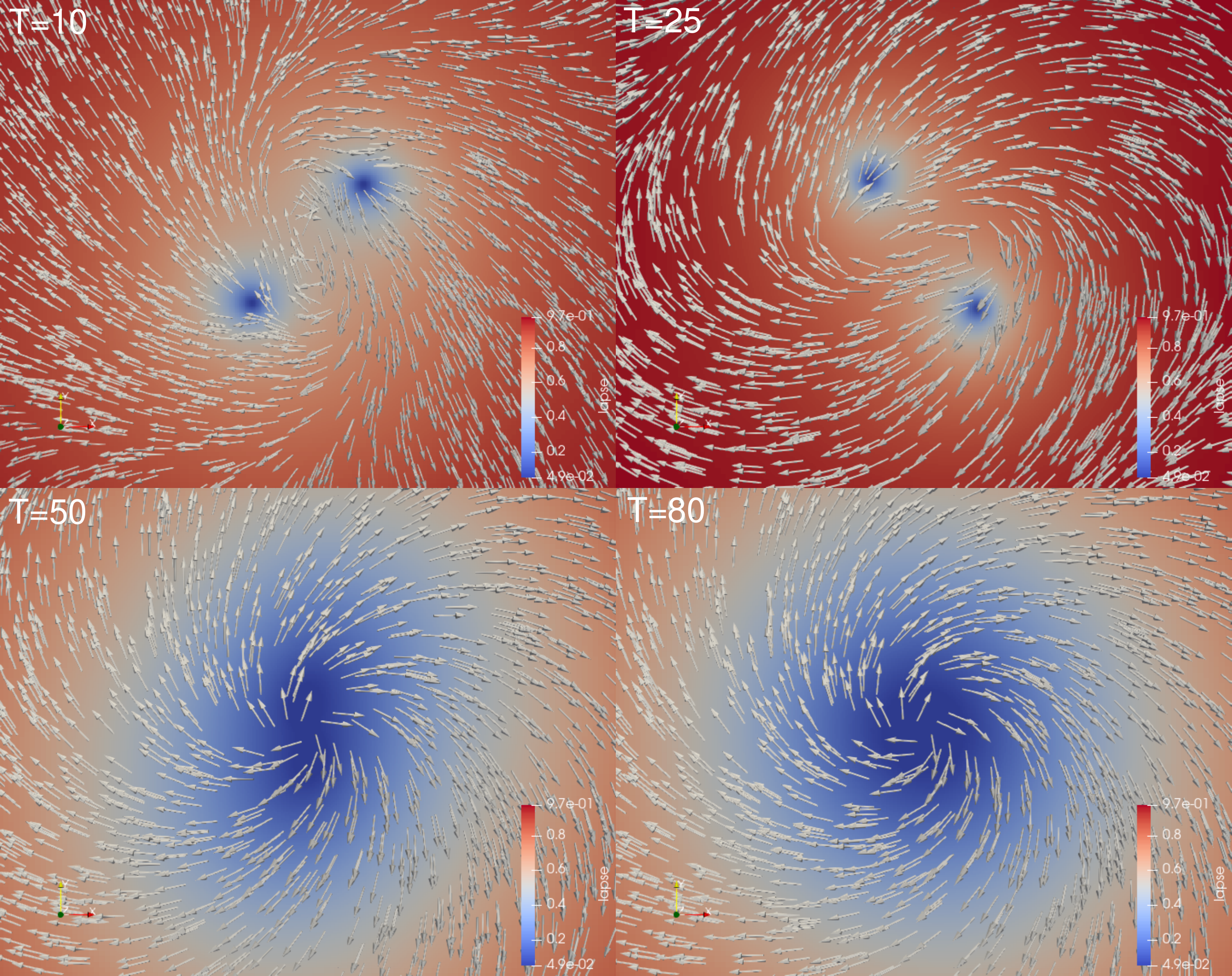}
  \caption{
    Snapshots of the gauge quantities, lapse $\alpha$ and shift vector
    $\beta^i$, for Figure~\ref{fig_bbh_snapshot} as color map or
    vector field respectively.
    \label{fig_bbh_alpha_beta}
  }
\end{figure*}

\begin{figure}
  \centering
  \includegraphics[width=0.5\textwidth]{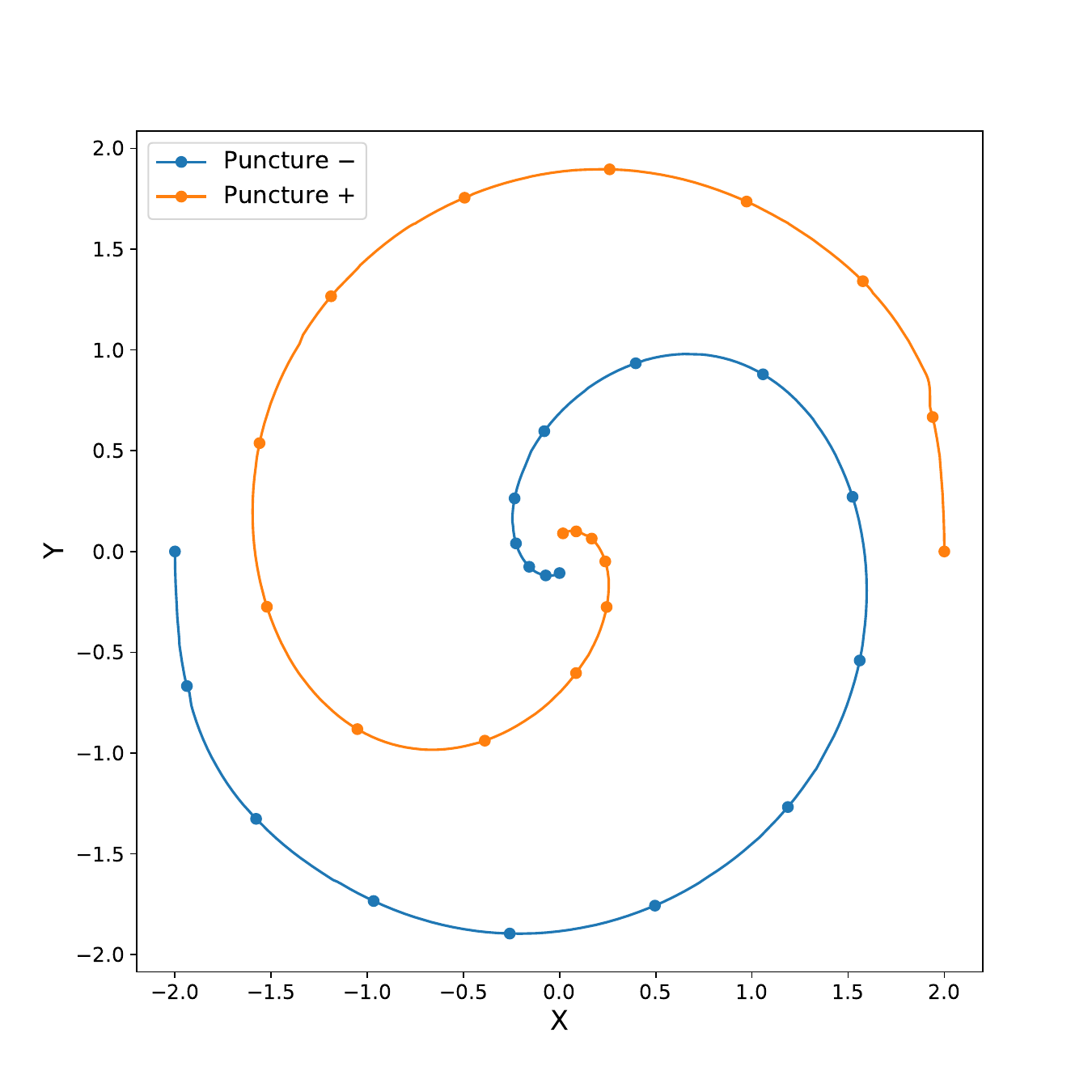}
  \caption{
    The trajectories of the two punctures in the binary black hole merger test. 
    The two black holes $\pm$ are initially located at
    $[x,y,z]=[2M,0,0],[-2M,0,0]$. \revision{The trajectories for the final ring-down stage were removed for clarity. The kink observed in the initial stage of the $+$ puncture arises from a minor inconsistency in data output as the tracer transitions between different subpartitions managed by separate MP threads. The symmetric trajectories that follow confirm that there is no underlying computational error.}
    \label{fig_bbh_puncture_tracker}
  }
\end{figure}

\begin{figure*}
    \centering
    \includegraphics[width=0.9\textwidth]{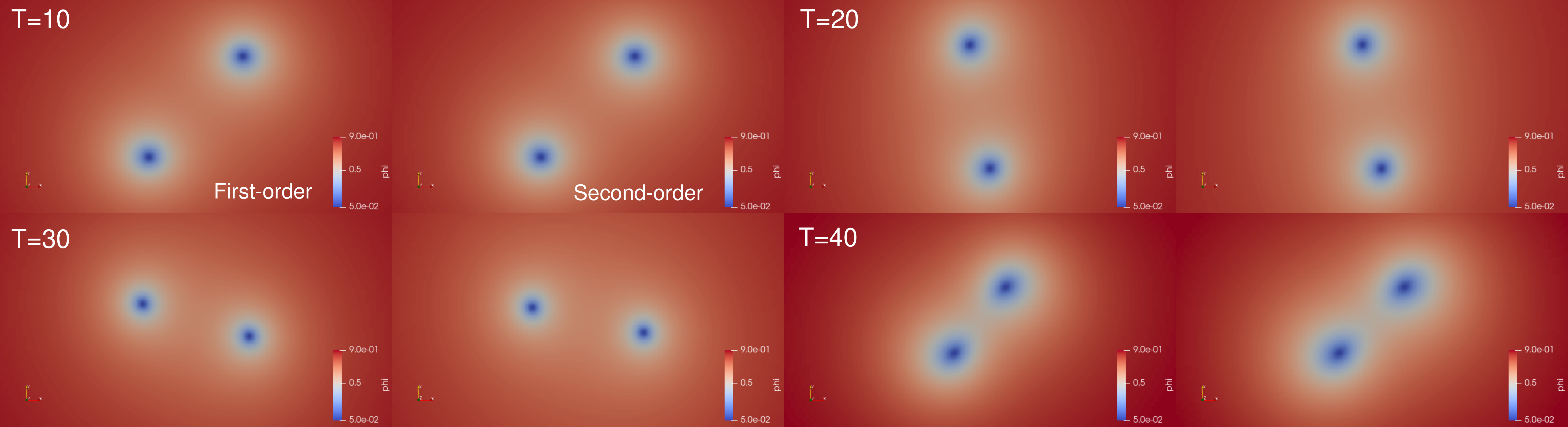}
  \caption{
    Evolution of the $\phi$ field in the central region at $t
    \in \{10,~20,~30,~40\}$.
    We compare the data stemming from a first-order formulation (left)
    to the second-order formulation (right).
    \label{fig_bbh_fo_so_GRC}
  }
\end{figure*}

\begin{figure*}
    \centering
    \includegraphics[width=0.9\textwidth]{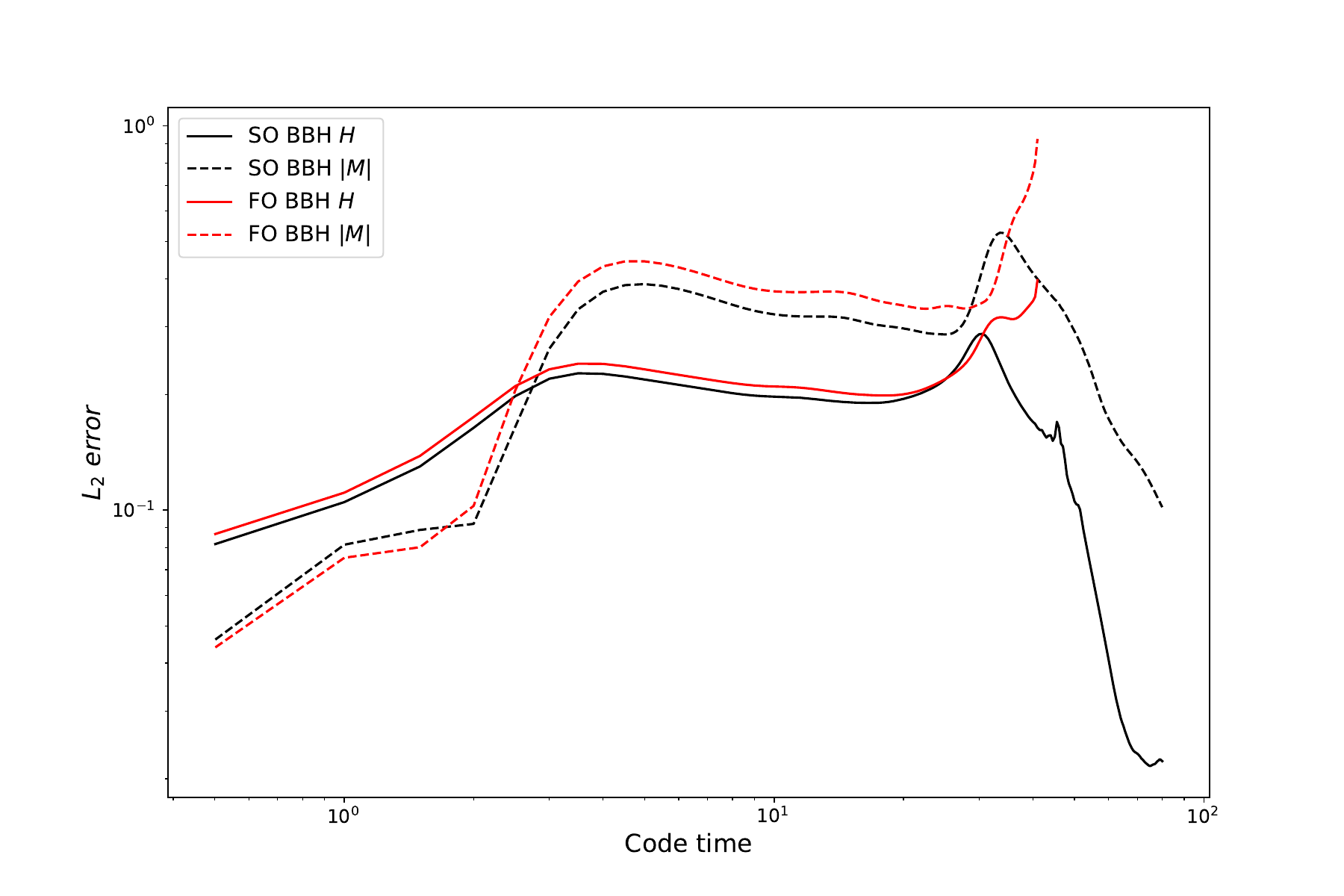}
  \caption{
    Evolution of the $L_2$ error of the Hamiltonian constraint $H$ and the magnitude of Momentum constraint $|M|$ for the binary black hole scenario for both formulations. The second-order formulation is stable throughout the evolution and shows a continuous damping of the violations after the merge. The first-order formulation can capture the dynamics of black holes correctly and yields only a slight increase in error during the rotation phase. However, it runs into a crash soon after the merge, when the system degenerates back to a single black hole scenario.
    \label{fig_bbh_L2_evo}
  }
\end{figure*}

%
%
The binary black holes complete one-and-a-half circular orbits before they merge
at a code time of $T \approx 70$. The system continues to evolve as a single
black hole until $T \approx 120$, where we terminate the simulation. The rotation and merging processes in the simulation are relatively smooth and subject to only limited violations of the constraints (Figures~\ref{fig_bbh_snapshot} and \ref{fig_bbh_alpha_beta}). As the punctures are rotating, the shift vector exhibits a spiral pattern. Its reversed direction at the puncture location indicates the velocity direction of the puncture, i.e., we can use it to inform \acronym's tracers from Section \ref{sec:code:tracer} (Figure~\ref{fig_bbh_puncture_tracker}). The spiral pattern in the post-merge stage shows that the remnant of the merged black holes carries a spin. Besides the actual physical data, some fluctuation starts to creep into the simulation. It stems from the resolution transitions and gets magnified in the post-merge phase. In line with previous experiments, the first-order formulation encounters instability after merging and soon runs into a crash. However, it can properly resolve the rotation of black holes before that. The dynamic behavior of the first-order and second-order
scheme are presented in Figure~\ref{fig_bbh_fo_so_GRC}, both schemes
have the same trajectories until the approaching punctures accumulate errors and start to affect the evolution. \revision{Furthermore, Figure~\ref{fig_bbh_L2_evo} shows the evolution of the $L_2$ error of the Hamiltonian constraint $H$ and the magnitude of Momentum constraint $|M|$ for both formulations. The first-order formulation can not stabilize the post-merge remnant as it degenerates back to a static single black hole. In contrast, the second-order formulation remains stable and shows a continuous damping of the constraints violation.}
%
%

\revisionTwo{To further validate the accuracy of our code, we conducted a benchmark comparison with GRChombo using the same initial physical conditions. Our simulation replicates the same dynamics with the published code, where the black holes complete approximately one and a half orbits before entering the ring-down phase. Minor differences in the puncture positions were observed over time, which may be from the difference in initial lapse conditions—GRChombo uses a conformal-factor-power initial lapse, while we use an averaged initial lapse. Notably, these differences decrease when higher resolutions or higher-order temporal schemes are employed in our simulation, suggesting a qualitative convergence (not shown). A more robust comparison should be made in terms of observables, such as gravitational waveforms. We plan to conduct a more comprehensive study with large-scale benchmarks in future work.}

Overall, the rotation and pre-merger phase are accurately captured in our benchmark, and the locations of the punctures are correctly tracked by the (inverse) shift vector guiding the tracers. The simulation of the dynamic behavior is stable, which supports the argument that any constraint violation is ``traveling'' through the domain, might be reflected at AMR boundaries, but eventually leaves the domain. We avoid the accumulation of numerical errors at specific locations as long as the black holes move.

\subsection{Context of Experimental Studies}

Our application successfully completes a rotating binary black hole
merger. It is able to handle both the approach phase as well as the actual ringdown. Our studies focus on the introduction of numerical ingredients within \acronym. 
\revision{They validate that the software is fully operational and prepared for more challenging benchmarks on larger scales, which will be crucial for further verifying its accuracy and performance.}
Here, we have to combine all ingredients and notably study a larger computational domain to avoid boundary artifacts (cmp.~squeeze-in effect in Figure~\ref{fig_bbh_fo_so_GRC}) and obtain realistic measurements. Further stationery tracers (cmp.~Section~\ref{sec:result:gw}) then can be used to sample the solution on a huge sphere around the area of interest. With such samples, we can finally calculate $\psi_4$ and its mode quantities.

Our data illustrates how delicate CCZ4 simulations react to numerical inaccuracies. We have seen two schools of thought in literature to tackle the intrinsic instabilities which are due to the nonlinear nature of the equations, their complexity, and notably pollution to the long-term quality of the numerical approximation:
The first school artificially damps equation components such that instabilities materializing in high-frequency solution components diffuse away. Kreiss-Oliger dissipation stands in this tradition, and we implemented it for both first-order and second-order formulation. We notice that these additional terms lack, to the best of our knowledge, physical motivation.
Groups seem to pick the damping calibration depending on their simulation setup, but
systematic studies on its impact are rare.
Typically, we find papers lacking systematic studies if this artificial term can pollute the long-term solution as well for some setups, e.g., with strong shock fronts.
It remains subject of future work to assess to which degree the magnitude of the KO dissipation and other damping components in the evolution system play a pivotal role in the quality of the simulation outcome, i.e., if larger or
certain choices introduce artificial long-term damping similar to the effects we documented for FV.

The other school of thought uses traditional numerical schemes (a limiter) to damp out instabilities. A posteriori limiting as used in ADER-DG methods \cite{Dumbser18} stands in this tradition. However, we notice that the subgrid Finite Volume methods that we use as a posteriori limiter is very costly, and we are not aware of systematic studies on how to couple the limiter spatially and temporally into the domain to avoid global over-dissipation.

Our data cannot conclusively answer if first-order or second-order
formulations of the underlying Einstein equations are superior. The second-order formulations fit more naturally to the PDEs of the numerical relativity and it is widely used by codes of other groups with Finite
Difference kernels. In our benchmarks, the second-order formulation prevents error accumulations in the derivative calculation and thus provides better stability in static black hole systems. Meanwhile, first-order formulations allow us to provide more generic implementations to deploy new solvers--indeed we can drop them into FV and DG formulations quite naturally---and we can realize them with smaller data overlaps and hence less synchronization effort. The halos for example can be chosen smaller, and we do not need additional auxiliary variable reconstruction steps. This is an important property for future large-scale simulations, as data movements dominate the compute cost. Indeed, our second-order \acronym\ variant is about a factor of two slower than its first-order cousin.

\acronym\ is not a finished software product but a tool suite offering some
important astrophysical simulation tools. Several sophisticated ingredients are not yet available with this release paper yet might have a substantial impact on simulation outcomes. Notably, the effects of higher-order interpolation along the AMR resolution boundaries have to be studied. We may assume that coefficient optimization as described in \cite{mccorquodale:2011:highorderIR} yields a smoother and more stable resolution transition in our code which might suppress the error accumulation for quasi-stationary setups. For the dynamic phases, we have yet to study the impact of dynamic adaptive mesh refinement on the solution quality, while techniques such as manual curl cleaning \cite{Dumbser:2020:curlcleaning} might help us to make the
simulation overall more stable and accurate.

\acronym\ provides the infrastructure to study various solver variants and
arrangements of ingredients. Its clear separation of concerns allows us to inject new components easily. Notably, we can even compose and couple them freely. It is subject of future work to study the impact of many numerical techniques, to understand them, and to construct simulation codes that are long-term stable and yield high-quality outputs. Our initial data suggests that this might be best achieved through a dynamic combination of ingredients and by giving up on the concept of a ``one solver for
numerical astrophysics''. It might be better to switch between different solvers and numerical techniques, balancing between computational efficiency (first-order vs.~second-order) and stability (higher order vs. low order discretizations).

\section{Discussion and Conclusion}
\label{sec:conclusion}

Our work introduces \acronym, a suite of solvers and related code modules designed for conducting numerical relativity simulations on \exahype\ 2. The latter is a rewrite of the \exahype\ PDE solver engine from \cite{ExaHyPE1}.
The rewrite has been motivated by the urge that we wanted to design a simulation
software base which realizes a clear separation of concerns on all levels and
yields a composable software stack such that we can relatively quickly study
new physical and numerical approaches in computational astrophysics. Our presentation focuses on the software design in the context of this application domain.

The software design of \acronym\ is guided by the concept of lowering, which is inspired by the compiler construction community: we aim for a very high-level API exposing the software to the user in Python. It allows the user to specify \emph{what} is to be computed. This high-level specification is then successively lowered into a numerical representation of the compute steps and from hereon into technical views which eventually are mapped onto plain C++ code. The mantra here is that the lowering and the underlying code base determine \emph{where} (on a parallel computer), \emph{when} (in which order) and \emph{how} (generic provision of some numerical ingredients) the specification is realized in the source code.

To implement this vision and to showcase the applicability and maturity of the
design, the \acronym\ endeavor had to add further building blocks to the
underlying \exahype\ engine, and we had to run an extensive set of numerical
experiments. \revision{The experiments demonstrate that our code is fully operational}, but also raise further questions, notably around the numerical stability of any composition of ingredients and the philosophical question if first- or second-order formulations are superior. Many aspects of these questions are intrinsically tied to this area of research and have been documented, either implicitly or explicitly, by other codes and groups, too. We consider it a strength of \acronym\ that we can quite flexibly study and swap different methodological ingredients to address these fundamental research questions, although the present work does not yet provide answers to many of them. \revision{The unique solver coupling features of \acronym\ make it an ideal tool for more advanced numerical relativity simulations. For instance, it allows the combination of a finite volume scheme for the matter field with a finite difference scheme for the spacetime background, leading to higher-quality simulations of relativistic hydrodynamics in scenarios such as black hole accretion and neutron star spacetimes. In the same way, undergoing extensions to the particle toolbox---currently used for the tracers---could enable general-relativistic smoothed particle hydrodynamics (SPH) simulations~\cite{Rosswog:2010b,Liptai:2019vyl,Magnall:2023tzm}. }

The success of any astrophysical code hinges upon the question of to which degree the code is able to scale up and to exploit modern machinery efficiently. They allow users to conduct astrophysics simulations with larger scale and higher accuracy. Our presentation largely neglects these aspects and focuses on software and domain challenges. However, we note that our strict separation of concerns and lowering approach yields a software architecture where supercomputing considerations can generically enter the software design in lower abstraction levels, i.e.~performance engineering can be realized without domain expertise. Further to that, our lowering formalism uses the notion of tasks, i.e.~expresses concurrency explicitly. We have started to publish work on how to exploit these properties---for example by offloading sets of tasks to GPUs~\cite{wille:2023:gpuoffloading}---and will make supercomputing aspects subject of future publications using the \acronym\ software architecture.
\section*{Acknowledgements}
\label{sec:acknowledge}
Part of the work was completed during the PhD studies of Han Zhang, funded by the Chinese Scholarship Council (CSC) - Durham University Joint PhD Studentship. We gratefully acknowledge the valuable discussions and assistance provided by our colleagues Holger Schulz, Anne Reinarz, Francesco Fambri, Mario Wille, and Dominic Charrier, as well as researchers from the {\sc GRChombo} group, Katy Clough, Eugene Lim and Lorenzo Rossi.

Our work has been supported by the UK's ExCALIBUR
programme through its cross-cutting project EX20-9 \textit{Exposing Parallelism: Task Parallelism}
(Grant ESA 10 CDEL) made by the Met Office and the EPSRC DDWG projects
\textit{PAX--HPC} (Gant EP/W026775/1) and 
\textit{An ExCALIBUR Multigrid Solver Toolbox for ExaHyPE} (EP/X019497/1). BL acknowledges additional support from ERC (ERC-StG-716522) and STFC (ST/I00162X/1, ST/P000541/1, ST/X001075/1) during this project.
Particular thanks are due to Intel's Academic Centre of
Excellence at Durham University.
This work has made use of Durham's Department of Computer Science NCC
cluster.
Development relied on 
the DiRAC@Durham facility managed by the Institute for Computational Cosmology
on behalf of the STFC DiRAC HPC Facility
(\href{www.dirac.ac.uk}{www.dirac.ac.uk}). The equipment was funded by BEIS capital funding via STFC capital grants ST/K00042X/1, ST/P002293/1, ST/R002371/1 and ST/S002502/1, Durham University and STFC operations grant ST/R000832/1. DiRAC is part of the National e-Infrastructure.

\bibliographystyle{elsarticle-num} 
\bibliography{ref}

\clearpage
\appendix

\section{Experiments}
\label{app:experiments}

The following section describes how to reproduce our results. We assume that we
rerun the experiments against the latest version of \peano\ which contains the
latest version of \exahype. However, all \acronym\ experiments here are made
with the tag version \texttt{2024ExaGRyPE}.

\begin{algorithm*}[htb]
  \caption{
    Clone repository, create autotools environment and build \peano's
    and \exahype's core libraries.
    \label{alg:experiments:clone-and-setup}
  }
  \begin{algorithmic}
    \State git clone https://gitlab.lrz.de/hpcsoftware/Peano
    \State cd Peano
    \State libtoolize; aclocal; autoconf; autoheader; cp src/config.h.in
    \State automake {-}{-}add-missing
    \State ./configure --enable-blockstructured --enable-particles
    --enable-exahype
    \State make
  \end{algorithmic}
\end{algorithm*}

The code base is available directly from our gitlab \cite{Peano_gitlab}.
\peano\ provides support for cmake and the autotools. 
We present the autotools configuration here
(Algorithm~\ref{alg:experiments:clone-and-setup}).

As Peano is a generic framework used for multiple projects, we have to enable
\acronym-specific extensions plus their dependencies manually.
In the code snippet, we enable the block-structured data management, \exahype\
and the particle toolbox which provides us with tracer facilities.
Setups might want to add \texttt{--with-multithreading=x}, where \texttt{x} can
be either \texttt{omp} (OpenMP), \texttt{tbb} Intel's Threading Building Blocks,
or \texttt{cpp} (native C++ threading).
Further to that, multi-node experiments require a build with
\texttt{--with-mpi=mpiicpc}, where the argument is the MPI C++ compiler wrapper
to be used.
GPU extensions are enabled via \texttt{--with-gpus=omp} if OpenMP 5 offloading
is to be used \cite{wille:2023:gpuoffloading}. 
The switch \texttt{--with-gpus=sycl} enables offloading via SYCL \cite{loi:2023:sycl}.

Specific \texttt{CXXFLAGS}, \texttt{LDFLAGS} and \texttt{LIBS} complement the
configuration, and vary depending on your working platform.
For initialization of current black hole experiments, the GSL library~\cite{Galassi:2009:GSL} math library is needed to support the TwoPuncture module from {\sc EinsteinToolkit}.
The \texttt{make} instruction in Algorithm~\ref{alg:experiments:clone-and-setup}
builds a set of static libraries which will be used by every \acronym\
application.

Every \acronym\ application is realized through a Python script, which uses
\exahype's Python API lowering into \peano's Python modules. 
They then generate exclusively C++ code plus a makefile.
The C++ code is linked against
the static \peano\ and \exahype\ modules as well as the astrophysics-specific \acronym\
source files. 
We eventually end up with one big executable
reflecting the \acronym\ script settings:
\peano\ realizes a strict ``we create one executable comprising all required
input settings'' paradigm, i.e.~from hereon no further input data are required.
The generated makefile parses the \texttt{configure} settings and adopts all the
configurations settings from there, i.e.~we do not have to re-specify
compilers, features or compiler and linker settings.

\subsection{Gauge Wave}

\begin{algorithm*}[htb]
  \caption{
    The Gauge Wave benchmark.
    \label{alg:experiments:gauge-wave}
  }
  \begin{algorithmic}
    \State cd application/exahype2/ccz4
    \State python3 ccz4.py -impl fd4-rk1-adaptive -s gauge -maxh 0.05 -minh 0.05 -ps 3 -plt 0.1 -et 0.5 -exn test --domain\_r 0.5 --ReSwi 0 -cfl 0.1 --KOSigma 8.0 -tracer 1
      \Comment Invokes \texttt{make} internally
    \State ./peano4\_test
    \State pvpython ../../../python/peano4/visualisation/render.py solution-CCZ4.peano-patch-file 
  \end{algorithmic}
\end{algorithm*}

The gauge wave benchmark simulates a simple standing wave in a domain with
periodic boundary conditions (Algorithm~\ref{alg:experiments:gauge-wave}). This setup utilizes the first order formulation, and by adding \texttt{-so} flag one can switch to the second order formulation.
The setup as specified uses tracer markers at script-specified points which track the solution in
these points.
The outcome is a single text table file (\texttt{.csv}).
We ship matplotlib scripts with the repository, which will produce
``seismograms'' at the tracer points.

\subsection{Single black hole}

\begin{algorithm*}[htb]
  \caption{
    The Single Black Hole benchmark
    \label{alg:experiments:single-black-hole-test}
  }
  \begin{algorithmic}
    \State cd application/exahype2/ccz4
    \State python3 ccz4.py -impl fd4-rk1-adaptive -s single-puncture -maxh 0.05 -minh 0.05 -ps 3 -plt 0 -et 0.01 -exn test --domain\_r 0.5 --ReSwi 1 -cfl 0.1 --KOSigma 8.0 -so
    \State ./peano4\_test
    \State pvpython ../../../python/peano4/visualisation/render.py solution-CCZ4.peano-patch-file 
  \end{algorithmic}
\end{algorithm*}
The Algorithm~\ref{alg:experiments:single-black-hole-test} builds a small benchmark of the single black hole scenario, which one can use for tests and profiling. The output is very likely to be unphysical as the resolution is quite low.

\begin{algorithm*}[htb]
  \caption{
    The Single Black Hole Productive Run
    \label{alg:experiments:single-black-hole-run}
  }
  \begin{algorithmic}
    \State cd application/exahype2/ccz4
    \State python3 ccz4.py -ext adm -impl fd4-rk1-adaptive -s single-puncture -maxh 0.4 -minh 0.04 -ps 6 -plt 0.5 -et 240 -exn sbh --domain\_r 9.0 --ReSwi 7 -cfl 0.1 -outdir /path/to/your/storage/ --KOSigma 8.0 -sommerfeld -so
      
    \State ./peano4\_sbh
    \State cd /path/to/your/storage/
    \State pvpython ../../../python/peano4/visualisation/render.py solution-CCZ4.peano-patch-file 
  \end{algorithmic}
\end{algorithm*}
The Algorithm~\ref{alg:experiments:single-black-hole-run} constructs the standard static single black hole simulation used to produce snapshots and results in the main content. This setup requires a significant amount of computational resources.

\subsection{Binary black hole}
\begin{algorithm*}[htb]
  \caption{
    The binary Black Hole Productive Run
    \label{alg:experiments:binary-black-hole-run}
  }
  \begin{algorithmic}
    \State cd application/exahype2/ccz4
    \State python3 ccz4.py -impl fd4-rk1-adaptive -s two-punctures -maxh 0.4 -minh 0.04 -ps 8 -plt 0.5 -et 120 -exn bbh --domain\_r 12.0 --ReSwi 6 -cfl 0.1 -outdir /path/to/your/storage/ --KOSigma 8.0 --BBHType 2 -sommerfeld -so -ext Psi4
    \State ./peano4\_bbh
    \State cd /path/to/your/storage/
    \State pvpython ../../../python/peano4/visualisation/render.py solution-CCZ4.peano-patch-file 
  \end{algorithmic}
\end{algorithm*}
Algorithm~\ref{alg:experiments:binary-black-hole-run} builds an application for production runs on the rotating binary black hole scenario, used to produce data for the main content. Similarly, this setup is computational-resource demanding.

\section{Spacetime Foliation and Evolution Equations}
\label{app:spacetimeFoliation}

This appendix gives an overview of the problem formulation of \acronym\ and also lists all the evolution equations used in the code. 

Throughout the paper, we have chosen the $(-,+,+,+)$ metric signature and followed the standard Einstein convention of summing over repeated indices. We use the Latin letters $a,b,\ldots$, for spacetime indices which run from 0 to 3, following the \emph{abstract index notation}. Tensor components are labelled by Greek letters $\mu,\nu,\ldots$, and indices running from 1 to 3 are labelled using the middle part of the Latin alphabet ($i,j,k,\ldots$) as usual. Finally, we will use the geometrized unit system 
where both the gravitational constant and the speed of light are set to unity, $c=G=1$.

The evolution equations for the spacetime utilized by \acronym\ are from the Z4 formulation of Einstein field equations, where an extra algebra dynamic field $Z_a$ and corresponding damping term are added into the system to enhance stability \cite{bona:03:Z4original, Gundlach:2005:Z4damping}:
\begin{flalign}
\begin{split}
    \label{eq_NR_z4_4d_field}
    R_{ab}-\frac{1}{2} g_{ab}R&+\nabla_{a} Z_{b}+\nabla_{b} Z_{a}-g_{ab}\nabla^c Z_c- \\ &\kappa_{1}[n_{a} Z_{b}+n_{b} Z_{a}+\kappa_{2} g_{ab} n_{c} Z^{c}]=8\pi T_{ab}.
    \end{split}
\end{flalign}
To solve the evolution equations, our code uses the standard ADM 3+1 spacetime foliation which cuts the spacetime into a set of three-dimensional spacelike hypersurfaces labelled by a global time scalar field $t(x^a)$. The spatial metric on those hypersurfaces is named $\gamma_{ab}$, and linked to the spacetime metric
\begin{equation}
    d s^{2}=-\alpha^{2} d t^{2}+\gamma_{i j}\left(d x^{i}+\beta^{i} d t\right)\left(d x^{j}+\beta^{j} d t\right),
\end{equation}
where $\gamma_{ij}$ are the spatial part of $\gamma_{ab}$. $\alpha$ and $\beta^i$ are called lapse functions and shift vector fields, representing four degrees of freedom of coordinates during the evolution. Their relations with the time axis $t^a$ are illustrated in Figure \ref{figure:theory:lapse_shift}.

The unit normal vector on the spacelike hypersurface $n^a$ is given by
\begin{equation}
    n^a:=-\alpha g^{ab} \nabla_b t,
\end{equation}
and the extrinsic curvature $K_{ab}$, a dynamic variable determining the evolution of the spatial structure, can be defined as the Lie derivatives of the spatial metric over $n^a$:
\begin{equation}
    K_{ab}:=-\frac{1}{2}\mathcal{L}_{\mathbf{n}} \gamma_{ab}.
\end{equation}
Similar to the spatial metric $\gamma_{ab}$, $K_{ab}$ is a spatial tensor. We therefore can only track the evolution of its spatial part, i.e., $K_{ij}$, in our simulations.  

One can already conduct simulations in this stage as $\gamma_{ij}, K_{ij}, \alpha, \beta^{i}$ with $Z^a$ field form a complete evolution system. \revision{We further follow~\cite{Alic12} and introduce a conformal factor $\phi$ and a conformal metric $\tilde{\gamma}_{ij}$, such that}
\begin{equation}
    \tilde{\gamma}_{ij}:=\phi^2 \gamma_{ij},~\phi=[\det(\gamma_{ij})]^{-1/6}.
\end{equation}
The second equation guarantees that $\tilde{\gamma}_{ij}$ has a unit determinant. We also decompose the extrinsic curvature into its trace part $K=K_{ij}\gamma^{ij}$ and traceless part $A_{ij}$, using the same scaling as for the metric:
\begin{equation}
    \label{eq_NR_conformal_decomposition_K}
    \tilde{A}_{ij}:=\phi^2 A_{ij}=\phi^2 (K_{ij}-\frac{1}{3}K \gamma_{ij}).
\end{equation}
The $Z^a$ represents four additional evolving variables, which are divided into its temporal component $\Theta:= -n_a Z^a=\alpha Z^0$ and spatial component $Z^i:=\gamma^a_b Z^b$. The latter is included in the evolution system by adding to the contracted Christoffel symbol:
\begin{equation}
    \label{eq_NR_CCZ4_Gamma^i}
    \hat{\Gamma}^{i} := \tilde{\Gamma}^{i}+2 \tilde{\gamma}^{i j} Z_{j}, ~\tilde{\Gamma}^{i} := \tilde{\gamma}^{j k} \tilde{\Gamma}_{j k}^{i},
\end{equation}
where $\tilde{\Gamma}^{i}_{jk}$ is the Christoffel symbol for the conformal metric $\tilde{\gamma}_{ij}$
\begin{equation}
    \tilde{\Gamma}^i_{jk}=\frac{1}{2}\tilde{\gamma}^{il}\left( \partial_j \tilde{\gamma}_{kl} +\partial_k \tilde{\gamma}_{jl}-\partial_l \tilde{\gamma}_{jk}\right).
\end{equation}
For the first-order formulation, \acronym\ further introduces the following auxiliary variables:
\begin{flalign}
\begin{split}
   &A_{i}:=\partial_{i} \alpha, \quad B_{k}^{i}:=\partial_{k} \beta^{i},
   \\
   &D_{k i j}:=\frac{1}{2} \partial_{k} \tilde{\gamma}_{i j}, \quad P_{i}:=\partial_{i} \phi.
   \label{eq_NR_foccz4_aux_variables}
   \end{split}
\end{flalign}
These variables satisfy the natural second-order constraints as
\begin{flalign}
\begin{split}
    &\partial_{k} A_{i}-\partial_{i} A_{k}=0,\quad\partial_{k} B_{l}^{i}-\partial_{l} B_{k}^{i}=0,\\ 
    &\partial_{k} D_{l i j}-\partial_{l} D_{k i j}=0,\quad\partial_{k} P_{i}-\partial_{i} P_{k}=0.
    \label{eq_NR_foccz4_so_cons_aux_variables}
   \end{split}
\end{flalign}
Because $\tilde{A}_{i j}$ is traceless $\gamma^{ij}\tilde{A}_{ij}=0$ and $\tilde{\gamma}=\det (\tilde{\gamma}_{ij})=1$, we also have 
\begin{equation}
    \partial_{k}\left(\tilde{\gamma}^{i j} \tilde{A}_{i j}\right)=\partial_{k} \tilde{\gamma}^{i j} \tilde{A}_{i j}+\tilde{\gamma}^{i j} \partial_{k} \tilde{A}_{i j}=0,
    \tilde{\gamma}^{i j} D_{k i j}=0. \label{eq_NR_foccz4_algebra_cons}
\end{equation}
The formulation of the evolution system in \acronym\ is given as the following nine equations for primary variables:
\begin{flalign}
    \begin{split}\label{eq_NR_foccz4_gamma_ij}
        &\partial_{t} \tilde{\gamma}_{i j}=\beta^{k} 2 D_{k i j}+\tilde{\gamma}_{k i} B_{j}^{k}+\tilde{\gamma}_{k j} B_{i}^{k}-\frac{2}{3} \tilde{\gamma}_{i j} B_{k}^{k}-\\ &\Lquad 2 \alpha\left(\tilde{A}_{i j}-\frac{1}{3} \tilde{\gamma}_{i j} \operatorname{tr} \tilde{A}\right)
        -\tau^{-1}(\tilde{\gamma}-1) \tilde{\gamma}_{i j},
    \end{split}
    \\
    &\partial_t\alpha = \beta^k A_k-\alpha^2 g(\alpha)\left(K-K_0-2c\Theta\right), \label{eq_NR_foccz4_lapse_gauge}
    \\
    &\partial_{t} \beta^{i}= \beta^{k} B_{k}^{i}+ f b^{i},
    \label{eq_NR_foccz4_shift_gauge}
    \\
    &\partial_t\phi = \beta^k P_k + \frac{1}{3}\phi\left(\alpha K - B^l_l\right),
    \\
    \begin{split}
        &\partial_{t} \tilde{A}_{i j}-\beta^{k} \partial_{k} \tilde{A}_{i j}-\phi^{2}\left[-\nabla_{i} \nabla_{j} \alpha+\alpha\left(R_{i j}+\nabla_{i} Z_{j}+\nabla_{j} Z_{i}\right)\right]^{\mathrm{TF}} \\
        &\quad\quad=\tilde{A}_{k i} B_{j}^{k}+\tilde{A}_{k j} B_{i}^{k}-\frac{2}{3} \tilde{A}_{i j} B_{k}^{k}+\alpha \tilde{A}_{i j}(K-2 \Theta c)- \\ &\Lquad 2 \alpha \tilde{A}_{i l} \tilde{\gamma}^{l m} \tilde{A}_{m j},
    \end{split}
    \\
    \begin{split}
        &\partial_{t} K-\beta^{k} \partial_{k} K+\nabla^{i} \nabla_{i} \alpha-\alpha\left(R+2 \nabla_{i} Z^{i}\right)\\
        &\Lquad=\alpha K(K-2 \Theta c)-3 \alpha \kappa_{1}\left(1+\kappa_{2}\right) \Theta,
    \end{split}
    \\
    \begin{split}
        &\partial_{t} \Theta-\beta^{k} \partial_{k} \Theta-\frac{1}{2} \alpha e^{2}\left(R+2 \nabla_{i} Z^{i}\right)\\
        &\quad =\frac{1}{2} \alpha e^{2}\left(\frac{2}{3} K^{2}-\tilde{A}_{i j} \tilde{A}^{i j}\right)-\alpha \Theta K c-Z^{i} A_{i}\\
        &\qquad-\alpha \kappa_{1}\left(2+\kappa_{2}\right) \Theta,
    \end{split}
    \\
    \begin{split}
        &\partial_{t} \hat{\Gamma}^{i}-\beta^{k} \partial_{k} \hat{\Gamma}^{i}-2 \alpha \tilde{\gamma}^{k i} \partial_{k} \Theta- \tilde{\gamma}^{k l} \partial_{(k} B_{l)}^{i}-s \frac{1}{3} \tilde{\gamma}^{i k} \partial_{(k} B_{l)}^{l}\\
        &\quad\quad-{\color{red}  2 \alpha \tilde{\gamma}^{i k} \tilde{\gamma}^{n m} \partial_{k} \tilde{A}_{n m}} +\frac{4}{3} \alpha \tilde{\gamma}^{i j} \partial_{j} K=\frac{2}{3} \tilde{\Gamma}^{i} B_{k}^{k}-\tilde{\Gamma}^{k} B_{k}^{i}\\
        &\quad\quad+2 \alpha\left(\tilde{\Gamma}_{j k}^{i} \tilde{A}^{j k}-3 \tilde{A}^{i j} \frac{P_{j}}{\phi}\right) -2 \alpha \tilde{\gamma}^{k i}\left(\Theta \frac{A_{k}}{\alpha}+\frac{2}{3} K Z_{k}\right) \\
        &\quad\quad-2 \tilde{A}^{i j} A_{j}-{\color{red} 4  \alpha \tilde{\gamma}^{i k} D_{k}^{n m} \tilde{A}_{n m}} -2 \alpha \kappa_{1} \tilde{\gamma}^{i j} Z_{j} \\
        &\quad\quad +2 \kappa_{3}\left(\frac{2}{3} \tilde{\gamma}^{i j} Z_{j} B_{k}^{k}-\tilde{\gamma}^{j k} Z_{j} B_{k}^{i}\right),
    \end{split}\\
    &\partial_{t} b^{i}- \beta^{k} \partial_{k} b^{i}=\partial_{t} \hat{\Gamma}^{i}-\beta^{k} \partial_{k} \hat{\Gamma}^{i}-\eta b^{i},\label{eq_NR_foccz4_b^i}
\end{flalign}
and four equations for the auxiliary variables, if the first-order scheme is used:
\begin{flalign}
    \begin{split}
        &\partial_t A_k-\beta^l\partial_l A_k +
        \alpha^2g(\alpha)\left(\partial_{k}K-\partial_{k}K_0-2c\partial_{k}\Theta\right)\\
        &\qquad+\textcolor{red}{\alpha g(\alpha) \tilde{\gamma}^{nm} \partial_{k} \tilde{A}_{nm}}=B^l_k A_l+\textcolor{red}{2\alpha g(\alpha)D_{k}^{\ nm}\tilde{A}_{nm}} \\
        &\quad\Lquad- \left[2{\alpha}g(\alpha)+\alpha^2g'(\alpha)\right]\left(K-K_0-2c\Theta\right)A_k, \label{eq_NR_foccz4_A_k}
    \end{split} \\
    \begin{split}
        &\partial_{t} B_{k}^{i}- \beta^{l} \partial_{l} B_{k}^{i}-f \partial_{k} b^{i}+{\color{red}\alpha^{2} \mu \frac{\tilde{\gamma}^{i j}}{\phi}\left(\partial_{k} P_{j}-\partial_{j} P_{k}\right)}\\
        &\Lquad-{\color{red}\alpha^{2} \mu \tilde{\gamma}^{i j} \tilde{\gamma}^{n l}\left(\partial_{k} D_{l j n}-\partial_{l} D_{k j n}\right)}= B_{k}^{l} B_{l}^{i}, \label{eq_NR_foccz4_B^i_j}
    \end{split}\\
    \begin{split}
        &\partial_{t} D_{k i j}-\beta^{l} \partial_{l} D_{k i j}+s\Big(-\frac{1}{2} \tilde{\gamma}_{m i} \partial_{(k} B_{j)}^{m}-\frac{1}{2} \tilde{\gamma}_{m j} \partial_{(k} B_{i)}^{m}\\
        &\quad+\frac{1}{3} \tilde{\gamma}_{i j} \partial_{(k} B_{m)}^{m}\Big)+\alpha \partial_{k} \tilde{A}_{i j}-{\color{red}\alpha \frac{1}{3} \tilde{\gamma}_{i j} \tilde{\gamma}^{n m} \partial_{k} \tilde{A}_{n m}}\\
        &\quad=B_{k}^{l} D_{l i j}+B_{j}^{l} D_{k l i}+B_{i}^{l} D_{k l j}-\frac{2}{3} B_{l}^{l} D_{k i j}\\
        &\quad-{\color{red}\alpha \frac{2}{3} \tilde{\gamma}_{i j} D_{k}^{n m} \tilde{A}_{n m}}- A_{k} \left(\tilde{A}_{i j}-\frac{1}{3} \tilde{\gamma}_{i j} \operatorname{tr} \tilde{A}\right),
    \end{split}\\
    \begin{split}\label{eq_NR_foccz4_P_i}
        &\partial_t P_k -\beta^l\partial_l P_k 
        - \frac{1}{3}\phi\left(\alpha\partial_{k}K - \partial_{(k}B_{l)}^l\right)-
        \textcolor{red}{\frac{1}{3} \alpha \phi \tilde{\gamma}^{nm} \partial_k \tilde{A}_{nm}} \\
        &\quad= B^l_k P_l + \frac{1}{3}\left(\alpha K - B^l_l\right)P_k + \frac{1}{3}\phi K A_k
     - \textcolor{red}{ \frac{2}{3}\alpha\phi D_k^{\ nm}\tilde{A}_{nm}}.
    \end{split}
\end{flalign}
Most of the symbols already appear in the context above and are self-explaining. The $\mathrm{TF}$ index means the Trace-Free part of the quantities, i.e., $R^{\mathrm{TF}}_{ij}=R_{ij}-\frac{1}{3}\gamma_{ij}\gamma^{kl} R_{kl}$. $\operatorname{tr} \tilde{A}= \gamma^{ij} \tilde{A}_{ij}$ is the trace of the conformal traceless extrinsic curvature. There are several red terms added to the evolution systems, using the constraints \eqref{eq_NR_foccz4_algebra_cons} above for a more symmetric characteristic matrix \cite{Dumbser18}. \revision{Those terms are only employed in the first formulation in \acronym\ and are removed in the second-order formulation}. The $g(\alpha)$ function is the one defined in the slicing condition in Section \ref{sec:theory:gauge}, and the ``1+log'' slicing corresponds to $g(\alpha)=2/\alpha$. Several new parameters are also introduced for numerical optimization: 
\begin{itemize}
    \item $\tau$ represents the relaxation time to enforce the algebraic constraints ($\operatorname{det} \tilde{\gamma}_{i j}=1$);
    \item $e$ is the cleaning speed for Hamiltonian constraint, following the idea of \cite{dedner:2002:DivergenceCleaning};
    \item $\mu>0$ in equation \eqref{eq_NR_foccz4_B^i_j} determines the effect of the terms from the constraints;
    \item $c$ controls the contribution from some algebraic source terms in Z4 systems. Its default value is 1 when the original CCZ4 is used \citep{Alic12}.
\end{itemize}
The standard hyperbolic Gamma driver shift gauge is given above but for some physical scenarios, we also provide the option of the static (zero) shift condition. 

The evolution equations above are not completed as we abbreviate some key quantities for readability. The whole system is closed with the evolving quantities using the following list of relations:
\begin{flalign}
    \partial_{k} \tilde{\gamma}^{i j}&=-2 \tilde{\gamma}^{i n} \tilde{\gamma}^{m j} D_{k n m}:=-2 D_{k}^{i j}\label{eq_NR_foccz4_rela1},
    \\
    \tilde{\Gamma}_{i j}^{k}&=\tilde{\gamma}^{k l}\left(D_{i j l}+D_{j i l}-D_{l i j}\right),
    \\
    \begin{split}\partial_{k} \tilde{\Gamma}_{i j}^{m}&=-2 D_{k}^{m l}\left(D_{i j l}+D_{j i l}-D_{l i j}\right)+\tilde{\gamma}^{m l}(\partial_{(k} D_{i) j l}
    \\&+\partial_{(k} D_{j) i l}-\partial_{(k} D_{l) i j}),\end{split}
    \\
    \Gamma^k_{ij} &= \tilde{\Gamma}^k_{ij} - \frac{1}{\phi}\tilde{\gamma}^{kl}\left(\tilde{\gamma}_{jl}P_i + \tilde{\gamma}_{il}P_j - \tilde{\gamma}_{ij}P_l\right),
    \\
    \begin{split}
        \partial_k\Gamma^m_{ij} =& -2D_k^{ml}\left(D_{ijl}+D_{jil}-D_{lij}\right)  \\&+\tilde{\gamma}^{ml}\left[\partial_{(k}D_{i)jl}+\partial_{(k}D_{j)il}-\partial_{(k}D_{l)ij}\right] \\
        & + \frac{2}{\phi}D_k^{ml}\left(\tilde{\gamma}_{jl} P_i+\tilde{\gamma}_{il} P_j-\tilde{\gamma}_{ij} P_l\right) \\&- \frac{2}{\phi}\tilde{\gamma}^{ml}\left(D_{kjl} P_i+D_{kil} P_j-D_{kij} P_l\right)\\
        & - \frac{1}{\phi}\tilde{\gamma}^{ml}\left[\tilde{\gamma}_{jl}\partial_{(k} P_{i)}+\tilde{\gamma}_{il}\partial_{(k} P_{j)}-\tilde{\gamma}_{ij}\partial_{(k} P_{l)}\right] \\  & + \frac{1}{\phi^2}\tilde{\gamma}^{ml}\left(\tilde{\gamma}_{jl} P_i P_k + \tilde{\gamma}_{il} P_j P_k - \tilde{\gamma}_{ij} P_k P_l\right),
    \end{split}\\
    R_{i k j}^{m}&=\partial_{k} \Gamma_{i j}^{m}-\partial_{j} \Gamma_{i k}^{m}+\Gamma_{i j}^{l} \Gamma_{l k}^{m}-\Gamma_{i k}^{l} \Gamma_{l j}^{m},~~~R_{i j}=R_{i m j}^{m},\\
    \nabla_i\nabla_j\alpha &= \partial_{(i} A_{j)} - \Gamma^k_{ij} A_k,\\
    \partial_{k} \tilde{\Gamma}^{i}&=-2 D_{k}^{j l} \tilde{\Gamma}_{j l}^{i}+\tilde{\gamma}^{j l} \partial_{k} \tilde{\Gamma}_{j l}^{i},\\
    Z_{i}&=\frac{1}{2} \tilde{\gamma}_{i j}\left(\hat{\Gamma}^{j}-\tilde{\Gamma}^{j}\right), ~~~ Z^{i}=\frac{1}{2} \phi^{2}\left(\hat{\Gamma}^{i}-\tilde{\Gamma}^{i}\right),\\
    \nabla_{i} Z_{j}&=D_{i j l}\left(\hat{\Gamma}^{l}-\tilde{\Gamma}^{l}\right)+\frac{1}{2} \tilde{\gamma}_{j l}\left(\partial_{i} \hat{\Gamma}^{l}-\partial_{i} \tilde{\Gamma}^{l}\right)-\Gamma_{i j}^{l} Z_{l},\\
    R+2 \nabla_{k} &Z^{k}=\phi^{2} \tilde{\gamma}^{i j}\left(R_{i j}+\nabla_{i} Z_{j}+\nabla_{j} Z_{i}\right)\label{eq_NR_foccz4_R_2nablaZ}.
\end{flalign}
Notice the second-order constraints \eqref{eq_NR_foccz4_so_cons_aux_variables} are used to ``symmetrize'' the spatial derivatives of the auxiliary variables in some equations and relations. In code practice, we also use a simplified expression for the modified Ricci tensor $R_{ij}+2\nabla_{(i}Z_{j)}$ from \cite{GRChombo2022}:
\begin{flalign}
    \begin{split}
        R_{ij} +& 2\nabla_{(i}Z_{j)} = \\
        &-\frac{1}{2} \tilde{\gamma}^{kl} \partial_k \partial_l \tilde{\gamma}_{ij} + \tilde{\gamma}_{k(i}\partial_{j)}\hat{\Gamma}^k + \frac{1}{2}\hat{\Gamma}^k\partial_k\tilde{\gamma}_{ij}  \\& + \tilde{\gamma}^{lm}\left(\tilde{\Gamma}^{k}_{li}\tilde{\Gamma}_{jkm} + \tilde{\Gamma}^{k}_{lj}\tilde{\Gamma}_{ikm} + \tilde{\Gamma}^{k}_{im}\tilde{\Gamma}_{klj}\right)\\
        &+ \frac{1}{\phi} \left(\tilde{\nabla}_i \tilde{\nabla}_j \phi+\tilde{\gamma}_{ij} \tilde{\gamma}^{kl} \tilde{\nabla}_k \tilde{\nabla}_l \phi \right) - \frac{2}{\phi^2}  \tilde{\gamma}_{ij} \tilde{\gamma}^{kl} \partial_k \phi \partial_l \phi\\
        &+ \frac{2}{\phi^3} Z^k \left(\tilde{\gamma}_{ik} \partial_j\phi + \tilde{\gamma}_{jk} \partial_i\phi - \tilde{\gamma}_{ij} \partial_k \phi\right),
    \end{split}
    \label{eq_NR_foccz4_modified_ricci}
\end{flalign}
with $\tilde{\nabla}_i\tilde{\nabla}_k\phi = \partial_i\partial_k\phi - \tilde{\Gamma}^k_{ij}\partial_k\phi$  the second \emph{conformal} covariant derivative of $\phi$ and $\tilde{\Gamma}_{kij} := \tilde{\gamma}_{kl} \tilde{\Gamma}^l_{ij}$  the newly-defined lowered Christoffel symbol.

There are two extra physical (ADM) constraints can be derived from the 3+1 foliation of \eqref{eq_NR_z4_4d_field}, which are the \emph{Hamiltonian Constraint}
\begin{equation}
    H:=R+\frac{2}{3}K^2 -\tiA_{ij}\tiA^{ij},
    \label{eq_hamilton_constraints}
\end{equation}
and \emph{Momentum Constraints}
\begin{equation}
    M^i:=\tiga^{kl}\left(\partial_k \tiA_{li}-2\tilde{\Gamma}^m_{l(i}\tiA_{k)m}-3\tiA_{ik}\frac{P_l}{\phi} \right)-\frac{2}{3}\partial_i K.
    \label{eq_momentum_constraints}
\end{equation}
The matter source terms are dropped as we currently focus on the vacuum solution. According to the Bianchi identities, if the constraints are zero at the initial hypersurface (which is guaranteed by the Bowen-York solution), they vanish through the whole evolution~\cite{frittelli:1997:constraintsPropagation}. Any deviation from zero can be seen as a result form the numerical errors. Therefore, they are widely used in numerical relativity code as accuracy metrics to assess the quality of the simulations. We examine the violation of the constraints in our simulation tests in Section~\ref{sec:result}.
\section{$\psi_4$ in \acronym}
\label{app:psi4}
The complex scalar field $\psi_4$ is written as
\begin{equation}
    \psi_4= ^{(4)}C_{abcd} k^a \bar{m}^b k^c \bar{m}^d,
\end{equation}
where $^{(4)}C_{abcd}$ is the four-dimensional Weyl tensor, and $k^a$ and $\bar{m}^a$ are two members of a null tetrad. A tetrad is a set of four null vectors that form a complete \revision{basis}, and we name them $l^a$, $k^a$, $m^a$ and $\bar{m}^a$. The tetrad is constructed such that the $l^a$ and $k^a$ are radial outgoing and ingoing vectors respectively, and only non-vanishing inner products between them are:
\begin{equation}
    -l^a k_a = m^a \bar{m}_a =1.
\end{equation}
In our code, the unit vectors in three spatial directions are calculated from a Gram-Schmidt orthonormalization of the following vectors \citep{brugmann:2008:movingP}:
\begin{eqnarray}
    v^i &=& [-y,x,0]  \to  v^i/\sqrt{L_{vv}},\\
    u^i &=&  [x,y,z]  \to  (u^i-v^i L_{vu})/\sqrt{L_{uu}},\\
    w^i &=&  \gamma^{ij}\epsilon_{jkl}v^k u^l  \to \nonumber\\
    &&\quad(w^i-v^i L_{vw}-u^i L_{uw})/\sqrt{L_{ww}},
\end{eqnarray}
where $L$ represents the inner product of corresponding vectors, i.e., $L_{vu}:=\gamma_{ij} v^i u^j$. Combining with the temporal unit vector, ${e^a_{\hat{t}}} = \left[\alpha^{-1}, -\alpha^{-1}\beta^i\right]$, the tetrad is given as
\begin{eqnarray}
    l^a &=& \frac{1}{\sqrt{2}}[\alpha^{-1},-\alpha^{-1}\beta^i+u^i],\\
    k^a &=& \frac{1}{\sqrt{2}}[\alpha^{-1},-\alpha^{-1}\beta^i-u^i],\\
    m^a &=& \frac{1}{\sqrt{2}}[0,w^i + i v^i], \\
    \bar{m}^a &=& \frac{1}{\sqrt{2}}[0,w^i - i v^i]\label{eq_NR_psi4_m_conj}.
\end{eqnarray}
On the other hand, the Weyl tensor can be simplified to~\cite{alcubierre:2008:EandB}
\begin{equation}
    \psi_4=(E_{ij}-iB_{ij})\bar{m}^i \bar{m}^j,\label{eq_NR_psi_code_implementation}
\end{equation}
where $E_{ij}$ and $B_{ij}$ are the spatial parts of the so-called electric and magnetic parts of the Weyl tensor. Their expression in the Z4 system is
\begin{flalign}
    &E_{ij} = \left(R_{ij}-K^m_{~i} K_{jm}+K_{ij}(K-\Theta)+D_{(i}Z_{j)}-4\pi S_{ij}\right)^{\mathrm{TF}},\label{eq_NR_psi4_E_ij}\\
    &B_{ij} = \frac{1}{2} \left(\epsilon_{mni}D^m K_j^{~n}
    +\epsilon_{mnj}D^m K_i^{~n} \right)^{\mathrm{TF}}.\label{eq_NR_psi4_B_ij}
\end{flalign}
$\mathrm{TF}$ is the trace-free index with the same definition as in Section~\ref{app:spacetimeFoliation}. Considering the systems only involve black holes, the source term in equation \eqref{eq_NR_psi4_E_ij} is zero. The final expression of $\psi_4$ is given by combining Eqs.~\eqref{eq_NR_psi4_E_ij}, \eqref{eq_NR_psi4_B_ij} and \eqref{eq_NR_psi4_m_conj}, putting them back in Eq.~\eqref{eq_NR_psi_code_implementation}.

\section{Formulation for the kernel FD4 Solver}
\label{app:solver}
The code utilizes the five-point stencil in one dimension to approximate the derivatives in fourth-order accuracy:
\begin{flalign}
\begin{split}
    \label{eq_exa_fd4_central_difference} \frac{\partial Q\left(x\right)}{\partial x}\bigg|_{x_0} = \frac{1}{12 \delta x}&\left[8\left(Q^{+}-Q^{-}\right)-\left(Q^{++}-Q^{--}\right)\right] \\
    &+ \mathcal{O}\left(\delta x^4\right),
\end{split}
\end{flalign}
where $x_0$ is the grid point of interest, the $+$ and $-$ upper indices of quantities indicate the relative position of the indexed quantities with the current grid point, and the number of signs specifies the distance, e.g., $Q^{++}:= Q(x_0+2\delta x)$, $Q^-:= Q(x_0-\delta x)$. We have omitted the vector symbol of the evolving variables $Q$ for simplicity.

Notice that we can apply this formulation to any spatial-dependent function, therefore the divergence of the flux is given as:
\begin{flalign}
\begin{split}
    \label{eq_exa_fd4_flux}
    \frac{\partial F}{\partial x}\bigg|_{x_0} = \frac{1}{12 \delta x}&\left[8\left(F^{+}-F^{-}\right)-\left(F^{++}-F^{--}\right)\right] \\
    &+ \mathcal{O}\left(\delta x^4\right).
\end{split}
\end{flalign}
Equipped with the relations \eqref{eq_exa_fd4_central_difference} and \eqref{eq_exa_fd4_flux}, we can derive the time stepping scheme of the fourth-order finite difference (FD4) solver from \eqref{equation:theory:pde}:
\begin{align}
    \vq&(t+\delta t)=\vq(t)+\delta t  S(\vq) 
    \\&- \sum_i \frac{\delta t}{12 \delta x_i}\left[8\left(F_i^{i+}-F_i^{i-}\right)-\left(F_i^{i++}-F_i^{i--}\right)\right] \nonumber\\-
    &\sum_i B_i \frac{\delta t}{12 \delta x_i}\left[8\left(Q^{i+}-Q^{i-}\right)-\left(Q^{i++}-Q^{i--}\right)\right], \nonumber 
\end{align}
where we have used the forward finite differences for the time derivatives, i.e. $\partial_t \vq = \left(\vq(t+\delta t)-\vq(t)\right)/\delta t$ for a timestep of size $\delta t$. The index $i$ goes from 1 to 3 and indicates the direction of the neighbouring points we are looking into. The five-point stencil of \eqref{eq_exa_fd4_central_difference} is the central difference and it is our preferred choice in most cases. However, some terms may need a lopsided stencil to capture specific features or behaviours, especially those that may lead to asymmetric patterns in the system, e.g. the advection terms of the CCZ4 system \revision{\citep{advection_term}} in numerical relativity, which appear in our application as well. The left and right lopsided stencils with also the fourth-order accuracy are
\begin{flalign}\label{eq_exa_fd4_lopsided1}
\begin{split}
    \frac{\partial Q\left(x\right)}{\partial x}\bigg|_{x_0} = \frac{1}{12 \delta x}&\left[-Q^{---} +6Q^{--} - 18 Q^- +  10Q + 3Q^{+}\right], \\ &\quad \text{if}~\beta^x>0
    \end{split}
\end{flalign}
\begin{flalign}\label{eq_exa_fd4_lopsided2}
\begin{split}
    \frac{\partial Q\left(x\right)}{\partial x}\bigg|_{x_0} = \frac{1}{12 \delta x}&\left[-3Q^- -10Q + 18 Q^+ -6Q^{++} +Q^{+++}\right], \\ & \quad \text{if}~\beta^x\leq 0.
    \end{split}
\end{flalign}


\noindent
To avoid numerical errors of high-frequency modes, we also introduce the standard Kreiss-Oliger (KO) dissipation at order $N = 3$ in the code as a low-pass filter. Its form in one dimension is~\cite{kreiss:1973:KOterms}:
\begin{flalign}
\begin{split}\label{eq_exa_fd4_ko}
    \textrm{KO}^{(3)} = \frac{\epsilon}{64 \delta x}(Q^{---}-&6Q^{--}+15Q^{-}-20Q \\& +15Q^{+}-6Q^{++}+Q^{+++}),
    \end{split}
\end{flalign}
where $\epsilon$ is a user-defined parameter to control the strength of the numerical dissipation. This term has a convergence speed $\sim \delta x ^5$ which can be easily verified by counting the Taylor expansion coefficients. The final numerical dissipation term is the sum over all directions of Eq.~\eqref{eq_exa_fd4_ko}.

\section{Matrices for Tensor Product Interpolation \& Restriction}
\label{app:amr}
The adaptive mesh in \exahype\ 2 is constructed by refining patches recursively in three partitions and creating new patches accordingly, and refinement transition of multiple levels is not allowed in the code. Therefore, the difference in patch length can only be a factor of three. As all patches in the domain have the same number of volumes, we, therefore, shall have $\delta x_\text{coarse}=3 \delta x_\text{fine}$.

\begin{figure}
    \centering
    \includegraphics[width=0.5\textwidth]{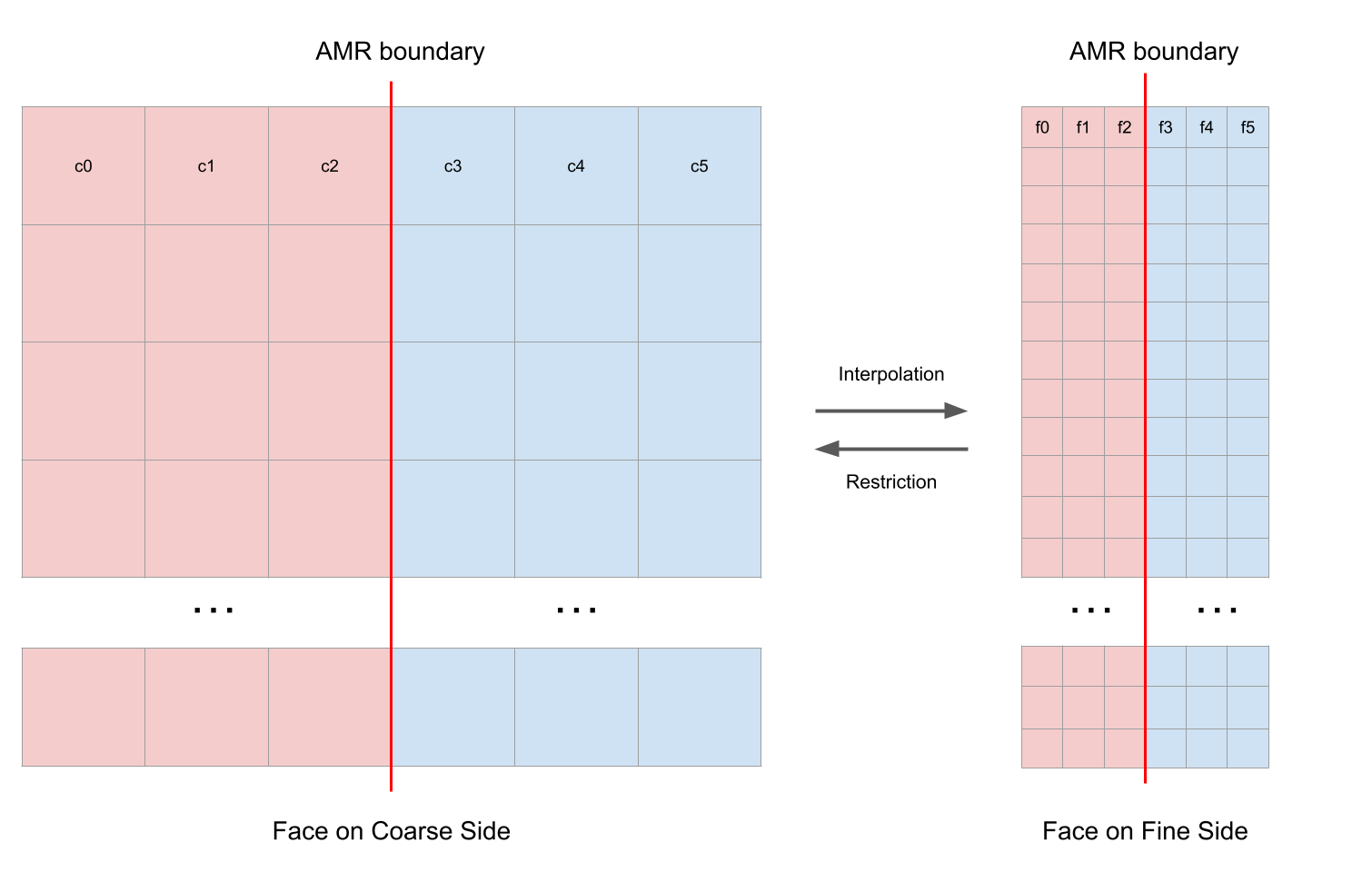}
    \caption{The illustration of the face structures on the coarse and fine side of a refinement transition boundary.  The mapping between them determines the strategy of the refinement transition in the application. For the interpolation, we need to calculate the outer half of the fine face (the light red squares on the right) using the information of the coarse face. On the other hand, we need to compute the outer half of the coarse face (the light blue squares on the left) utilizing the fine face for a restriction scheme. The layers of the faces are labeled from c0 to c5 and f0 to f5 respectively, it is not the actual enumeration we employed in the actual code.}
    \label{fig_exa_interpolation_map}
\end{figure}

For the interpolation scheme compatible with our default FD4 kernel, the outer half of a face contains three layers, i.e., layers f0, f1 and f2 in Figure \ref{fig_exa_interpolation_map}. The face of the coarse patch, on the other hand, inputs as a whole. Assuming our patches contain $p$ volumes per dimension, this scheme yields a map from $6p^2$ variables to $3\times (3p)^2$ variables. The current implementation of \acronym\ adopts the tensor product approach to perform this mapping, which decomposes the high-dimensional map into the product of three one-dimensional maps. Assuming the volumes in the coarse and fine faces are $Q_{i,j,k}^c$ and $Q_{i,j,k}^f$, the interpolation map can be written as
\begin{equation}
    Q^f_{i,j,k}=P^{x}_{il}P^y_{jm}P^z_{kn} Q^c_{l,m,n},
\end{equation}
where $P^x$, $P^y$, and $P^z$ are the matrices responsible for the corresponding one-dimensional map. As we shall have a symmetry for the two dimensions parallel to the face normal, the two matrices for mappings along them are identical. The matrix for the mappings parallel to the face normal and along the face normal are labelled as $P^{\parallel}$ and $P^{\perp}$, respectively.

Matrix $P^\parallel$ should have a size of $3p \times p$, as it maps $p$ coarse volumes to $3p$ fine volumes; on the other hand, matrix $P^\perp$ has a size of $3\times 6$, as it maps 6 coarse layers to 3 fine layers (only the outer half of the fine face needs the interpolation). The $P^\parallel$ and $P^\perp$ matrices of the trilinear interpolation scheme we currently adopt read as
\begin{equation}
P^\parallel=\left(
\begin{smallmatrix}
    4/3 & -1/3 & 0 & 0 & 0 & ...\\
    1 & 0 & 0 & 0 & 0 &...\\
    2/3 & 1/3 & 0 & 0 & 0 &...\\
    1/3 & 2/3 & 0 & 0 & 0 &...\\
    0 & 1 & 0 & 0 & 0 &...\\
    0 & 2/3 & 1/3 & 0 & 0 &...\\
    & & .... & &
\end{smallmatrix}\right),
\end{equation}
\begin{equation}
P^\perp=\left(
\begin{smallmatrix}
    0 & 1/3 & 2/3 & 0 & 0 & 0\\
    0 & 0 & 1 & 0 & 0 & 0\\
    0 & 0 & 2/3 & 1/3 & 0 & 0\\
\end{smallmatrix}\right).
\end{equation}
In this scheme, we apply a simple linear interpolation to compute the value in the fine volumes according to the relative positions of the closest two coarse volumes. The extrapolation is used for those volumes at the edge of the considered fine face. There is no such requirement in the case of $P^{\perp}$ as all three layers of the fine face are inside the coarse face. We also implement the matrices for the zero-order piecewise constant interpolations as
\begin{equation}
P_\text{const}^\parallel=\left(
\begin{smallmatrix}
    1 & 0 & 0 & 0 & 0 & ...\\
    1 & 0 & 0 & 0 & 0 &...\\
    1 & 0 & 0 & 0 & 0 &...\\
    0 & 1 & 0 & 0 & 0 &...\\
    0 & 1 & 0 & 0 & 0 &...\\
    0 & 1 & 0 & 0 & 0 &...\\
    & & .... & &
\end{smallmatrix}\right),
\end{equation}
\begin{equation}
P_\text{const}^\perp=\left(
\begin{smallmatrix}
    0 & 0 & 1 & 0 & 0 & 0\\
    0 & 0 & 1 & 0 & 0 & 0\\
    0 & 0 & 1 & 0 & 0 & 0\\
\end{smallmatrix}\right).
\end{equation}
This means that we use the value in the coarse volume (layer) if the targeted fine volume (layer) is within it. It is the most straightforward interpolation one may have.

The restriction, which is the reverse procedure of the interpolation, has a similar expression to interpolation:
\begin{equation}
    Q^c_{i,j,k}=R^{x}_{il}R^y_{jm}R^z_{kn} Q^f_{l,m,n},
\end{equation}
where we label the restriction matrix as $R$. Similarly, we define the matrix $R^\parallel$, which maps $3p$ fine volumes to $p$ coarse volumes, and the matrix $R^\perp$, mapping 6 fine layers to 3 coarse layers. 

In the code practices, we incorporate two versions of matrix $R^\parallel$: one for the injection (constant) scheme and the other for the average scheme. However, only one version of $R^\perp$ is implemented in the code as the corresponding scheme of zero-order accuracy yields significant numerical errors.

The matrix $R^\parallel$ for the injection scheme is
\begin{equation}
R_\text{injection}^\parallel=\left(
\begin{smallmatrix}
    0 & 1 & 0 & 0 & 0 & 0 & 0 & 0 & 0 &...\\
    0 & 0 & 0 & 0 & 1 & 0 & 0 & 0 & 0 &...\\
    0 & 0 & 0 & 0 & 0 & 0 & 0 & 1 & 0 &...\\
    0 & 0 & 0 & 0 & 0 & 0 & 0 & 0 & 0 &...\\
      &   &   &   & ... & &   &   & 
\end{smallmatrix}\right).
\end{equation}
In this scheme, we utilize the value in the central volume per three fine volumes for the coarse volume, as this central volume has the same central coordinate (in this parallel direction) as the restricted coarse volume. The matrix $R^\parallel$ for the average scheme is 
\begin{equation}
R_\text{average}^\parallel=\left(
\begin{smallmatrix}
    1/3 & 1/3 & 1/3 & 0 & 0 & 0 & 0 &...\\
    0 & 0 & 0 & 1/3 & 1/3 & 1/3 & 0 &...\\
    0 & 0 & 0 & 0 & 0 & 0 & 1/3 &...\\
      &   &   &   &   ... & &   &   &      
\end{smallmatrix}\right).
\end{equation}
The average of the three corresponding fine volumes is used to assign the coarse volume and provide improved accuracy. 

The version of matrix $R^\perp$ implemented in the code is
\begin{equation}
R^\perp=\left(
\begin{smallmatrix}
    0 & 0 & 0 & 1/3 & 1/3 & 1/3\\
    0 & 0 & 0 & 0 & -2 & 3\\
    0 & 0 & 0 & 0 & -5 & 6\\
\end{smallmatrix}\right).
\end{equation}
This matrix uses a combination of average and extrapolation for the map along the face normal. Notice the enumeration depends on the face orientation rather than relative position. So the right half of the matrix instead of the left half represents the inner part of the fine face (layers f3, f4 and f5). 





\end{document}